\documentclass[epj,twocolumn]{svjour}
\usepackage{epsfig,amssymb,amsfonts,bm,color}

\newcommand{\be}{\begin{equation}}
\newcommand{\ee}{\end{equation}}
\newcommand{\bea}{\begin{eqnarray}}
\newcommand{\eea}{\end{eqnarray}}
\newcommand{\beas}{\begin{eqnarray*}}
\newcommand{\eeas}{\end{eqnarray*}}
\newcommand{\ds}{\displaystyle}

\newcommand{\ov}{\bar}
\newcommand{\non}{\nonumber}

\newcommand{\pp}{\pi^+\pi^-}

\begin{document}

\title{Hadron physics potential of future high-luminosity $B$-factories at the $\Upsilon(5S)$ and above}

\author{A.G.~Drutskoy\inst{1,2} \and F.-K.~Guo\inst{3} \and F.J.~Llanes-Estrada\inst{4} \and
A.V.~Nefediev\inst{1,2,5,6} \and J.M.~Torres-Rincon\inst{4}}

\institute{Institute of Theoretical and Experimental Physics, 117218, Moscow, Russia \and
National Research Nuclear University MEPhI, 115409, Moscow, Russia \and
Helmholtz-Inst. f{\"u}r Strahlen-und Kernphysik \& Bethe Center for Theoretical Physics, Univ.
Bonn, D-53115 Bonn, Germany \and
Depto. Fisica Teorica I, Universidad Complutense Madrid, 28040 Madrid, Spain \and
Moscow Institute of Physics and Technology, 141700, Dolgoprudny, Moscow Region, Russia \and
All-Russia Research Institute of Automatics (VNIIA), 127055, Moscow, Russia
}

\date{Received: date / Revised version: date}
\abstract{We point out the physics opportunities of future high-luminosity $B$-factories
at the $\Upsilon(5S)$ resonance and above. Currently the two $B$-factories,
the SuperB factory in Tor Vergata, Italy and the Belle II factory in KEK,
Japan, are under development and are expected to start operation
in 2017 and 2016, respectively.
In this paper we discuss numerous interesting investigations,
which can be performed in the $e^+ e^-$ center-of-mass energy region
from the $\Upsilon(5S)$ and up to 11.5 GeV, where an efficient
data taking operation should be possible with the planned $B$-factories.
These studies include abundant $B_s$ production and decay properties; independent confirmation and if found, exhaustive exploration of Belle's claimed charged bottomonia;  clarification of puzzles of interquarkonium dipion transitions; extraction of the light quark mass ratio from hadronic $\Upsilon(5S)$ decays; analysis of quarkonium and exotic internal structure from open flavour decays, leading to severe $SU(3)$ symmetry violations; clarification of whether a hybrid state has similar mass to the $\Upsilon(5S)$ bottomonium, making it a double state;
searches for molecular/tetraquark states that should be more stable with heavy quarks;
completion of the table of positive-parity $B_J$ mesons and study of their basic properties; production of $\Lambda_b\bar{\Lambda}_b$ heavy baryon pairs, that, following weak decay, open vistas on the charmed baryon spectrum and new channels to study CP violation; confirmation or refutation of the deviation from pQCD of the pion transition form factor, by extending the $Q^2$ reach of current analysis; and possibly reaching the threshold for the production of triply-charmed baryons. If, in addition, the future colliders can be later upgraded to 12.5  GeV, then the possibility of copious production of $B_c\bar{B}_c$ pairs opens, entailing new studies of CP violation and improved, independent tests of the CKM picture (through determination of $V_{bc}$), and of effective theories for heavy quarks.}

\maketitle

\tableofcontents

\section{Introduction}

In the last years we have witnessed a renaissance of hadron spectroscopy caused by
a series of discoveries made at $B$-factories. Before $B$-factories, most
of the known states in the spectrum of charmonium and bottomonium
lay below the open-flavour thresholds, a well-developed phenomenology based on
potential quark models was quite successful in describing properties of these
states, and, in particular, threshold effects could be cast into constant mass
shifts. However the situation changed dramatically after $B$-factories started
operation and data collection. A lot of new states were observed,
the majority of them lying above the open-flavour thresholds in the spectra of
both charmonium and bottomonium. Many such states in fact are close to various
thresholds, so that the threshold effects appear to be the dominant feature, and a
drastic departure from the Breit--Wigner form in line shapes is expected,
especially when the coupling to the corresponding open-flavour channels is in an
$S$ wave. At the same time, a lot of theoretical ideas were put forward to explain
the data.  As a result, while some experimental observations remain not yet well
understood theoretically, a large number of theoretical predictions do call for
improvements in the experimental situation, which are expected to be achieved at
the next-generation $B$-factories, such as SuperB and Belle II.

This paper is mostly devoted to the heavy quarkonium studies at the centre-of-mass energy above
the $\Upsilon$(4S). We will review available experimental results and different
theoretical models applicable in the heavy quarkonium physics with a significant
emphasis on unresolved theoretical issues. The potentially
important experimental studies in this area, feasible for the future high-luminosity $B$-factories, will also be
discussed.
In particular, the SuperB experiment plans to collect during several years of running about 75~ab$^{-1}$
at the $\Upsilon$(4S) and a few ab$^{-1}$ at the $\Upsilon$(5S) and, probably,
at the $\Upsilon$(6S) and around.
With expected integrated luminosity
significantly larger than previously available at $B$-factories,
such experiment will become a valuable source of experimental data which not only will allow
to improve statistics for previously studied reactions, but it will enable to study
various rare decays and reactions, not achievable at present.

A large part of our discussion will be the isoscalar vector
hidden-bottom state that resides in this region, with mass and width, according to
recent measurements \cite{Aubert:2008ab},
\be
\label{eq:Mass5S} M=10876\pm
2~\mbox{MeV},\quad \Gamma=43\pm 4~\mbox{MeV}.
\ee
In the framework of the quark
model, this state is a conventional $n^{2S+1}L_J=5^3S_1$ $b\bar{b}$ meson, where
$n$ is the radial quantum number, while $S$, $L$, and $J$ denote the quark spin,
the  quark--antiquark angular momentum, and the total spin, respectively\footnote{$S$-$D$ wave mixing is
discussed below in chapers~\ref{raddec} and \ref{Rydberg}.}. In what
follows we therefore refer to the corresponding state as $\Upsilon(5S)$.
However, we discuss in addition other possible non-$b\bar{b}$ vector resonances
residing in this energy region.

We will then extend the discussion to larger energies, particularly the $\Upsilon(6S)$ resonance, but also several interesting thresholds expected above it.
To keep the article within a manageable size we have limited ourselves to brief comments on most of the topics. We hope that the references given will be a reasonable starting point for the interested reader.

Measurements that can be mostly conducted at the $\Upsilon(4S)$, and much of the flavour discussion involving $B_s$ decays, have been the object of other studies and
different theoretical approaches, so we have purposefully avoided dwelling on them here.
We have included some discussion on quarkonium states, but the reader may want to refer to the quarkonium working group report~\cite{Brambilla:2010cs} for extended coverage of many topics.

\section{Closed-flavour analysis at {\boldmath $\Upsilon(5S)$}}

\subsection{Dipion transitions from {\boldmath $\Upsilon(5S)$}}

The $\Upsilon(5S)$, well above open-bottom threshold, decays ma\-in\-ly to open-flavour
channels. Closed-flavour decays are however also easily observable, and those with
the largest branching fractions are dipion transitions into other bottomonium
states. They are an important source of information on the nature of this state.
Two types of dipion transitions
can be identified: transitions without spin flip
($\Upsilon(5S)\to\Upsilon(nS)\pi\pi$ with $n<5$) and transitions with spin flip
($\Upsilon(5S)\to h_b(nP)\pi\pi$ with $n=1,2$).

\subsubsection{Dipion transitions without {\boldmath $b$}-quark spin flip}

\begin{figure}[t]
\begin{center}
\includegraphics*[width=0.48\textwidth]{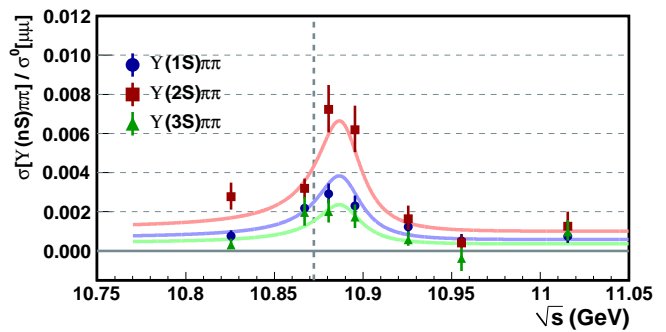}
\end{center}
\caption{The ratio $\sigma[e^+e^- \to \Upsilon(nS)\pi^+ \pi^-]/\sigma[e^+e^- \to \mu^+\mu^-]$  for $n=1,2,3$, as measured by Belle \cite{Chen:2008xia}.
The vertical dashed line marks the position of the inclusive hadronic cross section maximum.}\label{fig:pipibelle}
\end{figure}

\begin{figure*}[ht]
\begin{center}
\includegraphics*[width=0.3\textwidth]{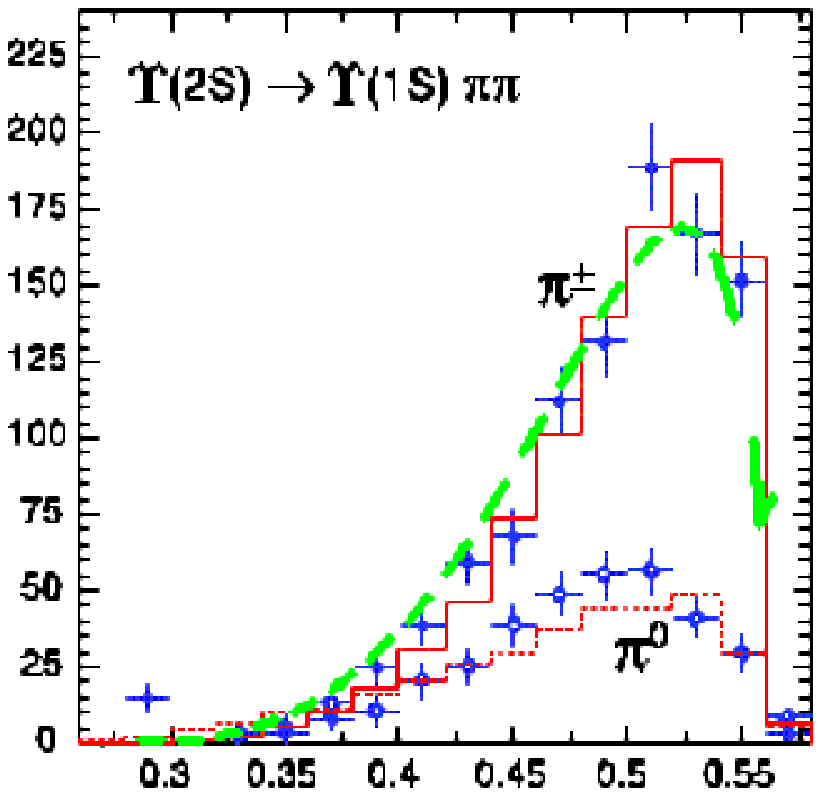}
\includegraphics*[width=0.3\textwidth]{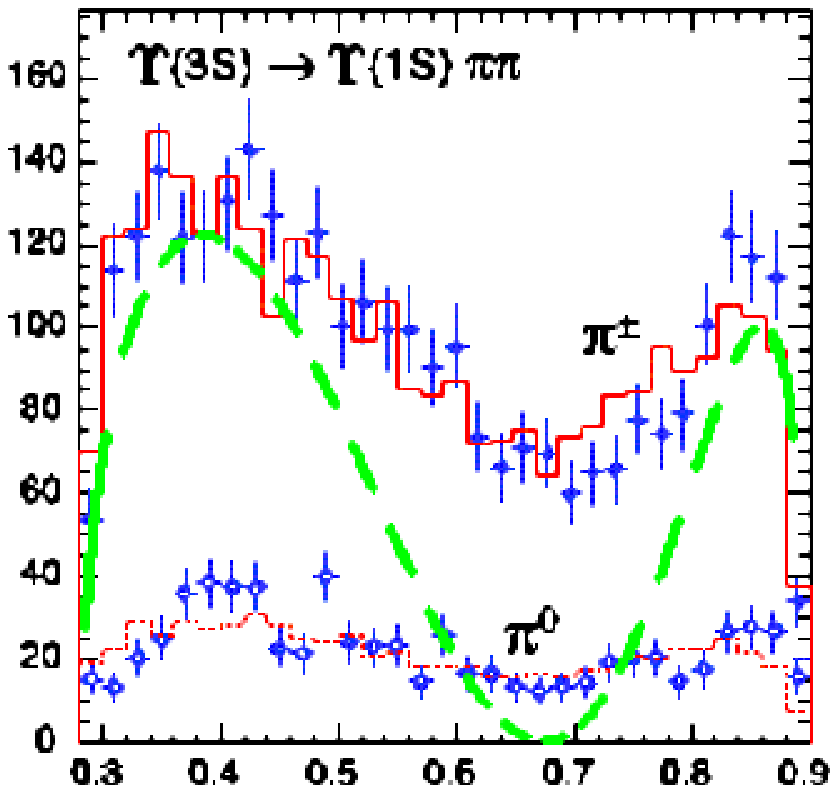}
\includegraphics*[width=0.3\textwidth]{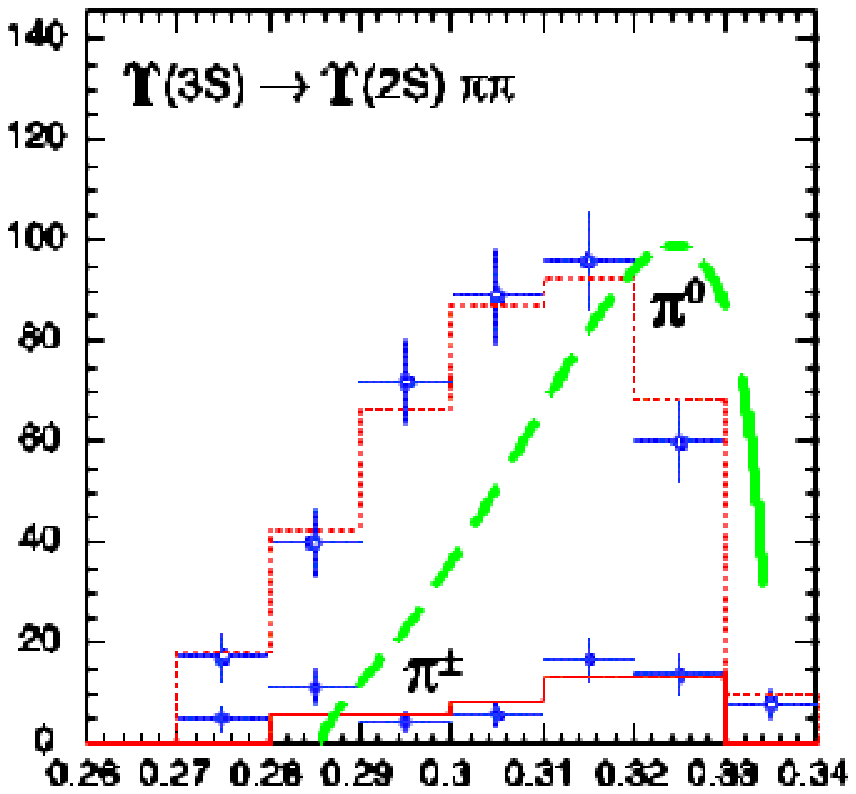}
\end{center}
\caption{CLEO Collaboration data for dipion transitions from $\Upsilon(2S)$ and
$\Upsilon(3S)$ states \cite{CLEO2007}. Theoretical results from \cite{Simonov:2008qy} are shown with the dashed green line.}\label{fig:3Spipiexp}
\end{figure*}

Dipion transitions $e^+e^- \to \Upsilon(nS)\pi^+ \pi^-$ with $n=1,2,3$ measured by
Belle \cite{Chen:2008xia} peak at the same centre-of-mass energy (see
Fig.~\ref{fig:pipibelle}), \be \mu=[10888.4_{-2.6}^{+2.7}\pm 1.2]~\mbox{MeV},\quad
\Gamma=[30.7^{+8.3}_{-7.0}\pm 3.1]~\mbox{MeV}, \ee and the deviation of this
quantity (more than $2\sigma$) from the maximum position in the hadronic cross
section (vertical dashed line in Fig.~\ref{fig:pipibelle}) has been
discussed theoretically assuming different interpretations. In particular,
$B^{(*)}$-$\bar B^{(*)}$ rescattering was suggested in \cite{Meng:2008dd} as
a possible mechanism responsible for the peak shift in the $\Upsilon(nS)\pp$ dipion
invariant mass distribution.

Another interesting feature of the distribution shown in Fig.~\ref{fig:pipibelle}
is an almost zero cross section level outside the $\Upsilon(5S)$ resonance
region. It indicates that the $\Upsilon(nS)\pp$ states cannot be produced
in the continuum, but only through the resonance formation
mechanism. In contrast, the $b\bar{b}$ production cross section level
is rather high at all energies below the $B\ov{B}$ mass threshold.
This difference remains to be explained theoretically.
Measurements of the $B^*\ov{B}^*$ exclusive channel cross
section as a function of the centre-of-mass energy may shed some light on this
difference as well as on the discussed above mass deviation.

Furthermore, Belle reported \cite{Abe:2007tk} observation of an unexpectedly large
$\Upsilon(nS)\pp$ ($n=1,2,3$) production at the $\Upsilon(5S)$ energy comparing with one
at the $\Upsilon(4S)$. In
particular, the measured values
\bea
&&\Gamma[\Upsilon(5S)\to\Upsilon(1S)\pp]=[0.59\pm 0.04\pm 0.09]~\mbox{MeV},\nonumber\\
&&\Gamma[\Upsilon(5S)\to\Upsilon(2S)\pp]=[0.85\pm 0.07\pm 0.16]~\mbox{MeV},\nonumber\\
&&\Gamma[\Upsilon(5S)\to\Upsilon(3S)\pp]=[0.52^{+0.20}_{-0.17}\pm 0.10]~\mbox{MeV},
\eea
appear to be by about two orders of magnitude larger than
similar dipion decays of $\Upsilon(nS)$, with $n<4$~\footnote{Typical values of
the widths for the decays $\Upsilon(nS)\to\Upsilon(n'S)\pp$, with $n'<n<4$, appear
to be of order a few keV \cite{Kuang:1981se,CLEO2007,Aubert:2006bm}, in agreement
with theoretical predictions \cite{Simonov:2008ci,Simonov:2008qy}.}. A
tetraquark explanation of such an anomalous production suggests that the final states
are produced from a tetraquark with a mass of 10890~MeV, rather than from the
$\Upsilon(5S)$~\cite{tetraquark2,Ali:2010pq}. Another possible mechanism is related
to strong interference between the direct production and the final state
interactions~\cite{Chen:2011qx,Chen:2011zv}.

\begin{figure*}[ht]
\begin{center}
\begin{tabular}{c}
\includegraphics*[width=0.35\textwidth]{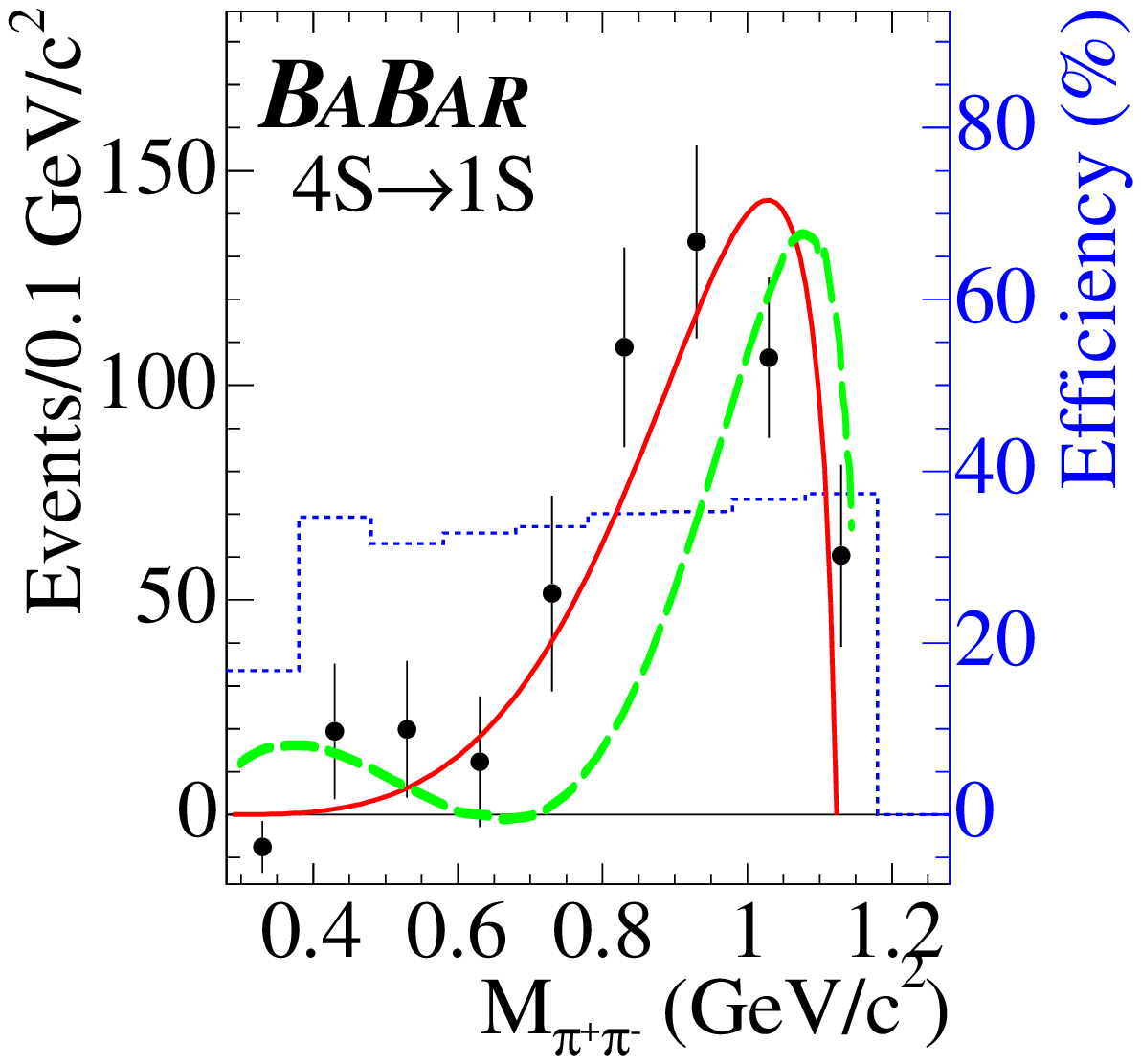}\hspace*{10mm}
\includegraphics*[width=0.35\textwidth]{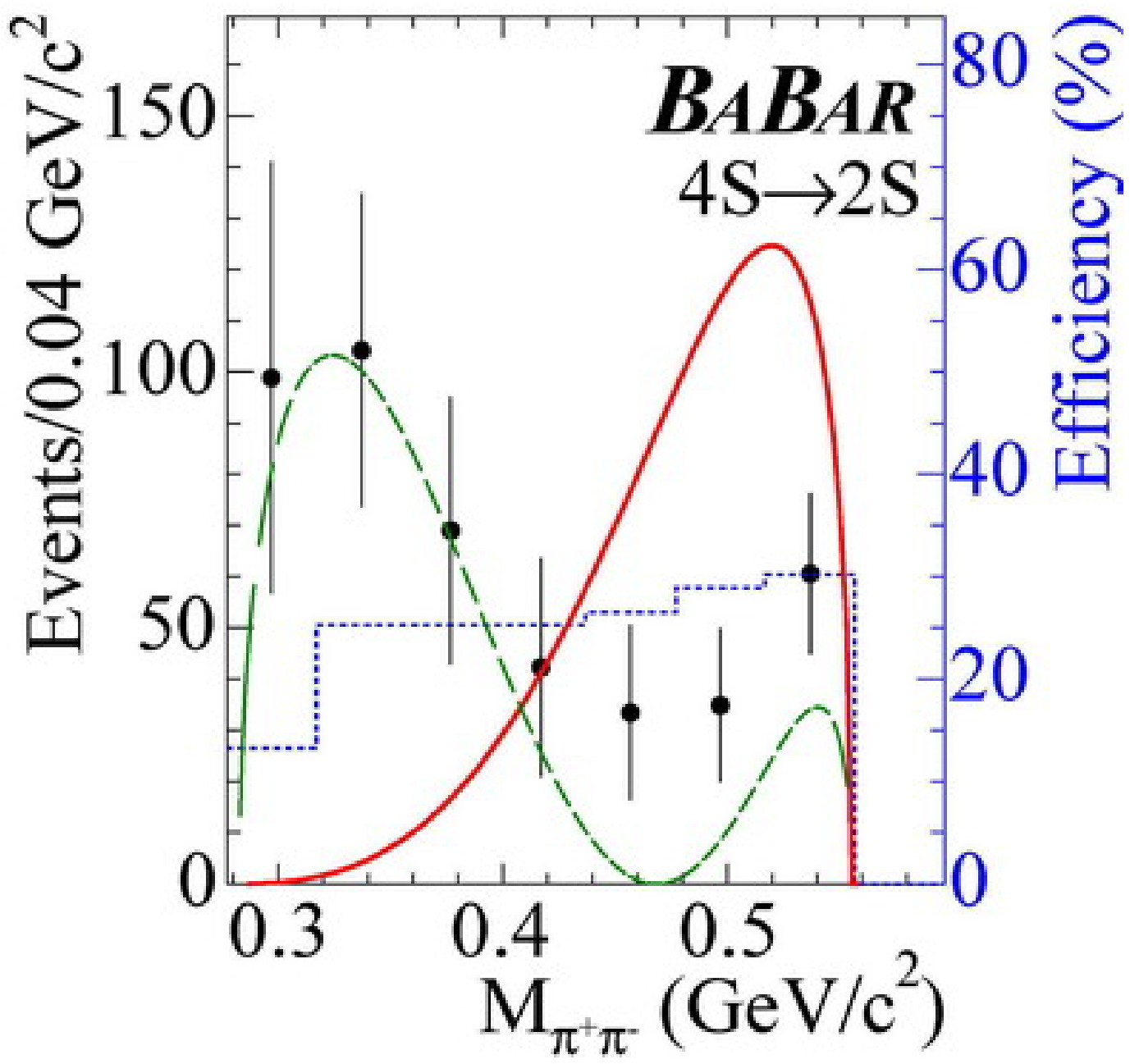}
\end{tabular}
\end{center}
\caption{BABAR Collaboration data for dipion transitions from $\Upsilon(4S)$ state \cite{Aubert:2006bm}. Theoretical results from \cite{Simonov:2008qy} are shown with the dashed green line.}\label{fig:4Spipiexp}
\end{figure*}

\begin{figure*}[ht]
\begin{center}
\begin{tabular}{c}
\includegraphics*[width=0.3\textwidth]{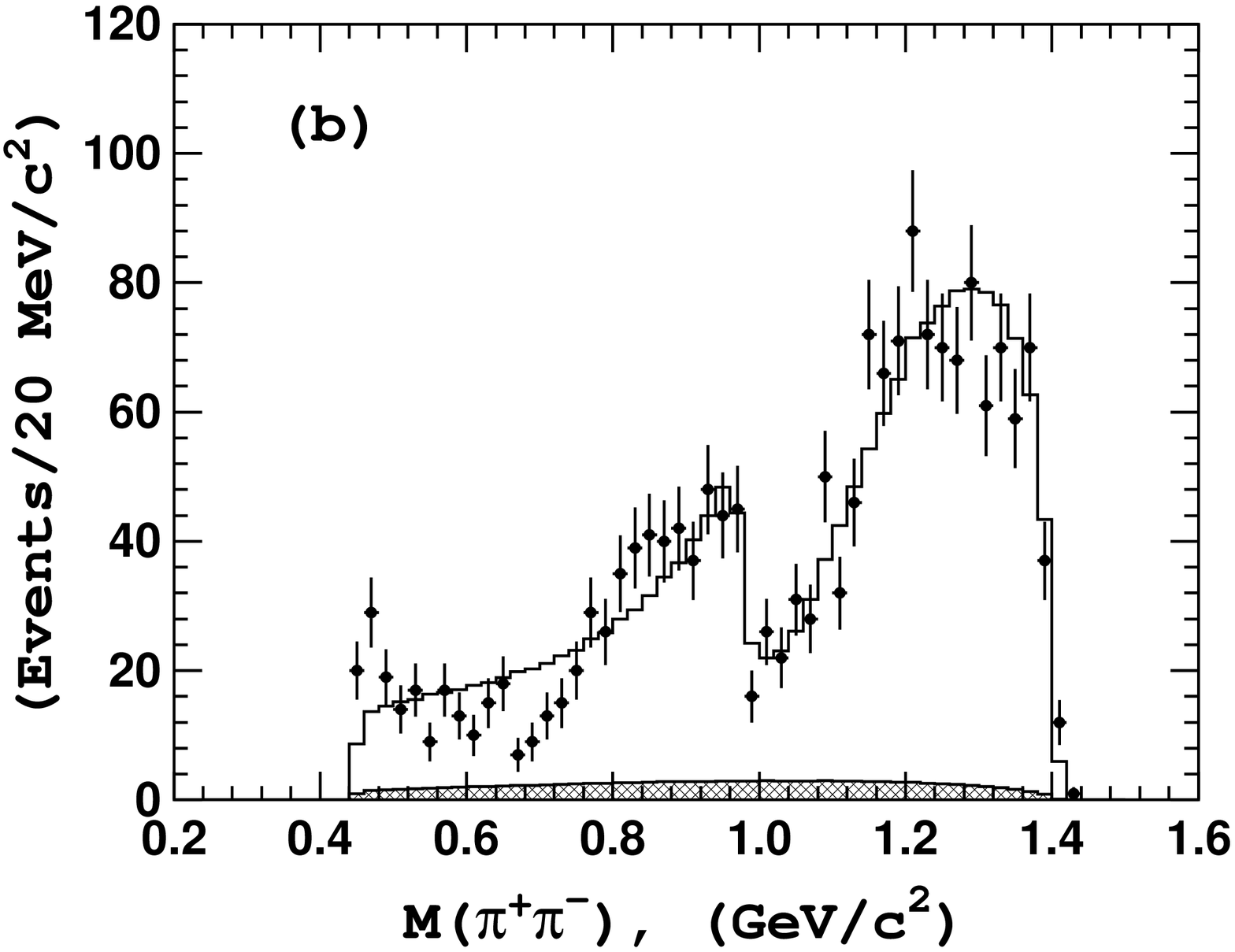}
\includegraphics*[width=0.3\textwidth]{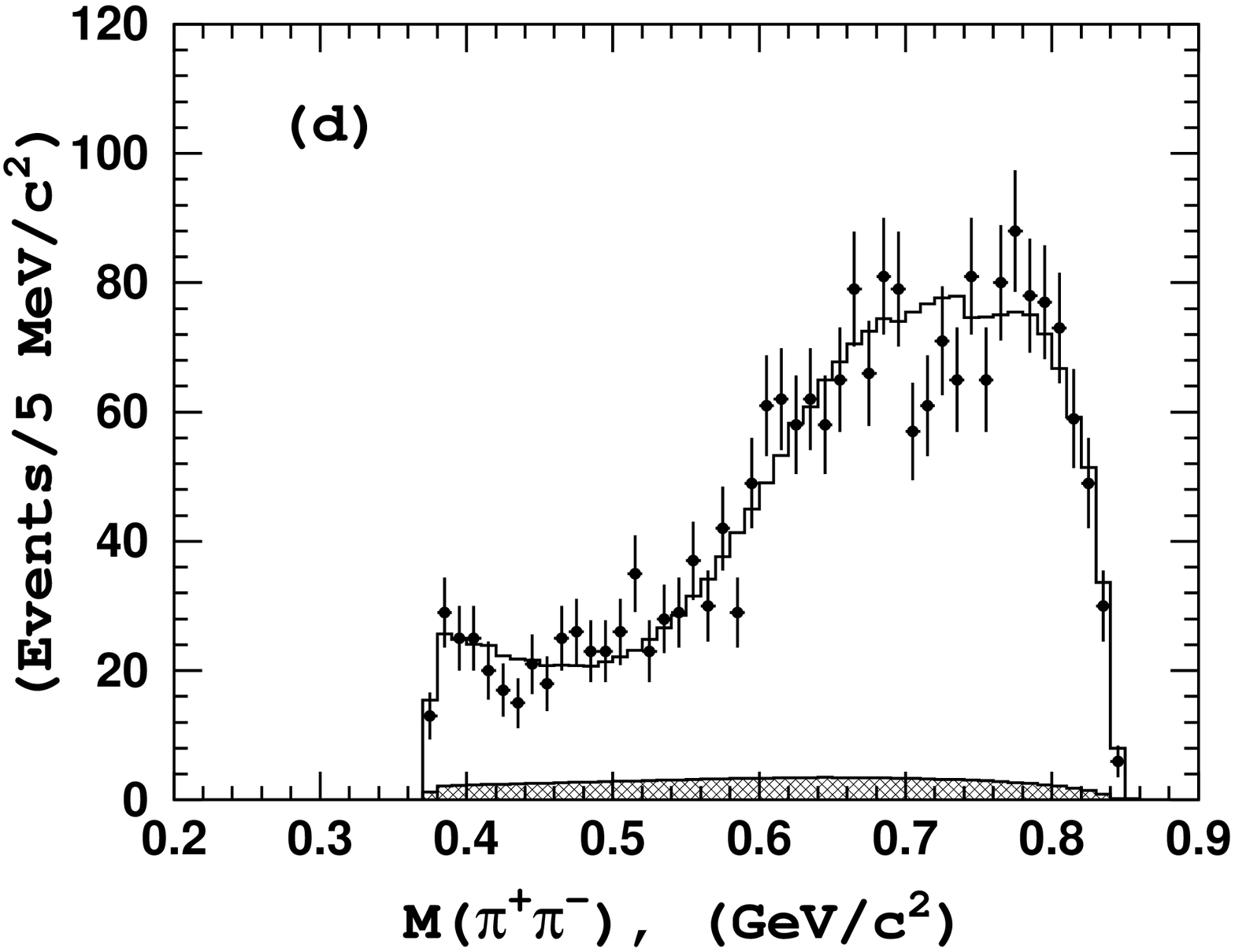}
\includegraphics*[width=0.3\textwidth]{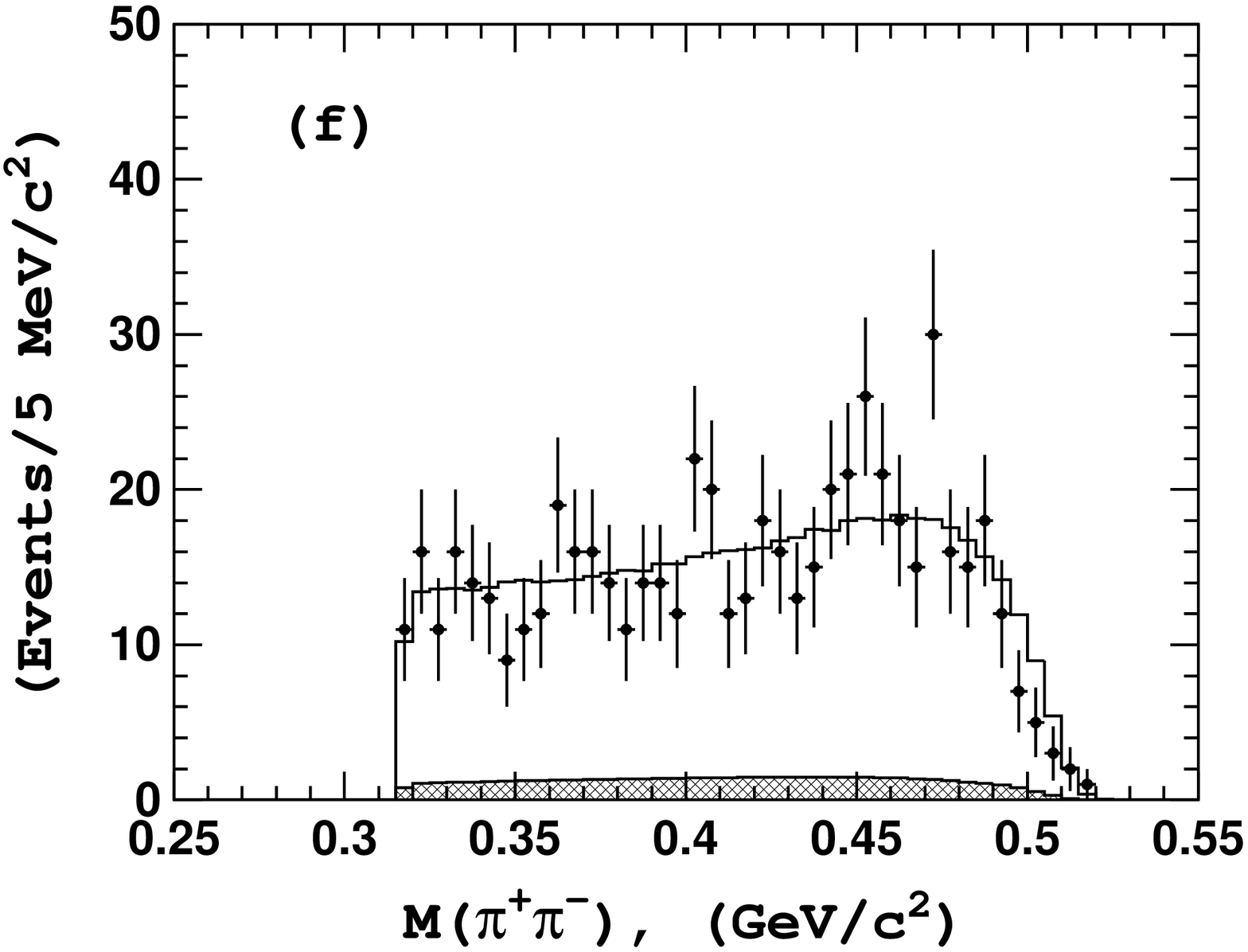}
\end{tabular}
\end{center}
\caption{Belle Collaboration data for dipion transitions from $\Upsilon(5S)$ state to $\Upsilon(1S)\pp$ (first plot),
$\Upsilon(2S)\pp$ (second plot), and $\Upsilon(3S)\pp$ (third plot)
\cite{Adachi:2011gj,Belle:2011aa}.}\label{fig:5Spipiexp}
\end{figure*}

Another question related to the dipion transitions from $\Upsilon(nS)$ states is
the  shape of the dipion invariant mass spectrum. Before the observation of the
$\Upsilon(4S)$ and $\Upsilon(5S)$ dipion decays, there was a question about the
peculiar dipion invariant mass spectrum of the $\Upsilon(3S)\to\Upsilon(1S)\pp$
transition \cite{CLEO2007,CLEO1994}. In analogous heavy quarkonium transitions,
such as $\psi'\to J/\psi\pp$ and $\Upsilon(2S)\to\Upsilon(1S)\pp$, the $\pp$
invariant mass distribution shows a single broad peak towards the higher end of the
phase space (see, for example, the first plot in Fig.~\ref{fig:3Spipiexp}). In the
$\Upsilon(3S)\to\Upsilon(1S)\pp$ transition, however, the structure is quite
different, with two bumps showing up (the second plot in Fig.~\ref{fig:3Spipiexp}).
Similar double-bump structures were observed later in higher $\Upsilon(nS)$ dipion
transitions (see Figs.~\ref{fig:4Spipiexp} and \ref{fig:5Spipiexp}). In particular,
the $\pp$ invariant mass distribution for the $\Upsilon(4S)\to\Upsilon(1S)\pp$
transition was measured first by the Belle Collaboration~\cite{Belle4S},
and later on by the BaBar Collaboration, together with the $\Upsilon(4S)\to\Upsilon(2S)\pp$
~\cite{Aubert:2006bm}. Finally, measurements of the
$\Upsilon(5S)\to\Upsilon(nS)\pp$ transitions, with $n<4$, were reported by the
Belle Collaboration~\cite{Abe:2007tk,Adachi:2011gj,Belle:2011aa}.

Compounding this problem, one should notice that the kinematically
allowed phase space is quite different in various transitions. For
$\Upsilon(2S)\to\Upsilon(1S)\pp$, the dipion invariant mass is limited up to about
560~MeV, similarly to the analogous charmonium transition $\psi(2S)\to J/\psi\pp$. On
the other hand, most of the transitions having a more complicated dipion invariant
mass distribution have larger phase space, with the exception of the
$\Upsilon(4S)\to\Upsilon(2S)\pp$.

 From the theory point of view, these double peak
structures were studied by a number of authors, for reviews,
see~\cite{VoloshinReview,KuangReview,Guo2004}. The problem was further investigated
in recent years (see, for
example,~\cite{Simonov:2008ci,Simonov:2008qy,VoloshinNew}). Among various
explanations, it is interesting to notice the proposal of an isovector exotic
$b\bar b q\bar q$ state in the bottomonium mass region~\cite{Guo2004,X1,X2}.

A small bump close to the lower end of the mass distribution in the
$\Upsilon(4S)\to\Upsilon(1S)\pp$ was noticed in~\cite{Guo2006}. There has been no
agreement whether the $S$-wave $\pi\pi$ final state interaction (in agreement with
the strong $f_0(500)$ (or $\sigma$) signal in the reaction $J/\psi\to
\omega\pi\pi$~\cite{Ablikim:2004qna}), or rather a relativistic correction, is the
key to understanding the structure in the region 400-600 MeV.

\begin{figure*}[t]
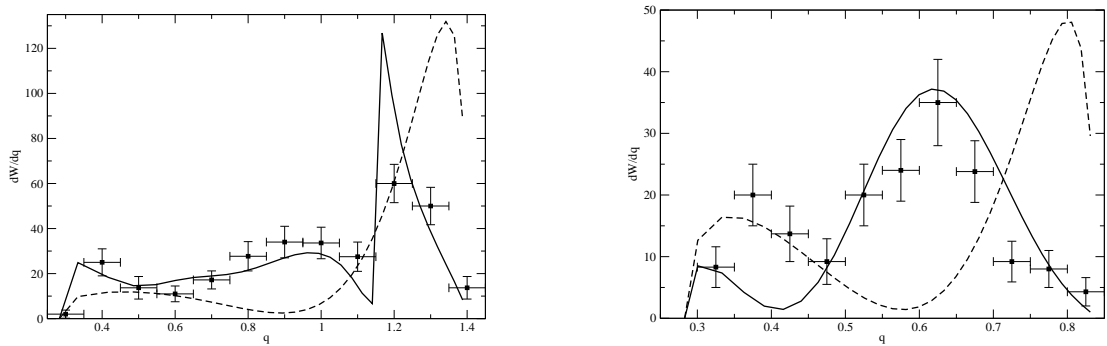

\begin{center}
\begin{tabular}{ccc}
\includegraphics*[width=0.35\textwidth]{FIGS.DIR/q51.eps}&\hspace*{10mm}&\includegraphics*[width=0.35\textwidth]{FIGS.DIR/q52.eps}
\end{tabular}
\end{center}
\caption{The dipion invariant mass spectrum for the decays $\Upsilon(5S)\to\Upsilon(nS)\pp$ with $n=1,2$ ($\Gamma_{\pp}=\int(dw/dq)dq$, $q\equiv M_{\pp}$). Data are from \cite{Abe:2007tk}, while theoretical curves are from \cite{Simonov:2008ci}. The solid line corresponds to the FSI taken into account properly.}\label{fig:wpipiex2p}
\end{figure*}

When the emitted pions are soft,
an effective chiral Lagrangian can be used, with first steps taken already at the
time of the charm discovery~\cite{Brown:1975dz}. At lowest order each
$\Upsilon(n)\to \Upsilon(m) \pi\pi$ channel requires four unknown
parameters~\cite{Mannel:1995jt} that cannot be measured elsewhere, so that the
predictive power is very limited.

Another issue is the importance of open-bottom coupled
channels~\cite{Lipkin:1988tg,Moxhay:1988ri,Zhou:1990ik}. They  cannot be simply
neglected in the effective field theory, because the gap between heavy quarkonium
and open-flavour heavy meson thresholds is often small compared with the hard scale
of 1~GeV. Assuming these thresholds are more important than the contact terms
described by the chiral Lagrangian\footnote{This could be checked using the
nonrelativistic effective field theory proposed in \cite{FK:NREFT1,FK:NREFT2}.}, one may construct a predictive
formalism.

\begin{figure*}[t]
\begin{center}
\begin{tabular}{ccc}
\includegraphics*[width=0.35\textwidth]{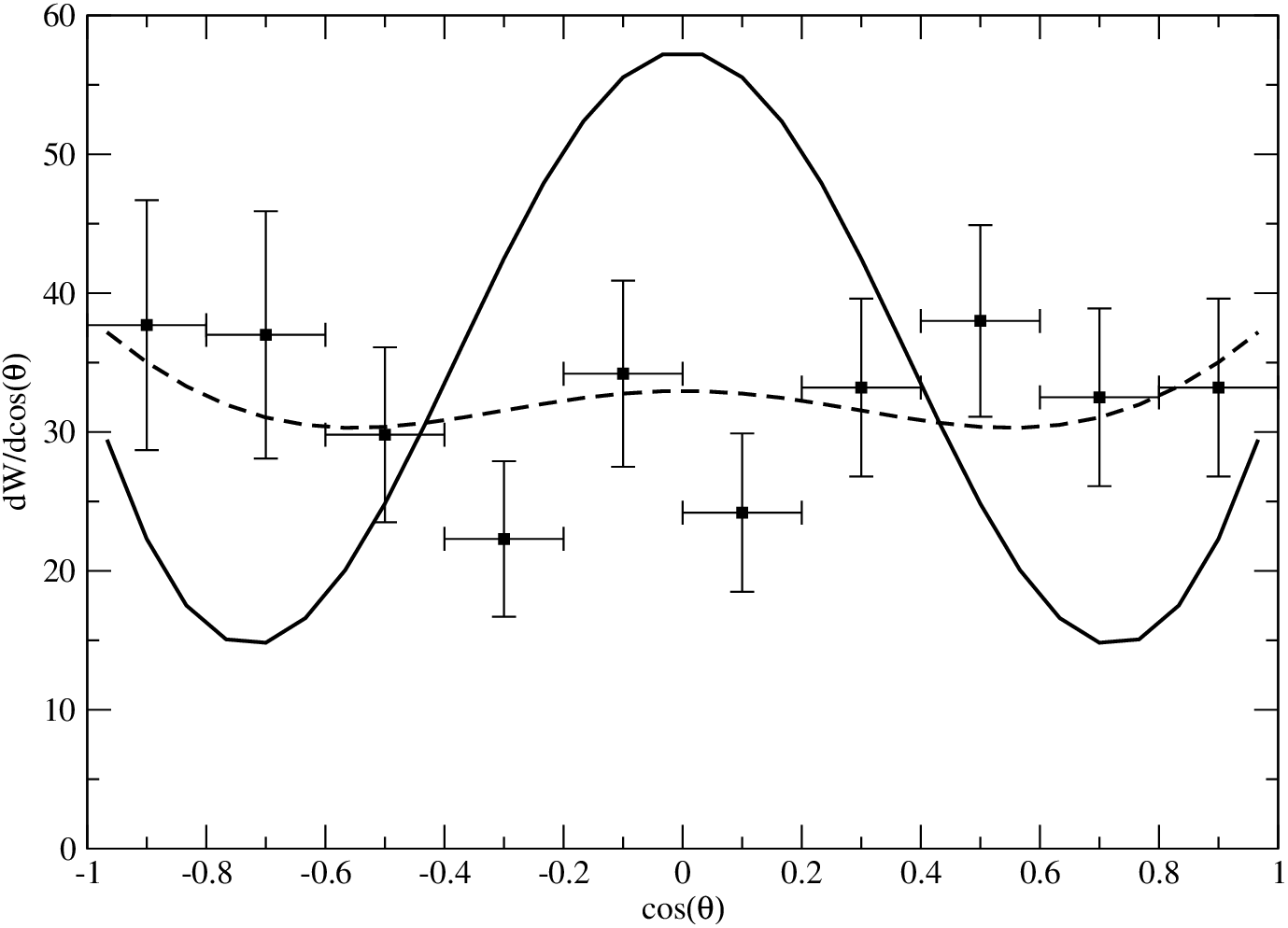}&\hspace*{10mm}&
\includegraphics*[width=0.35\textwidth]{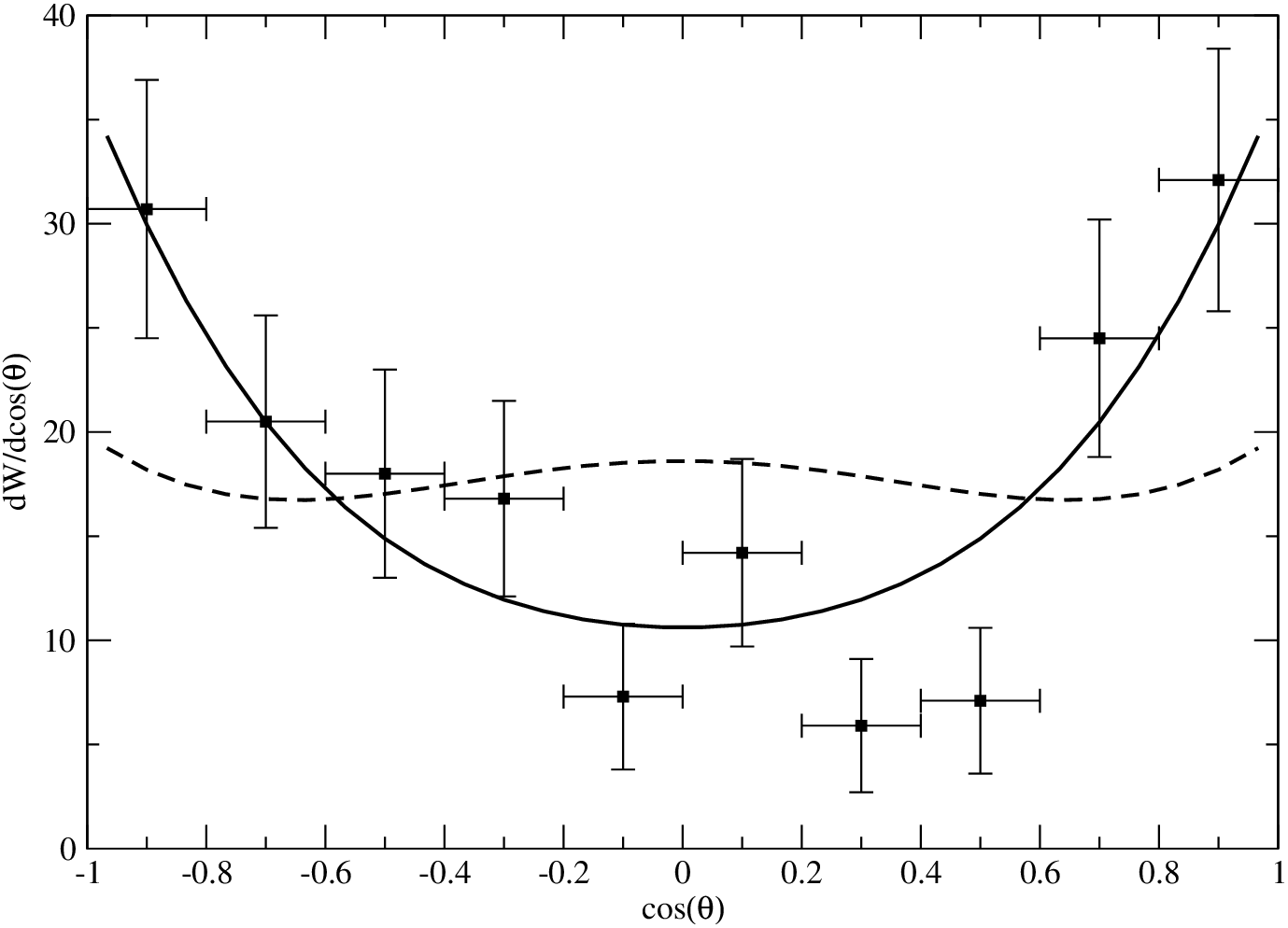}
\end{tabular}
\end{center}
\caption{The angular distributions $dw/d\cos\theta$ for the transitions $\Upsilon(5S)\to\Upsilon(1S)\pp$ (upper plot) and $\Upsilon(5S)\to\Upsilon(2S)\pp$ (adapted from \cite{Simonov:2008ci}) compared to the data from \cite{Abe:2007tk}. Theoretical predictions are given without FSI (dashed line) and with FSI (solid line) included.}\label{fig:tpipiexp}
\end{figure*}

In fact, theorists have provided~\cite{Simonov:2008ci,Simonov:2008qy} a
parameter-free description of the dipion spectra stressing the role of the Adler Zero Requirement (AZR) imposed
on the amplitude, as well as that of the Final State Interaction (FSI). Predictions of the approach \cite{Simonov:2008ci,Simonov:2008qy}
are shown in Figs.~\ref{fig:3Spipiexp} and \ref{fig:4Spipiexp} with the green
dashed line, as well as in Fig.~\ref{fig:wpipiex2p}. In particular, dips in the
$\pp$ spectrum are explained by the AZR which suppresses the corresponding
amplitude at small pion momenta. In the meantime, the peak structure at the lower
end of the dipion invariant mass spectrum (see Fig.~\ref{fig:wpipiex2p}) is
explained in \cite{Simonov:2008ci} to be due to an interplay of the FSI and AZR.
Then the angular distributions $dw/d\cos\theta$ ($\theta$ is the angle between the
initial $\Upsilon$ and the $\pi^+$) are presented in
\cite{Simonov:2008ci,Simonov:2008qy} for various $\Upsilon(nS)\to\Upsilon(n'S)\pp$
transitions and are compared with the data from \cite{Abe:2007tk,CLEO2007} (see
Fig.~\ref{fig:tpipiexp}).

\begin{figure}[t]
\begin{center}
\begin{tabular}{c}
\includegraphics*[width=6.5cm]{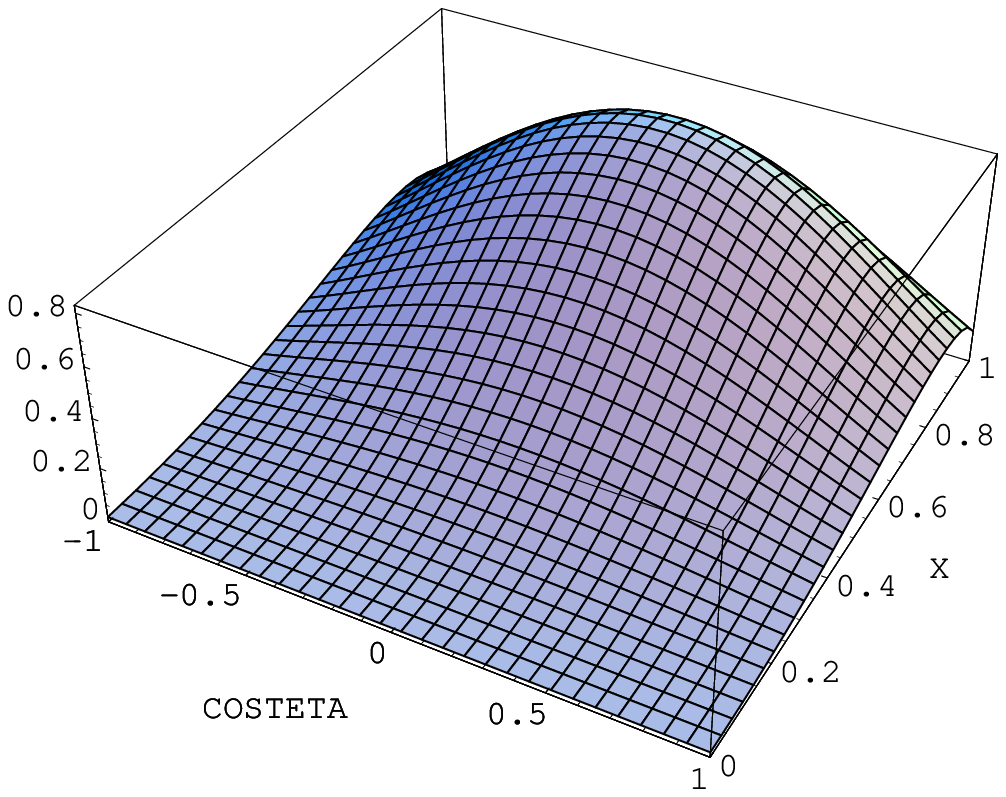} \\
\includegraphics*[width=6.5cm]{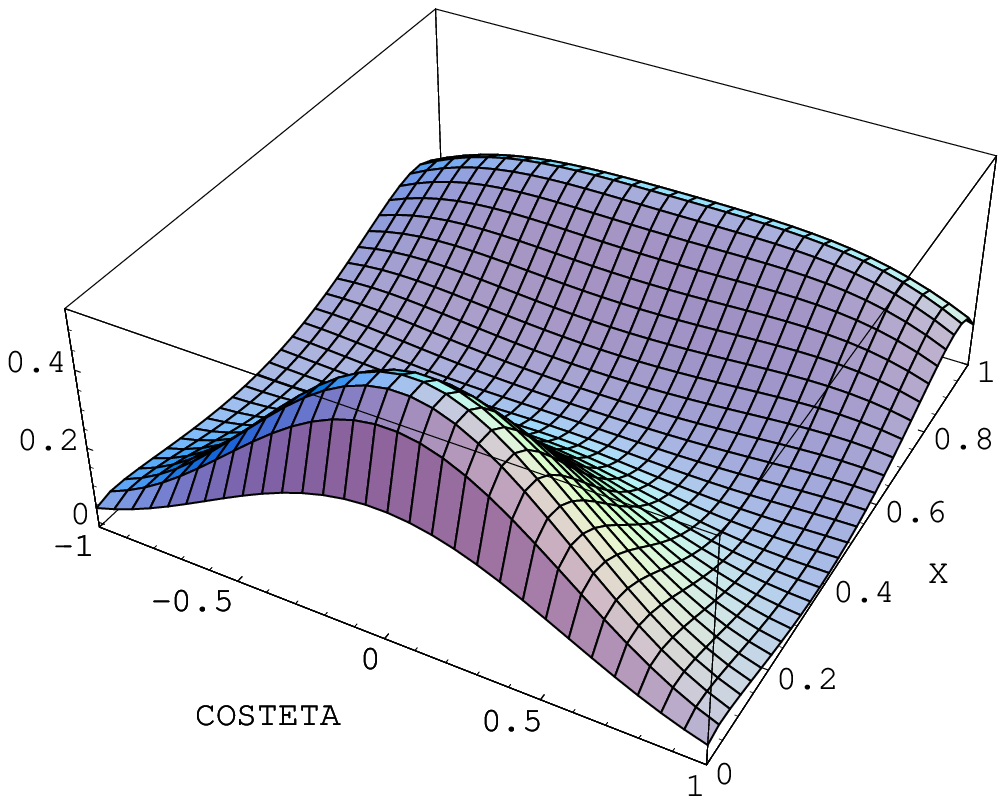}
\end{tabular}
\end{center}
\caption{The differential decay probability $dw/(dxd\cos\theta)$ for the transitions $\Upsilon(2S)\to\Upsilon(1S)\pp$ (upper plot) and $\Upsilon(3S)\to\Upsilon(1S)\pp$ (adapted from \cite{Simonov:2008qy}). Further examples can be found in \cite{Simonov:2008qy}.}\label{fig:wpipiexp}
\end{figure}

In addition, three-dimensional plots for the differential decay
probability $dw/(dxd\cos\theta)$ for various dipion $\Upsilon(nS)\to\Upsilon(n'S)\pp$ decays
were suggested and presented in \cite{Simonov:2008qy} (see examples shown in
Fig.~\ref{fig:wpipiexp}), where the dimensionless variable $x\in [0,1]$ is defined
as
$$
x=\frac{M_{\pp}^2-4m_\pi^2}{[M(\Upsilon(nS))-M(\Upsilon(n'S))]^2-4m_\pi^2}.
$$

While general features of the data have been correctly captured, there remain
details to be understood. In the second panel of Fig.~\ref{fig:3Spipiexp}, the dip
predicted in the $\Upsilon(3S)\to\Upsilon(1S)\pi\pi$ is deeper than experimental
data. A similar result was also found in the effective Lagrangian formalism taking
into account the $\pi\pi$ FSI~\cite{Guo2004}, where the dip was made consistent
with the data with the help of an additional theorised isovector $b\bar b q\bar q$
state. Indeed two exotic isovector structures $Z_b(10610)^\pm$ and $Z_b(10650)^\pm$
have been reported in $\Upsilon(5S)$ decays, see Sec.~\ref{Zbstories}, in all of
the $\Upsilon(1S)\pi$, $\Upsilon(2S)\pi$ and $\Upsilon(3S)\pi$ dipion modes. It is
important therefore to revisit the lower $\Upsilon$ transitions, including now the
effect of possible $Z_b$ states, and to perform a systematic study of all the data
for $\Upsilon(nS)\to\Upsilon(mS)\pp$ transitions to understand the true mechanism
of these decays.

Experimentally, it would be very helpful to measure the dipion and also the
$\Upsilon(nS)\pi$ invariant mass distributions in the
$\Upsilon(5S)\to\Upsilon(nS)\pp$ precisely, so that one can extract the information
of coupling of the $Z_b$ states to the $\Upsilon(nS)\pi$. The information can then
be fed back to the lower $\Upsilon(nS)$ dipion transitions, eventually providing a
deeper understanding of these puzzling decays.

The future high-luminosity $B$-factories could contri\-bu\-te more precise data on the low\-er end of the $\pp$ mass
distribution, where Belle data are
somewhat insufficient, in order to observe these nontrivial structures in the $\Upsilon(5S)$ dipion transitions.
Moreover, this is also a region where the more ``normal" dipion
transitions such as the $\Upsilon(2S)\to\Upsilon(1S)\pp$ show striking differences with the ``abnormal" ones like the $\Upsilon(3S)\to\Upsilon(1S)\pp$: the lower end is rather suppressed for the former while it is enhanced for the latter.

Next, particular attention should be paid to the region around 1~GeV (in
$\Upsilon(1S)\pp$ since the other final states do not have enough phase space).
Detailed knowledge in this 1~GeV region would be helpful in understanding the
nature of both the $\Upsilon(5S)$ and the light scalar $f_0(980)$. The transition
$\Upsilon(5S)\to\Upsilon(1S)\pp$ offers a rather unique possibility for studying
light scalar mesons in the sense that both the initial and final particles (besides
the pions) are $SU(3)$ flavour singlet states. In the $SU(3)$ limit, the production
couplings should be equal for up, down, and strange flavours. Hence, this transition
is able to provide information on the light scalar mesons complementary to those
from the decays such as $J/\psi\to\omega\pp$ and $J/\psi\to\phi\pp$.

Precise measurements of the
$\Upsilon(5S)\to\Upsilon(1S) K^+K^-$ transition in the region near the $K^+K^-$ threshold would also be quite useful, because the $f_0(980)$ couples strongly to two kaons.
The rate for this reaction has been found to be about 9 times smaller than the dipion channel, with a partial width of about 0.067 MeV~\cite{Abe:2007tk}.

Additional important information on the dipion
tran\-sitions can be obtained with a $B$-factory running at the $\Upsilon(6S)$.
The decay widths, dipion invariant mass spectra and angular distributions in the
$\Upsilon(6S)\to\Upsilon(nS)\pp$ transitions, with $n \leq 4$, should be compared
to ones obtained at the $\Upsilon(5S)$. Due to its larger mass, the open-beauty mass
threshold effects should be smaller than for the $\Upsilon(5S)$ and
contributions from different processes can be somewhat separated.

\subsubsection{Dipion transitions with {\boldmath $b$}-quark spin flip}

Nominally, inverting the spin of the heavy quark in a decay is a process suppressed by a factor $\Lambda_{\rm QCD}/m_b$.
Data however do not always make it obvious. There is a recent Belle Collaboration observation of the $h_b(1P)$ and $h_b(2P)$ states produced via $e^+e^-\to h_b(nP)\pp$ in the $\Upsilon(5S)$ region \cite{Adachi:2011ji} (with the data sample of 121.4 fb$^{-1}$ collected near the $\Upsilon(5S)$ peak).

It raises yet another question related to dipion $\Upsilon(5S)$ transitions
since the reported ratios
\begin{eqnarray}
\frac{\Gamma(\Upsilon(5S)\to h_b(1P)\pi^+\pi^-)}{\Gamma(\Upsilon(5S)\to \Upsilon(2S)\pi^+\pi^-)}\approx 0.4,\nonumber
\\ 
\frac{\Gamma(\Upsilon(5S)\to h_b(2P)\pi^+\pi^-)}
{\Gamma(\Upsilon(5S)\to \Upsilon(2S)\pi^+\pi^-)}\approx 0.8,
\label{2Svshb}
\end{eqnarray}
are unexpectedly large. Indeed, while the $b\bar{b}$ pair is in the spin-triplet state in the vectors $\Upsilon(nS)$, it appears to be in spin-singlet state in the axials $h_b(nP)$. This implies that, unlike the $\Upsilon(5S)\to\Upsilon(2S)\pp$ transition, proceeding without heavy-quark spin flip, naively such a flip must occur in $\Upsilon(5S)\to h_b(nP)\pp$, so that the amplitude for the latter process is expected to be
suppressed  by $\Lambda_{\rm QCD}/m_b$ when compared with that for the former transition.

Summarising this subsection devoted to dipion transitions, it is important to notice
that, with the data sample of about 1~ab$^{-1}$ collected by a future $B$-factory near the $\Upsilon(5S)$, not only
high-statistics data will be available for dipion decays with charged pions, but
similar decays with two neutral particles (pions and/or $\eta$'s) in the final
state should be readily accessible. Yet another piece of critical
information can be obtained with a few hundred ~fb$^{-1}$ taken at the $\Upsilon(6S)$.

\subsection{{\boldmath $Z_b$} states} \label{Zbstories}

In 2011 the Belle Collaboration announced the first observation of two charged
bottomonium-like states, $Z_b(10610)$ and $Z_b'(10650)$, in five different decay channels of $\Upsilon(5S)$ ($\pi^\pm\Upsilon(nS)$, $n=1,2,3$ and $\pi^\pm h_b(mP)$, $m=1,2$)~\cite{Adachi:2011gj,Belle:2011aa},
with averaged masses and widths \cite{Belle:2011aa}
\bea
M_{Z_b}=10607.2\pm2.0~\mbox{MeV},\; \Gamma_{Z_b}=18.4\pm 2.4~\mbox{MeV},\label{Zb1exp}\\
M_{Z_b'}=10652.2\pm 1.5~\mbox{MeV},\;\Gamma_{Z_b'}=11.5\pm 2.2~\mbox{MeV}.\label{Zb2exp}
\eea

These structures are produced from, and reconstructed in, conventional $b\ov{b}$
states, so it is natural to assume that they contain a $b\ov{b}$ quark pair.
However, for $Z_b$'s being electrically charged, a pure $b\ov{b}$ assignment is discarded.
These states are still in want of confirmation by other experiments and their
nature is not well understood yet. It is clear however that they provide an
excellent test ground for QCD. In particular, these newly observed states may be
relevant for understanding the dipion $\Upsilon(nS)$, in particular
$\Upsilon(5S)$, transitions. Given the proximity of the observed $Z_b$ states to
two-body thresholds $B\ov{B}^*(10604.5\pm0.4~{\rm MeV})$\footnote{Here and in
what follows proper combinations with a given $C$-parity are understood, for
example, $B\ov{B}^*$ is a shorthand notation for
$\frac{1}{\sqrt{2}}(B\ov{B}^*+\ov{B}B^*)$.} and $B^*\ov{B}^*(10650.4\pm0.6~{\rm
MeV})$, a molecular interpretation of the $Z_b$'s was suggested shortly after the
Belle announcement \cite{Bondar:2011ev}. Although the four-quark interpretation of the $Z_b$'s
still requires a theoretical explanation of their large production cross section at the $\Upsilon(5S)$ energies, this idea immediately motivated many
theoretical efforts (see, for example,
\cite{Yang:2011rp,Sun:2011uh,Nieves:2011zz,Cleven:2011gp,Bugg:2011jr,Ohkoda:2011vj,Danilkin:2011sh,Ali:2011ug,Li:2012wf})
and a number of predictions were made following this conjecture. Below we
discuss them briefly.

If an exact heavy-quark symmetry is assumed (implying the limit $m_b\to\infty$),  a
proper basis to consider $S$-wave $B^{(*)}\ov{B}^{(*)}$ pairs consists of the
direct product of the eigenstates of the spin operators for the heavy-quark pair
and the light-quark pair, $S_{b\bar{b}} \otimes S_{q\bar{q}}$, with
$S_{b\bar{b}}=0,1$ and $S_{q\bar{q}}=0,1$. Then the wave functions of the $Z_b$
states can be represented as \cite{Bondar:2011ev}
\bea
1^+(1^+)&Z_b':&
\frac{1}{\sqrt{2}}\left(0_{b\bar{b}}\otimes 1_{q\bar{q}}-1_{b\bar{b}}\otimes
0_{q\bar{q}}\right),
\nonumber\\[-3mm]
\label{Zss}\\[-3mm]
1^+(1^+)&Z_b :&\frac{1}{\sqrt{2}}\left(0_{b\bar{b}}\otimes
1_{q\bar{q}}+1_{b\bar{b}}\otimes 0_{q\bar{q}}\right), \nonumber
\eea
where, in the
first column, the quantum numbers are qu\-o\-ted in the form $I^G(J^P)$  with $I$ and
$J$ being the isospin and total spin of the state, and with the superindices denoting its
$G$- and $P$-parity, respectively.

\begin{figure}[ht]
\begin{center}
\includegraphics[width=6cm]{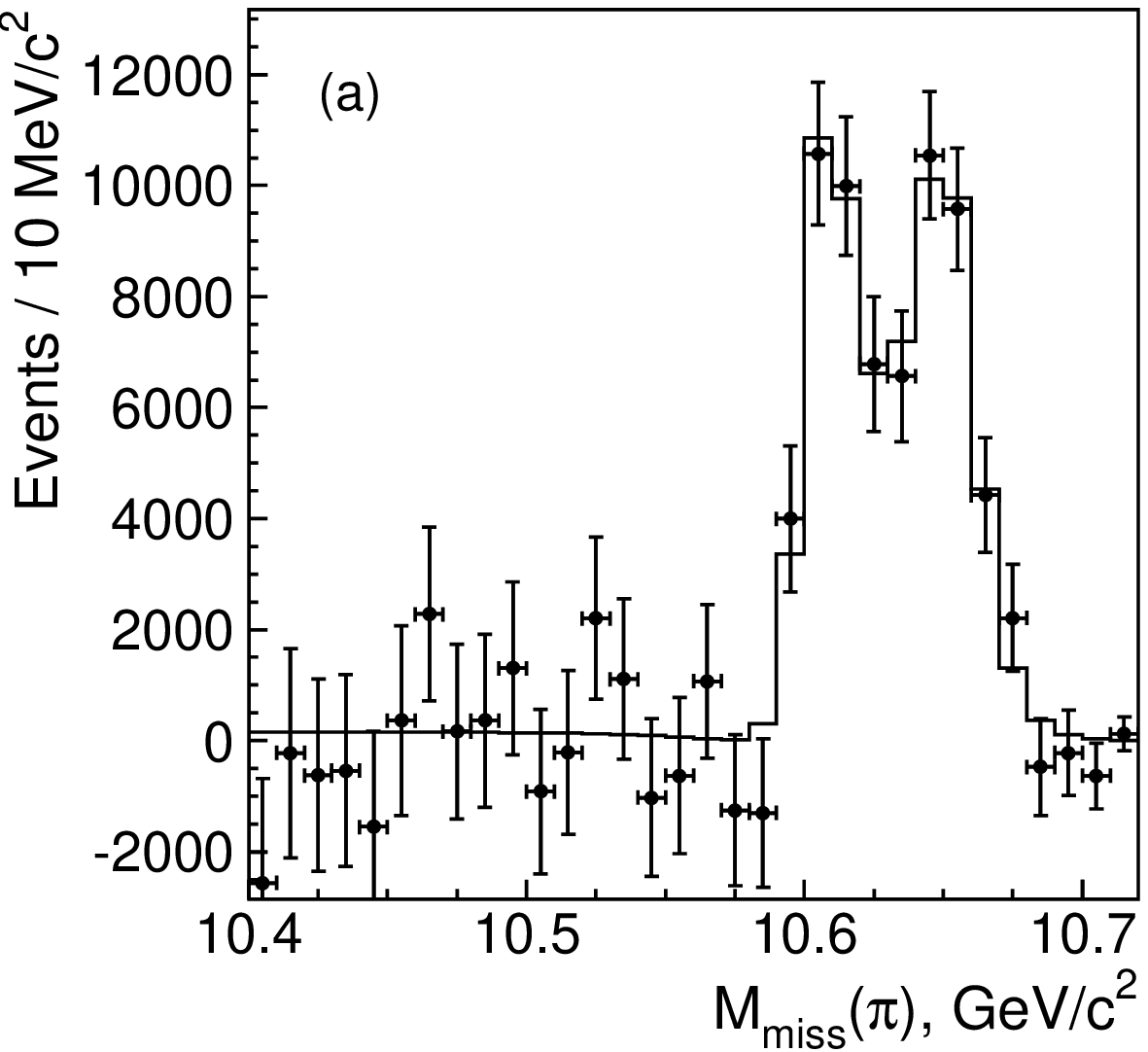}
\includegraphics[width=6cm]{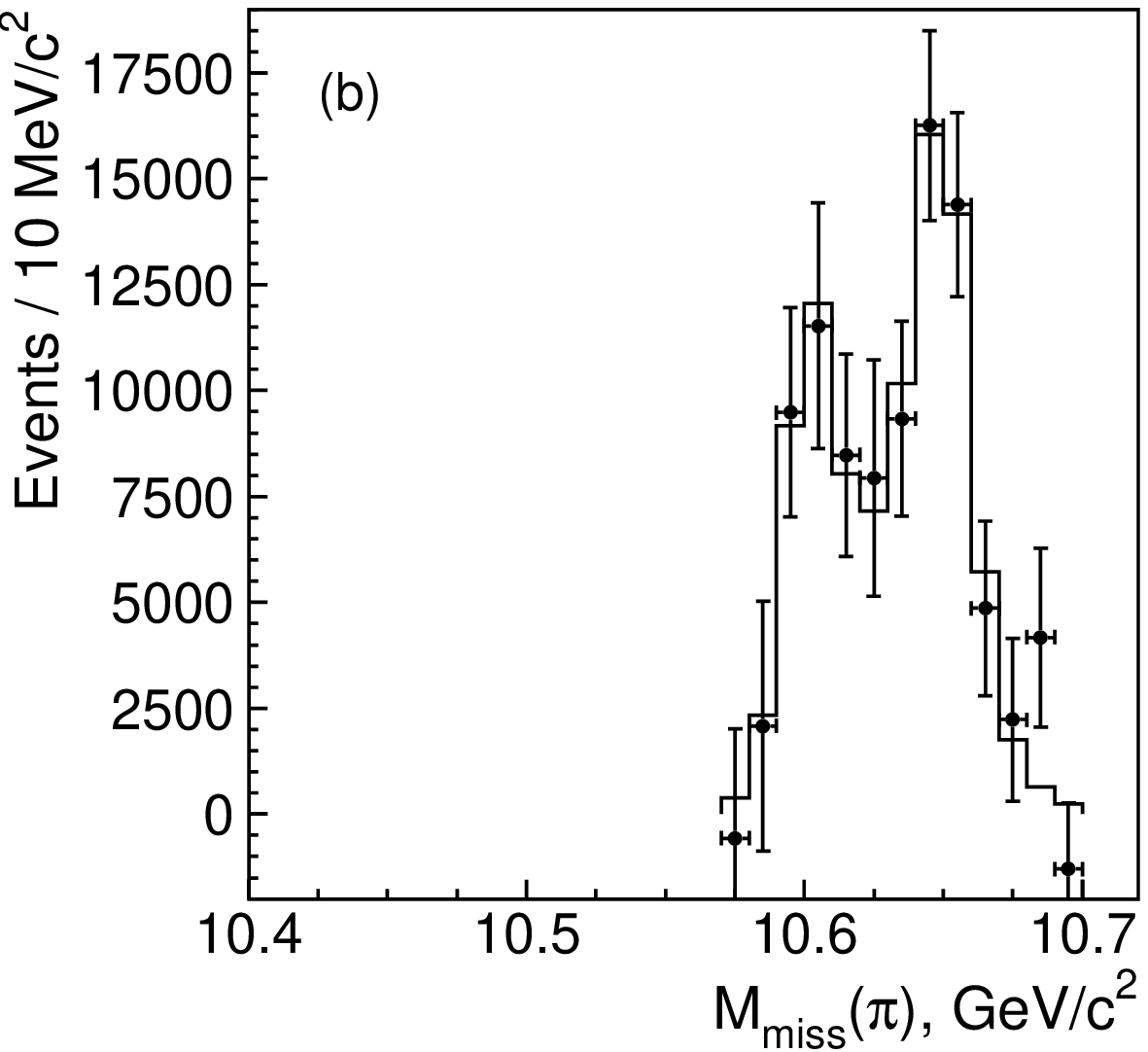}
\end{center}
\caption{The (a) $\Upsilon(5S)\to h_b(1P)\pp$ and (b) $\Upsilon(5S)\to h_b(2P)\pp$ yields as a function of the pion missing mass (adapted from \cite{Belle:2011aa}).}\label{fig:h1h2}
\end{figure}

\begin{figure}[ht]
\begin{center}
\includegraphics[width=8cm]{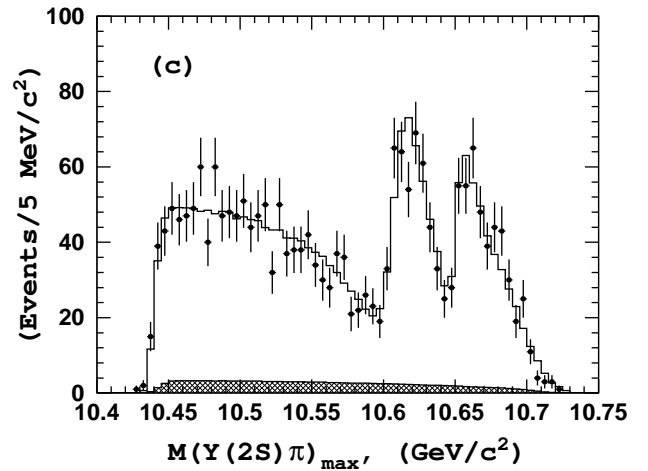}
\end{center}
\caption{One-dimensional projection of the data for the decay $\Upsilon(5S)\to \Upsilon(2S)\pp$ (adapted from
\cite{Belle:2011aa}).}\label{fig:y2s}
\end{figure}

This provides a possible explanation for the anomalous ratios (\ref{2Svshb}) as
no suppression is expected for the processes proceeding through the $Z_b(Z_b')\pi$
intermediate states, as cascades \cite{Bondar:2011ev}:
\begin{eqnarray*}
\Upsilon(5S)&\to& Z_b(Z_b')\pi\to h_b(1P)(h_b(2P))\pi^+\pi^-,\\
\Upsilon(5S)&\to& Z_b(Z_b')\pi\to \Upsilon(nS)\pi^+\pi^-,
\end{eqnarray*}
because components with both heavy-quark spin orientations have equal weight in the
$Z_b$'s wave functions. Besides that, representation (\ref{Zss}) predicts the
interference pattern for the amplitudes of the processes proceeding through the
$Z_b$ and $Z_b'$. In particular, for the reactions $\Upsilon(5S)\to h_b(nP)\pp$
($n=1,2$), because of the different relative sign between the two components of the
wave function, Eq.~(\ref{Zss}) predicts constructive interference between the $Z_b$
peaks and destructive interference outside this region \cite{Bondar:2011ev}. In the
meantime, the interference pattern is quite opposite for the reactions
$\Upsilon(5S)\to \Upsilon(nS)\pp$. This prediction is consistent with the recent
Belle data \cite{Belle:2011aa} (small dip between peaks in Fig.~\ref{fig:h1h2}
versus a well-pronounced dip between peaks in Fig.~\ref{fig:y2s}).

Furthermore, it is argued in \cite{Chen:2011zv} that results of model calculations
both for the dipion invariant mass spectrum as well as for the angular distribution
$$d\Gamma(\Upsilon(5S)\to\Upsilon(2S)\pp)/d\cos\theta$$
 can be reconciled with the
data, if the $Z_b$'s are included into the analysis as intermediate states.

\begin{figure}[t]
\begin{center}
\includegraphics[width=0.48\textwidth]{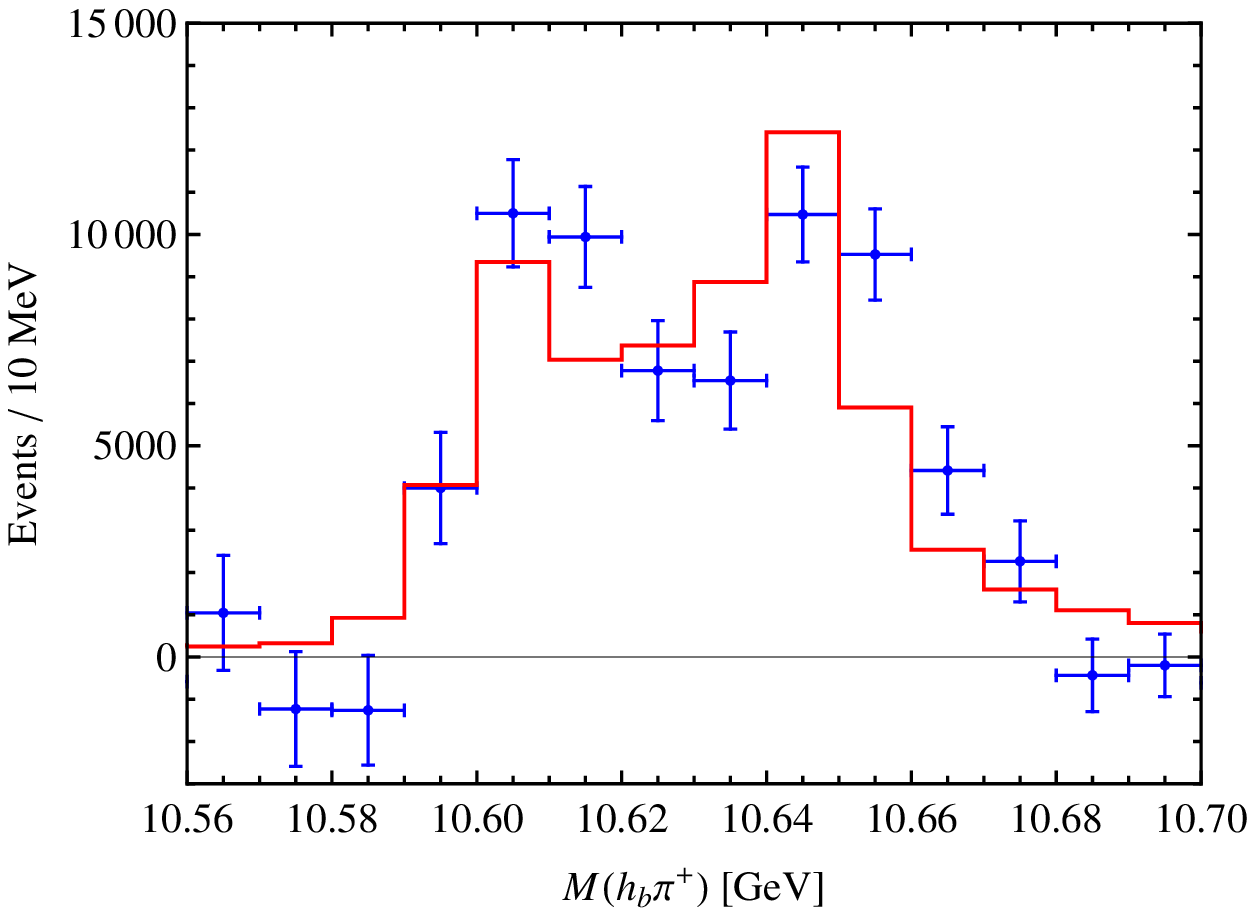}\hfill
\includegraphics[width=0.48\textwidth]{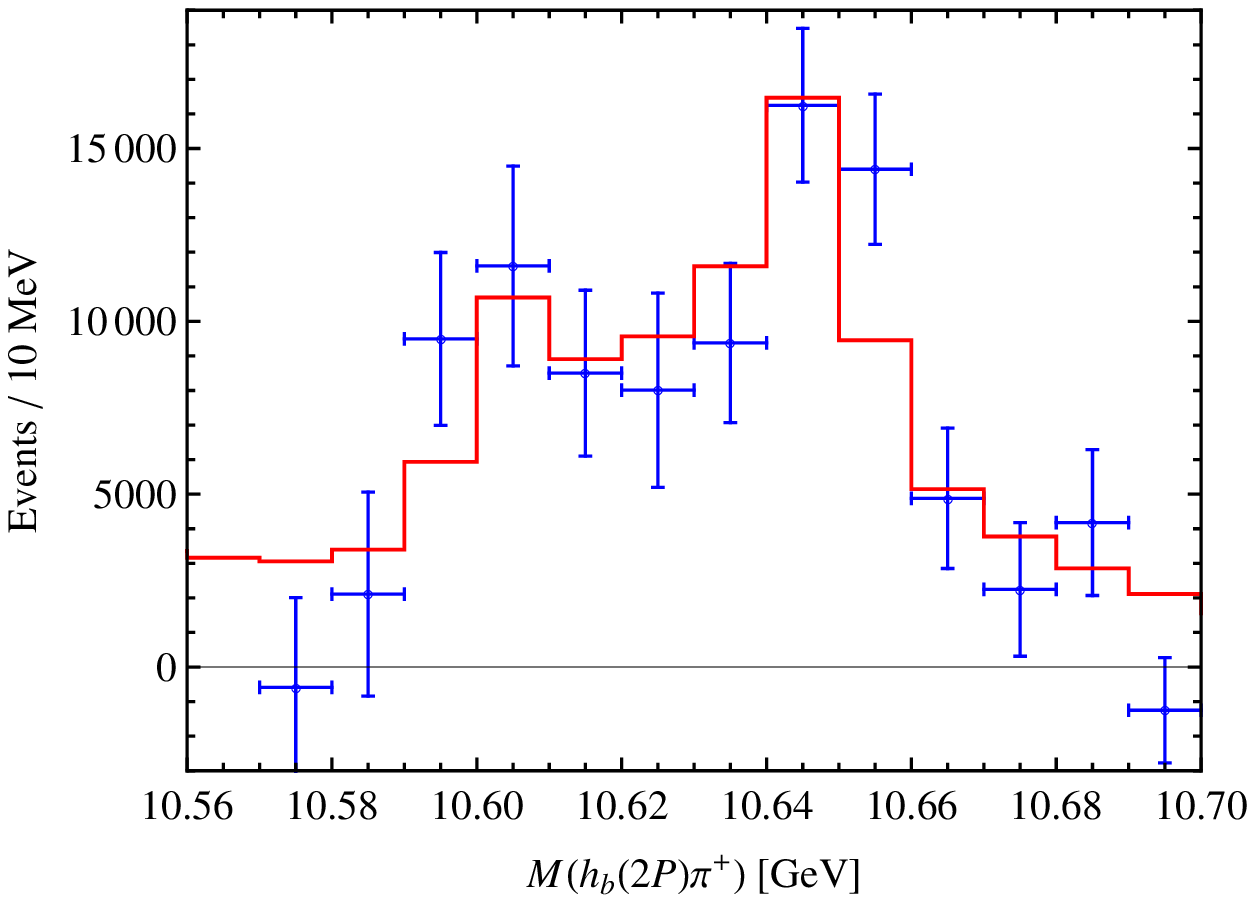}
\caption{Comparison of the invariant mass spectra of $h_b\pi^+$ and $h_b(2P)\pi^+$
calculated in \cite{Cleven:2011gp} with the measured missing mass spectra
$MM(\pi)$. \label{fig:Zbfit}}
\end{center}
\end{figure}

According to the updated analysis \cite{Belle:2011aa} (in line with the original
report \cite{Adachi:2011gj}), the Breit-Wigner masses of the $Z_b$ and $Z_b'$ states lie just above the $B\ov{B}^*$ and $B^*\ov{B}^*$ thresholds, respectively.
Theoretical work~\cite{Cleven:2011gp}, taking into account
the coupling to the intermediate $B\bar B^*$ and $B^*\bar B^*$ states,
shows that $Z_b$ masses slightly below the corresponding thresholds are also consistent
with the data in the $h_b(1P)\pp$ and $h_b(2P)\pp$ (see the comparison in
Fig.~\ref{fig:Zbfit}).
In any case, given the proximity of the $Z_b$ states to these two-body $S$-wave
thresholds, the latter are expected to have a strong impact on the properties of
the $Z_b$'s. In particular, the analysis in \cite{Cleven:2011gp} is based on the
assumption that the $Z_b$'s are shallow bound states. An alternative interpretation
of their resonant nature suggested in~\cite{Bugg:2011jr} is based on the
observation that the Belle data are consistent with the $Z_b$'s as threshold cusps.

A coupled-channel approach to hadron spectroscopy near the open-bottom thresholds was developed in \cite{Ohkoda:2011vj}.
In particular, meson exchange potentials between $B^{(*)}$ mesons at threshold were derived and the corresponding Schr{\"o}dinger equations were solved numerically. As a result, the masses of the twin $Z_b$ resonances
were reproduced and a number of other possible bound and/or resonant states in other channels were predicted. Decay modes suggested for the experimental searches of the predicted states can be found in \cite{Ohkoda:2011vj}.

\begin{figure}[t]
\begin{center}
\includegraphics*[width=6cm]{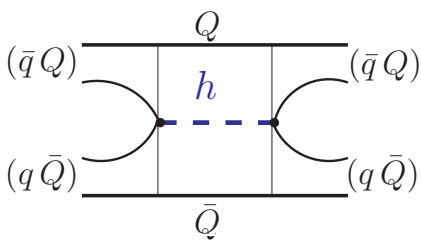}
\end{center}
\caption{The fundamental building block (kernel to be iterated) of the coupled-channel scheme in Eq.~(\ref{couplchan}) \cite{Danilkin:2011sh}.}\label{fig:yua}
\end{figure}

An alternative explanation for the $Z_b$ states is proposed in \cite{Danilkin:2011sh}, which employs a coupled-channel scheme (see Fig.~\ref{fig:yua})
\be
(q\bar{Q})(Q\bar{q})\leftrightarrow(Q\bar{Q})h,
\label{couplchan}
\ee
with $Q$ and $q$ standing for the heavy ($b$) and light quark, respectively, and with $h$ denoting a light hadron
(pion). Such an approach is based on the concept of the QCD string breaking with pion emission (further details can be
found in \cite{simonovpions}). In particular, for the $\Upsilon(nS)$ dipion decays this scheme implies multiple
iterations of the basic building block $B^{(*)}\ov{B}^{(*)}\leftrightarrow\Upsilon(nS)\pi$. The calculated production
rates for the transitions
$\Upsilon(5S) \to \Upsilon(nS)\pp$ ($n=1,2,3$) reveal peaks at the $B \bar B^*$ and $B^* \bar B^*$ thresholds. This computation does not assume any direct interaction between $B^{(*)}$ mesons, so that the $Z_b$ singularities appear solely due to the interchannel coupling (\ref{couplchan}). Similar calculations of the $\Upsilon(5S)$ dipion transitions into $h_b(nP)\pp$ final state constitute a challenge being addressed by theorists with these tools.

Additional measurements of the Belle Collaboration on the $Z_b$'s have very
recently been reported~\cite{Adachi:2012cx}. Two additional observation modes of the $Z_b$ states seem to have been
found. They are $\Upsilon(5S)\to \bar B B^*\pi (B \bar B^*\pi)$
and $\Upsilon(5S)\to B^* \bar B^*\pi$, and the $Z_b(10610)$ and $Z_b(10650)$ were
found in the missing mass spectra of the pion in the $\bar B B^*\pi$ $(B \bar
B^*\pi)$ and $B^* \bar B^*\pi$ final states, respectively. Assuming that the
$\Upsilon(1S,2S,3S)\pi$, $h_b(1P,2P)\pi$ and $B^{(*)}\bar B^*+c.c.$ modes saturate
the decays of the $Z_b$ states, the reported branching fractions are listed in
Table~\ref{tab:zbbf}.
\begin{table}
\begin{center}
\begin{tabular}{|l|cc|}
\hline
Mode                & $Z_b(10610)^+$ & $Z_b(10650)^+$  \\
\hline
$\Upsilon(1S)\pi^+$ & $0.32\pm 0.09$ & $0.24\pm 0.07$\\
$\Upsilon(2S)\pi^+$ & $4.38\pm 1.21$ & $2.40\pm 0.63$\\
$\Upsilon(3S)\pi^+$ & $2.15\pm 0.56$ & $1.64\pm 0.40$ \\
$h_b(1P)\pi^+$      & $2.81\pm 1.10$ & $7.43\pm 2.70$\\
$h_b(2P)\pi^+$      & $4.34\pm 2.07$ & $14.8\pm 6.2$ \\
$B^+\bar B^{*0}+c.c.$ & $86.0\pm 3.6$& $-$ \\
$B^{*+}\bar B^{*0}$ & $-$            & $73.4\pm 7.0$ \\
\hline
\end{tabular}
\caption{Branching fractions in per cent for the $Z_b$
states~\cite{Adachi:2012cx}. \label{tab:zbbf}}
\end{center}
\end{table}
Furthermore, evidence for the neutral partners of the charged $Z_b$'s was
observed in the $\Upsilon(5S)\to\Upsilon(2S)\pi^0\pi^0$. Clearly, the models
proposed for the $Z_b$ states have to be confronted with the new data.

Additional insight on the nature of near-threshold resonances can be obtained
from an unbiased analysis of the data in the region below threshold. Indeed, in the
appropriate mode, the bound state should reveal itself as a peak below the
corresponding open-bottom thresholds in the line shape. In particular, data on
radiative decays of the $Z_b$ states could potentially confirm or rule out their
bound-state interpretation. To this end, high-statistics and high-resolution data
in the region below the nominal $B^{(*)}\ov{B}^*$ thresholds is needed to study the
radiative transitions $Z_b(10610)\to B\ov{B}\gamma$, $Z_b(10650)\to B\ov{B}\gamma$,
and $Z_b(10650)\to B^*\ov{B}\gamma$. As mentioned above, bound states would reveal
themselves as below-threshold peaks which, given the expected high-statistics data
from a future $B$-factory, can potentially be observed.

To summarise, a data sample expected to be collected by a high-luminosity $B$-factory with the known
production reaction should be sufficient to perform a high-statistics analysis, to
exclude alternative interpretations, and to fix the parameters of the $Z_b$ states
to a high accuracy, in particular their position relative to the corresponding
$B^{(*)}\ov{B}^*$ thresholds. A sophisticated line shape form, respecting
unitarity and accounting for threshold vicinity, should be used for the data analysis.
The production of the $Z_b$'s can be further investigated at the
$\Upsilon(6S)$ energy, and the additional information would be helpful in
identifying their nature.

\subsection{{\boldmath $W_{bJ}$} states}

The idea put forward in \cite{Bondar:2011ev} that the $Z_b$ and $Z_b'$ states can be explained as
molecule-type structures residing at the corresponding $B^{(*)}\ov{B}^*$ thresholds was extended further in
\cite{Bondar:2011ev,Voloshin:2011qa,Mehen:2011yh} and a possible existence of a few sibling states, denoted as
$W_{bJ}^{(\prime)}$ with $J=0,1,2$ and defined by four orthogonal combinations,
\bea
1^-(2^+)&W_{b2}: & 1_{b\bar{b}} \otimes 1_{q\bar{q}}\Big|_{J = 2},\nonumber\\
1^-(1^+)&W_{b1}: & 1_{b\bar{b}} \otimes 1_{q\bar{q}}\Big|_{J = 1}, \nonumber\\[-3mm]
\label{Wss}\\[-3mm]
1^-(0^+)&W_{b0}': & \frac{\sqrt{3}}{2}\,0_{b\bar{b}} \otimes 0_{q\bar{q}} + \frac{1}{2}\,1_{b\bar{b}} \otimes
1_{q\bar{q}}\Big|_{J = 0},\nonumber\\
1^-(0^+)&W_{b0}: & \frac{\sqrt{3}}{2}\,1_{b\bar{b}} \otimes 1_{q\bar{q}}\Big|_{J =
0} - \frac{1}{2}\,0_{b\bar{b}} \otimes 0_{q\bar{q}},\nonumber
\eea
was predicted
and their properties were outlined. The me\-sonic components of the $W_{b2}$,
$W_{b1}$, $W_{b0}'$ and $W_{b0}$ are the $B^*\bar B^*$, $(B\bar B^*+B^*\bar
B)/\sqrt{2}$, $B^* \bar B^*$ and $B\bar B$, respectively.

It has to be noticed that, since the binding mechanisms responsible for the formation of the $Z_b$'s are still obscure,
only
model-dependent conclusions can be made concerning the existence or nonexistence of the $W_b$'s. In particular,
if the binding mechanism operates in the $S_{q\bar{q}} = 1$ channel (or in both $S_{q\bar{q}} = 0$ and $S_{q\bar{q}} =
1$ channels), then all four $W_b$'s are expected to exist. However if this mechanism operates only in the $S_{q\bar{q}}
= 0$ channel, then only two sibling states ($W_{b0}$ and $W_{b0}'$) are predicted. In what follows it is assumed that
all four $W_b$'s exist and the corresponding predictions found in the literature are quoted and discussed.

To proceed, it is convenient to adopt the classification scheme suggested in
\cite{Voloshin:2011qa} exploiting the heaviness of the $b$ quark, that implies
that its spin is decoupled from the remaining degrees of freedom, so that a
convenient basis can be built in terms of the states $H\otimes SLB$, where $H$
stands for the heavy pair spin ($S_H=0,1$), while $SLB$ denotes all remaining
degrees of freedom, like angular momentum, light quark spins, etc. In particular,
this basis matches naturally the decompositions (\ref{Zss}) and (\ref{Wss}) above,
if $SLB$ is associated with the light-quark total spin $S_{q\bar{q}}$
\cite{Voloshin:2011qa}. Since the $H$ degree of freedom plays the role of a
spectator in all decay processes, it is sufficient to just pick up the term with the appropriate $H\otimes SLB$
structure in the
final state in order to extract the corresponding decay amplitude. Although a detailed microscopic theory is needed to
predict each individual decay width, relative coefficients between different decay widths can be evaluated solely using
these simple symmetry-based considerations.
Such predictions are \cite{Voloshin:2011qa}
\bea
&&\Gamma[Z_b]=\Gamma[Z_b'],\label{ZZwidth}\\
&&\Gamma[W_{b1}]=\Gamma[W_{b2}]=\frac{3}{2}\Gamma[W_{b0}]-\frac{1}{2}\Gamma[W_{b0}'],
\label{WWwidth}
\eea
and \cite{Mehen:2011yh}
\bea
&&\Gamma[Z_b]=\Gamma[Z_b']=\frac{1}{2}\left(\Gamma[W_{b0}] +\Gamma[W_{b0}']\right).\label{ZWwidth}
\eea
Relation (\ref{ZZwidth}) can be verified with the help of the Belle measurement of the $Z_b$'s widths (equations
(\ref{Zb1exp}) and (\ref{Zb2exp}) above) and indeed it is approximately fulfilled.

Further predictions for specific decay channels follow from the structure of
the wave functions (\ref{Zss}) and (\ref{Wss}), as explained above. In particular,
\cite{Voloshin:2011qa}
\bea
\Gamma[W_{b0}\to\Upsilon\rho]:\Gamma[W_{b0}'\to\Upsilon\rho]:\Gamma[W_{b1}\to\Upsilon\rho]\nonumber\\
:\Gamma[W_{b2}\to\Upsilon\rho]=\frac34:\frac14:1:1 \label{rel0}
\eea
or, similarly~\cite{Mehen:2011yh}
\bea
\Gamma[W_{b0}\to\eta_b l]:\Gamma[W_{b0}'\to\eta_b l]:\Gamma[Z_b\to\Upsilon l]\nonumber\\
:\Gamma[Z_b'\to\Upsilon l]=\frac12:\frac32:1:1, \label{rel1}
\eea
\bea
\Gamma[W_{b0}\to\chi_{b1}l]:\Gamma[W_{b0}'\to\chi_{b1}l]:\Gamma[Z_b\to h_b l]\nonumber\\
:\Gamma[Z_b'\to h_b l]=\frac32:\frac12:1:1,\label{rel2}
\eea
where $l$ denotes a
suitable configuration of light hadrons. It should be noted however that relations
(\ref{rel0})-(\ref{rel2}) rely only on the structure of the amplitudes in the
strict heavy-quark symmetry limit. These predictions acquire corrections due to the
hyperfine splitting, conveniently parameterised  in terms of the difference
$M_{Z_b'}-M_{Z_b}$ \cite{Voloshin:2011qa,Mehen:2011yh} as well as due to the
difference in kinematic phase space. With these effects taken into account and
within an effective Lagrangian technique with binding contact interaction, the
following predictions were obtained in \cite{Mehen:2011yh}:
\beas
&\Gamma[W_{b0}\to\pi\eta_b(3S)]:\Gamma[W_{b0}'\to\pi\eta_b(3S)]&:
\nonumber \\
&\Gamma[Z_b\to\pi\Upsilon(3S)]:\Gamma[Z_b'\to \pi\Upsilon(3S)]
&\nonumber\\
&=0.26:2.0:0.62:1 \quad (\lambda_\Upsilon\to 0),&\\
&=0.12:2.1:0.41:1 \quad (|\lambda_\Upsilon|\to\infty),& \nonumber
\eeas
where $\lambda_\Upsilon$ is a certain combination of various coupling constants appearing in the effective Lagrangian
(see \cite{Mehen:2011yh}).

\subsection{{\boldmath $X_b$} and {\boldmath $Y_b$} states}
\label{sec:xbyb}

As argued in \cite{Bondar:2011ev,Voloshin:2012yq}, isovector $Z_b$ and $W_{bJ}$
states may possess isoscalar C-odd and C-even partners, also residing at the
$B^{(*)}\ov{B}^{(*)}$ thresholds. There is no general consensus in the literature
concerning the naming scheme for such states, so, for the sake of definiteness, we
refer to the C-even partners as $X_b$'s (in analogy with the C-even $X(3872)$
charmonium state) while we use the notation $Y_b$'s for C-odd states\footnote{The
authors of \cite{Bondar:2011ev} follow this scheme, however the notation $X_b$ is
used in \cite{Voloshin:2012yq} for C-odd states too.}. It should be noticed that,
while C-even states can be  structured as $B\ov{B}$,
$\frac{1}{\sqrt{2}}(B\ov{B}^*+\ov{B}B^*)$ or $B^*\ov{B}^*$, only two structures,
$\frac{1}{\sqrt{2}}(B\ov{B}^* - \ov{B}B^*)$ and $B^*\ov{B}^*$, are allowed for the
C-odd states. As a result $X_b$'s can have the $J^{PC}$ quantum numbers $0^{++}$,
$1^{++}$, and $2^{++}$ while only the $1^{+-}$ option is available for $Y_b$'s.
Then, similarly to the famous $X(3872)$ charmonium, $X_b$'s and $Y_b$'s can mix
with the conventional $^3P_J$ and $^1P_1$ bottomonium states, respectively, that
makes possible to produce such states through their quarkonium (compact)
components in high-energy collisions. Then $Y_b$'s states, should they exist and
be produced in an experiment, could be sought for, for example, using the decays
$Y_b\to\Upsilon(nS)\eta^{(\prime)}$, $Y_b\to\Upsilon(nS)\pi\pi$, or
$Y_b\to\Upsilon(nS)K\ov{K}$ \cite{Bondar:2011ev}. Similarly, $X_b$'s might be seen
in the mode $X_b\to\Upsilon(1S)\omega$ \cite{Bondar:2011ev}.

Finally, it is argued in \cite{Voloshin:2012yq} that, while the C-odd isosca\-lar
$Y_b$ states are not  directly accessible in the $\Upsilon(5S)$ decays (neither in
radiative decays nor in hadronic transitions), studies of the ratio $R_c/R_n$ (of
its deviation from unity) of the yields for pairs of charged and neutral $B^{(*)}$
mesons in the decays $\Upsilon(5S)\to\pi^0 B^{(*)}\ov{B}^{(*)}$ may give an insight
on the interaction of $B^{(*)}$ mesons in the isoscalar channel. In particular, in
presence of a near-threshold isoscalar resonance, the isospin breaking Coulomb
effect provides a specific behaviour of this ratio which can potentially be studied
by the future $B$-factories.

The search for the $W_b,X_b,Y_b$ partners of the putative exotic
$Z_b$ states could also be conducted with any data taken at the $\Upsilon(6S)$ resonance,
which would facilitate additional phase space for the search. Spin splittings for
states involving light quarks (as the $Z_b$'s have due to their electric charge)
are of order one or two hundred MeV, as for example the $h_1(1170)$-$f_2(1270)$
splitting in the light meson spectrum. Since the $Z_b$ candidates have masses in
the 10605-10650~MeV range, the $\Upsilon(5S)$ at 10876~MeV may be too light as a starting point to
reconstruct decay chains containing the $W_b$, $X_b$, and $Y_b$ partners. The $\Upsilon(6S)$ state at
11020~MeV, on the other hand, offers the safety of the additional phase space, that
will allow setting of more extensive exclusion limits and offer larger
discovery potential.

\subsection{Radiative decays of {\boldmath $\Upsilon(5S)$}}\label{raddec}

A recent BaBar analysis using $(111\pm 1)\times 10^6$ $\Upsilon(3S)$ and $(89\pm 1)\times 10^6$ $\Upsilon(2S)$ events
\cite{Lees:2011mx} allowed one to observe $\Upsilon(3S)\to \gamma \chi_{b0,2}(1P)$ decay and to make precise
measurements of the branching fractions for the $\chi_{b1,2}(1P,2P)\to \gamma\Upsilon(1S)$ and
$\chi_{b1,2}(2P)\to\gamma\Upsilon(2S)$ decays. A search for the $\eta_b(1S)$ and $\eta_b(2S)$ states was performed in
the decays $\Upsilon(nS)\to\gamma\eta_b(n'S)$ $(n=2,3;n'<n)$,
however the significance of the result obtained was insufficient to draw conclusions regarding the $\eta_b(nS)$ masses,
so that more data are needed.

Data expected with future high-luminosity $B$-factories at the $\Upsilon(5S)$ energy will allow precise studies of the
radiative decays $\Upsilon(5S)\to\gamma\eta_b(nS)$ and $\Upsilon(5S)\to\gamma\chi_{b0,1,2}(mP)$ which meets challenges
provided by the lattice and model theoretical calculations of various $\eta_b$ and $\chi_b$ meson properties (hyperfine
splittings with the $\Upsilon(nS)$ states, total widths, various decay branching fractions, and so on).

Meanwhile the Belle Collaboration announced \cite{Mizuk:2012pb} a successful identification (with the significance of
15$\sigma$) of the $\eta_b(1S)$ state in the decay chain $\Upsilon(5S)\to Z_b^+\pi^-\to h_b(1P)\pi^+\pi^-\to
\gamma\eta_b\pi^+\pi^-$. The
measured hyperfine splitting $M(\Upsilon(1S))-M(\eta_b(1S))$ appeared to be lower than the world average by about 10
MeV,
and this improves the agreement of the data with lattice simulations as well as with pNRQCD theoretical predictions.
Remarkable prog\-ress was achieved by using the large sample of data for the $h_b(1P)$ state collected at the energy of
the $\Upsilon(5S)$ resonance. Related searches for $h_b(2P)$ radiative decays are expected soon.

Generally, a missing mass method for $\gamma$, $\pi^0$, and $\eta$ can be applied with the higher statistics expected
at the future $B$-factories, however it will require longer chains with additional requirements, like the decay
$\eta_b(nS)\to
\Upsilon(mS)\gamma$ with $m<n$.

\begin{figure}
\includegraphics[width=8.6cm]{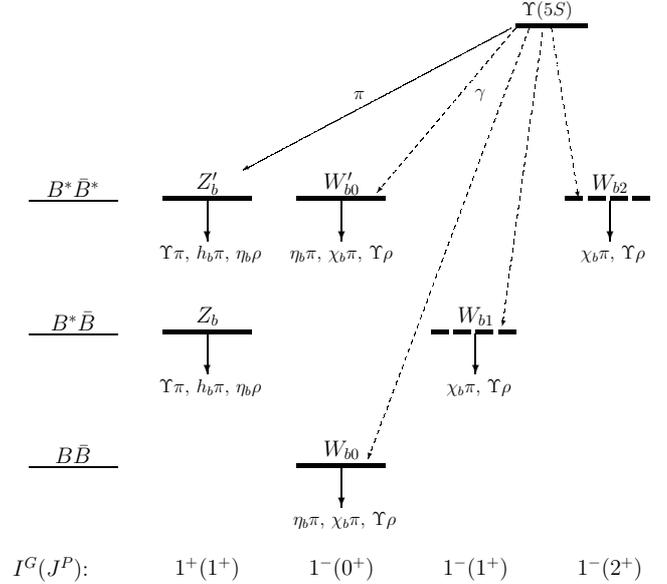}
\caption{Schematic representation for various closed--flavour decay chains
of the $\Upsilon(5S)$ decays, including radiative transitions with conjectured, new $W_{bJ}$ states created at
intermediate stages (adapted from \cite{Voloshin:2011qa}).}\label{fig:raddec:Upsilon}
\end{figure}

The conjecture of the existence of $W_{bJ}$'s, negative $G$-parity isovector
partners of the $Z_b$'s, opens extra opportunities for the future $B$-factories in what the
$\Upsilon(5S)$ radiative decays are concerned. Indeed, because of the $G$-parity, a
natural way these $W_{bJ}$ states could be produced from the $\Upsilon(5S)$ are
radiative decays of the latter, as shown in Fig.~\ref{fig:raddec:Upsilon} (adapted
from \cite{Voloshin:2011qa}). Although such radiative decays are suppressed by the
fine structure constant $\alpha=1/137$ as compared, for example, to
$\Upsilon(5S)\to Z_b^{(\prime)}\pi$ decays, the statistics expected at the future
$B$-factories should be sufficient to observe such decays.

$\Upsilon(5S)$ is produced in the $e^+e^-$ annihilation, so that there are two possible $b\ov{b}$ configurations, $S$-
and $D$-wave, that when combined with spin yield the production quantum numbers $1^{--}$. In the nonrelativistic
approximation for $b$ quarks only the $S$-wave contribution is retained. Then only the $^3S_1$ component of the
$\Upsilon(5S)$ wave function is considered which has the following decomposition in terms of the $H\otimes SLB$ states:
$1^-_H\otimes 0^+_{SLB}$, where the $H$ component comes from the total $b\bar{b}$ spin while the $SLB$ component
accounts for the angular momentum \cite{Voloshin:2011qa}. Such identification leads, in particular, to the following
prediction \cite{Voloshin:2011qa}:
\bea
&\Gamma(\Upsilon(5S)\to W_{b0}\gamma):\Gamma(\Upsilon(5S)\to W_{b0}'\gamma):&
\nonumber \\
& \Gamma(\Upsilon(5S)\to W_{b1}\gamma)
:\Gamma(\Upsilon(5S)\to W_{b2}\gamma)&
\nonumber \\
=&\ds\frac34\omega_0^3:\frac14\omega_2^3:3\omega_1^3:5\omega_2^3\approx 8.5:1:21:20,&\label{Wgammas}
\eea
where $\omega_0\approx 305$ MeV, $\omega_1\approx 260$ MeV, and $\omega_2 \approx 215$ MeV are the photon energies in
the corresponding transitions.

In Fig.~\ref{fig:raddec:Upsilon} various cascade reactions for the $\Upsilon(5S)$ decays are shown graphically,
specifically its radiative transitions with $W_{bJ}$'s created in the intermediate state. Measurements of such decay
chains may become one of the most promising tasks for the future $B$-factories.

\subsection{Extraction of the light quark mass ratio}

It was noticed recently that the light-quark mass ratio $\frac{m_u}{m_d}$ can be extracted
to high accuracy from data for bottomonium decays~\cite{FK}. Specifically, the decays
$\Upsilon(4S)\to h_b(1P)\pi^0$ and $\Upsilon(4S)\to h_b(1P)\eta$ are useful
because the bottom meson loops are highly suppressed in these reactions \cite{FK}.
Similarly, decays of $\Upsilon(5S)$ into $h_b(nP)\pi^0(\eta)$ should provide the
same information on the $\frac{m_u}{m_d}$ ratio. Having a high-statistics measurement from
a $B$-factory would be useful to estimate the systematics.

To be specific, a combination of the light quark mass
ratios~\cite{Donoghue:1992ac,Donoghue:1993ha}
\bea
r_{\rm DW} &\equiv& \frac{m_d-m_u}{m_d+m_u}
\frac{m_s+\hat{m}}{m_s-\hat{m}} \non\\
&=& \frac{4}{3\sqrt{3}} r_{G\tilde G} \frac{F_\pi}{F_\eta}
\frac{F_K^2M_K^2-F_\pi^2M_\pi^2}{F_\pi^2M_\pi^2} (1-\delta_{\rm GMO}) \non\\
&& \times \left[1+\frac{4L_{14}}{F_\pi^2} (M_\eta^2-M_\pi^2)\right]
\\
&=& 10.59 \, (1+132.1 \, L_{14}) \, r_{G\tilde G}\non
\eea
can be related to the ratio of the branching fractions of the decays of the
$\Upsilon(5S)\to h_b(nP)\pi^0(\eta)$. Here, $\hat{m}$ is the averaged mass of the
up and down quarks, $F_{\pi,K,\eta}$ are the meson decay constants, $\delta_{\rm
GMO}=-0.06$ denotes deviation from the Gell-Mann--Okubo relation among the light
meson masses, $L_{14}=(2.3\pm1.1)\times10^{-3}$ is a low-energy constant in the
$\mathcal{O}(p^4)$ chiral Lagrangian~\cite{Donoghue:1992ac,Donoghue:1993ha}, and
$r_{G\tilde G}$ is a ratio of the gluonic matrix elements $\langle 0| G\tilde{G}
|\pi^0\rangle / \langle 0|G\tilde{G}|\eta\rangle$. The relation reads
\bea
\label{eq:Rpi0eta} \frac{\Gamma(\Upsilon(5S)\to
h_b(nP)\pi^0)}{\Gamma(\Upsilon(5S)\to h_b(nP)\eta)} = r_{G\tilde G}^2
\left|{\vec{q}_\pi\over\vec{q}_\eta}\right|.
\eea
Then with additional information on the strange quark mass from elsewhere, one is
able to extract the ratio $\frac{m_u}{m_d}$. However, as will be argued below, only the
decays into the $h_b(2P)$ can be used.

The light quark mass ratio has to be extracted from isospin breaking and $SU(3)$
breaking processes. Heavy quarkonium transitions with an emission of a pion or
$\eta$ are of potential use. However, coupled-channel effects from the intermediate
heavy meson loops are often important in the transitions between two heavy
quarkonia. In that case, a large amount of the isospin breaking is provided by the
mass difference between the charged and neutral heavy mesons, and hence such decays
(for example, $\psi'\to J/\psi\pi^0(\eta)$~\cite{FK:NREFT1}) are not suitable for
the extraction of the $\frac{m_u}{m_d}$ ratio. Therefore, one must be sure that the heavy
meson loops give a small contribution to the transition amplitudes so they do
not invalidate the quark mass ratio extraction. An advantage of bottomonium
transitions is that the $B$ meson mass difference $M_{B^0}-M_{B^+}$ is
about 10 times smaller than the quark mass difference $m_d-m_u$, so that the
bottom meson loops in the isospin breaking bottomonium transitions are indeed
suppressed. Furthermore, because $|M_{b\bar b}-2 M_B|\ll M_B$, the bottomed-meson
velocity $v$ is small and the system is nonrelativistic. Therefore a power counting
approach in the framework of nonrelativistic effective field theory can be used to
analyse $\eta$-emitting transitions~\cite{FK:NREFT2}. It turns out that the bottom
meson loops are always important in transitions between two $S$-wave as well as two
$P$-wave bottomonia. However, for the single-$\eta$ transition amplitude between a
$P$-wave and a $S$-wave bottomonium, the effect of the loops can be estimated
as~\cite{FK:NREFT2,FK:NREFT3} \be \frac{{\cal A}^{\rm loop}}{{\cal A}^{\rm tree}}
\sim \frac{\vec{q}^2}{v^3M_B^2}, \ee where $\vec{q}$ is the three-momentum of the
$\eta$ in the rest frame of the decaying bottomonium and $M_B$ is the mass of the
intermediate bottom meson. Decay of $\Upsilon(5S)$ provides enough phase space to
produce both $h_b(1P)\eta$ and $h_b(2P)\eta$ final state. For the decay
$\Upsilon(5S)\to h_b(1P)\eta$
$$
|\vec{q}|=771~\mbox{MeV},\quad v\approx0.5,\quad\mbox{so that}\quad\frac{\vec{q}^2}{v^3M_B^2}\sim 0.7,
$$
which implies that loops are not negligible and this mode cannot be used for the quark mass ratio extraction. In the
meantime, for the decay mode $\Upsilon(5S)\to h_b(2P)\eta$,
$$
|\vec{q}|=274~\mbox{MeV},\quad v\approx0.3,\quad\mbox{so that}\quad\frac{\vec{q}^2}{v^3M_B^2}\sim 0.2.
$$
This ensures considerable suppression of the bottom meson loops (uncertainty arising
due to the loop contributions is of the order of 20\%). Hence, this mode, together with
its counter-part mode $\Upsilon(5S)\to h_b(2P)\pi^0$, is suitable for the quark
mass ratio extraction.

\section{Open-flavour analysis and {\boldmath $\Upsilon(5S)$} internal structure}

\begin{table*}
\begin{center}
\begin{tabular}{|l|c c c c c c|}
\hline
Mode    & $B\bar B$         & $B^*\bar B+c.c.$         & $B^* \bar B^*$    & $B_s \bar B_s$    & $B_s^*\bar B_s+c.c.$  
& $B_s^*\bar B_s^*$ \\
\hline
BF, $\%$& $5.5\pm 1.0$ & $13.7 \pm 1.6$ & $38.1 \pm 3.4$ & $0.5\pm 0.5$ & $1.5\pm 0.7$ &
$17.9\pm 2.8$\\
\hline
\end{tabular}
\caption{PDG values for the branching fractions (BF) of two-body $\Upsilon(5S)$ decays.  \label{table:2bodybranch}}
\end{center}
\end{table*}

Important information can be also obtained from various $\Upsilon(5S)$ decays
to open-flavour final states, where the $b$-$\bar{b}$ pair separate into different final state mesons.
Among them, decays to two-body final states are of a special interest because these decays can be precisely measured
experimentally and reliably calculated theoretically.
Good measurements are also important as benchmarks for the high-energy nuclear physics experimental programme where
op\-en-flavour decays of heavy quarkonia are used as probe of high-density nuclear matter~\cite{Friman:2002fs}.
The branching fractions (BF) for all six available open-flavour decays
$\Upsilon(5S)\to B_x \ov{B}_x$ have been measured~\cite{PDG} and the results are shown in Table~\ref{table:2bodybranch}.

\subsection{Problems with light-quark $SU(3)$ flavour symmetry}

Open-flavour hadron decays entailing light quarks are
often fruitfully analysed with the help of $SU(3)$ flavour symmetry. The assumption underpinning effective theory
analysis is that hadrons are pointlike particles at small energy scales, and $SU(3)$ symmetry relates the couplings of
the parent state to the various decay channels.

With the $\Upsilon(5S)$ being a $SU(3)$ flavour singlet (as natural for a $b\bar b$ state),
$SU(3)$ is badly broken by the branching fractions shown in Table~\ref{table:2bodybranch} (as will be
demonstrated below in this chapter).

Had $SU(3)$ flavour symmetry been exact,
the couplings of the $\Upsilon(5S)$ to the $B^0\bar B^0$, $B^+B^-$ and $B_s^0\bar
B_s^0$ would all have been the same.
In practice one would expect some deviations in the branching fractions due to phase space differences (the strange
quark being somewhat heavier). But these will turn out to be larger than expected, see Eq.~(\ref{BFdeviations}) below.

Furthermore, a naive application of heavy-quark spin symmetry would  entail that the
couplings to the vector bottom mesons would also be the same.
Therefore, all six open-flavour decays should be related to each other, with their decay amplitudes proportional to the
same coupling constant.
Thus, all difference in the branching fractions should be related to the decay kinematics.

In detailed calculations,
because the $\Upsilon(5S)$ width is large when compared to the
difference between its mass and the thresholds of certain channels, such as $B_s\bar
B_s^*$ and $B_s^* \bar B_s^*$, one should take into account the finite-width effect
when calculating its decay widths into the open-bottom mesons.
A typical decay formula reads
\bea
\ds\Gamma = \frac1{W}\int_{w_{\rm min}^2}^{(M_\Upsilon+2\Gamma_\Upsilon)^2} \!\!ds\,
\frac{(2\pi)^4}{2\sqrt{s}} \int d\Phi_2 |\mathcal{A}|^2 \frac1{\pi}\nonumber\\[-2mm]
\\[-2mm]
\ds\times{\rm
Im}\left(\frac{-1}{s-M_\Upsilon^2+iM_\Upsilon\Gamma_\Upsilon}\right),\nonumber \eea
where $M_\Upsilon$ and $\Gamma_\Upsilon$ are the mass and width of the
$\Upsilon(5S)$, respectively, see Eq.~(\ref{eq:Mass5S}), $w_{\rm
min}=\max(M_\Upsilon-2\Gamma_\Upsilon,m_1+m_2)$ with $m_1$ and $m_2$ the masses of
the bottom mesons in the final state,
$\int d\Phi_2$ denotes the two-body phase space,
and $\mathcal{A}$ is the decay amplitude. The factor $1/W$
is included in order to normalise the spectral function of the $\Upsilon(5S)$, and
it can be calculated as
$$
W = \int_{w_{\rm min}^2}^{(M_\Upsilon+2\Gamma_\Upsilon)^2}
\!\!ds\, \frac1{\pi}{\rm
Im}\left(\frac{-1}{s-M_\Upsilon^2+iM_\Upsilon\Gamma_\Upsilon}\right)\ .
$$

Thus, in the $SU(3)$ limit, the ratios of the branching fractions of all the
open-bottom decay modes can be obtained by eliminating the coupling
implicit in $|\mathcal{A}|^2$, assuming that it is the same for every channel,
\bea\label{BFdeviations}
&B\bar B: [B^*\bar B+c.c.] : B^*\bar B^* : B_s \bar B_s : [B^*_s\bar B_s+c.c.] : B^*_s\bar B^*_s \non &\\
&= 1:3.21:12.95:0.16:0.36:0.78.&
\eea

However, slightly different ratios result from the
measurements shown in Table~\ref{table:2bodybranch}:
\bea
&B\bar B: [B^*\bar B+c.c.] : B^*\bar B^* : B_s \bar B_s : [B^*_s\bar B_s+c.c.] : B^*_s\bar B^*_s &\non \\
&=1:(2.49\pm0.54):(6.93\pm1.40):(0.09\pm0.09) &\nonumber \\
&:(0.27\pm0.14):(3.25\pm0.78).&
\eea
The largest deviations from the $SU(3)$ limit occur in the vector
channels $B^*\bar B^*$ and $B^*_s\bar B^*_s$, and they occur in opposite
directions. So, the experimental value
$$
B^*_s\bar B^*_s:B^*\bar B^*=0.47\pm0.08
$$
is one-order-of-magnitude larger than the $SU(3)$ value of $0.06$. This
means that the coupling of the $\Upsilon(5S)$ to the strange vector channel
$B^*_s\bar B^*_s$ is stronger than that of the non-strange vector channel $B^*\bar
B^*$ by a factor of 2. Normally, $SU(3)$ breaking does not exceed
30\%.

Such a huge $SU(3)$ breaking must have a dynamical origin.
One possibility is that the $\Upsilon(5S)$ cannot be taken as a pure light flavour
singlet. The possibility of an isospin-1 state is ruled out by the decays of
$\Upsilon(5S)$ into the bottom-strange mesons. It would be more subtle if the
structure around the $\Upsilon(5S)$ contained a sizable tetraquark
component~\cite{tetraquark2,Ali:2010pq,tetraquark3}, see Sec.~\ref{tetrasec}, with
the light quark part of the wave function being  purely an isoscalar ($a(|u\bar
u+d\bar d\rangle )+b|s\bar s\rangle $).

Another possibility that we address in the next subsection, is that the internal
structure of the parent bottomonium state enhances the kinematic differences
between light quarks \emph{at the coupling level}.

\begin{figure*}[ht]
\begin{center}
\includegraphics[width=10cm]{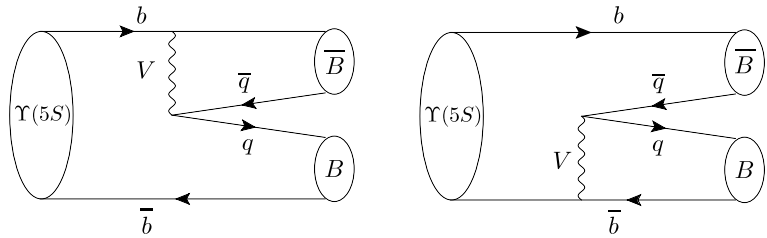}
\end{center}
\caption{Diagrams used to calculate the open-flavour decays of $\Upsilon(5S)$ in \cite{Hwang:2008kb}.}\label{fig:5Sdiag}
\end{figure*}

\subsection{Internal {\boldmath $b\bar{b}$} structure.}

\begin{figure*}[ht]
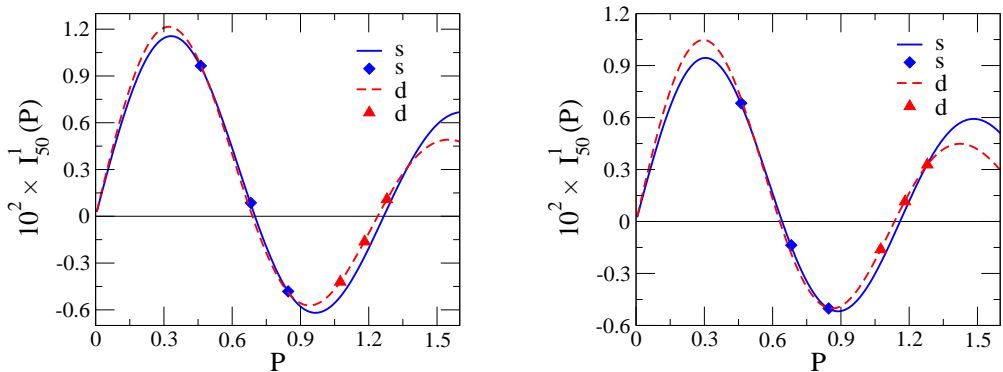

\begin{center}
\includegraphics*[width=6cm]{FIGS.DIR/Position_InL_A.eps}\hspace*{10mm}
\includegraphics*[width=6cm]{FIGS.DIR/Position_InL_B.eps}
\end{center}
\caption{The position of the final momentum $P$ of the $B$ meson for each mode
in the $\Upsilon (5S)$ rest frame for two different sets of model parameters. The decay amplitude $A_{5S}$ is related to
the quantity $I_{50}^1$ through a simple numerical factor (see \cite{Hwang:2008kb} for the details).}\label{fig:5SP}
\end{figure*}

We now turn to the structure of the parent state.
In \cite{Hwang:2008kb}, the branching fractions in Table~\ref{table:2bodybranch} were calculated in a model with the
Hamiltonian (:: stand for normal ordering)
\bea
H_I=\frac12\sum_{k=1}^8\int d^3x d^3y
:\left[\psi^{\dagger}(\vec{x})\frac{\lambda_k}{2}\psi(\vec{x})\right] V(\vec{x}-\vec{y})\nonumber\\[-2mm]
\\[-2mm]
\times\left[\psi^{\dagger}(\vec{y})\frac{\lambda_k}{2}\psi(\vec{y})\right]:\nonumber
\eea
given by the quark colour densities interacting via the Cornell potential
\be
V(r)=-{\kappa\over r}+{r\over a^2}.
\ee
The corresponding diagrams are shown in Fig.~\ref{fig:5Sdiag}. It was noticed then
that the decay amplitude $A_{5S}(P)$ is an oscillating function of $P$ (the
centre-of-mass momentum of the $B$-mesons in the final state), so that its value
changes dramatically from one decay channel to another (see Fig.~\ref{fig:5SP}),
that may provide an explanation of the data presented in
Table~\ref{table:2bodybranch}.

In \cite{felipe,TorresRincon:2010fu} the formalisation of these statements, to
reduce model dependence, exploited the analogy between open-flavour $\Upsilon(5S)$
decays and the Franck-Condon factorisation in molecular physics (therefore the term
``heavy quark fluorescence''  used). The central idea of the approach is a
so-called ``velocity superselection rule'' encompassed, for example, in the
leading-order of such effective theories for QCD with heavy quarks as potential
Non-Relativistic QCD (pNRQCD) and Heavy Quark Effective Theory (HQET). This rule
states that the heavy quark does not change its velocity upon emitting or
interacting with the light degrees of freedom, such as light quarks, gluons, pions,
and so on, with a momentum of $\mathcal{O}(\Lambda_{\rm QCD})$. This entails
that the momentum distribution of the heavy mesons in open-flavour decays
should be proportional to the momentum distribution of the heavy quarks inside the
parent hadrons, thus giving a window to their internal
structure~\cite{felipe,TorresRincon:2010fu}.

In mathematical terms, the momentum $k$ and velocity $v$ of the final $B_x$ meson are related to those of the $b$ quark
as
$$
k=k_b+\mathcal{O}(\Lambda_{QCD}),\quad v=v_b+\mathcal{O}(\Lambda_{QCD}/M),
$$
so that one has a leading-order factorisation theorem
\bea
\mathcal{A}_{\Upsilon\to B_x\ov{B}_x}(k) &=& \int q^2 dq \psi_\Upsilon(q)\langle
b\bar{b}(q)|H_{\rm int}|B_x\bar{B}_x(k)\rangle \non\\
&\simeq& F(k)\psi_\Upsilon(k),
\eea
in analogy to the above-mentioned Franck-Condon
factorisation in molecular physics.

\begin{figure*}[t]
\begin{center}
\includegraphics[width=6.8cm]{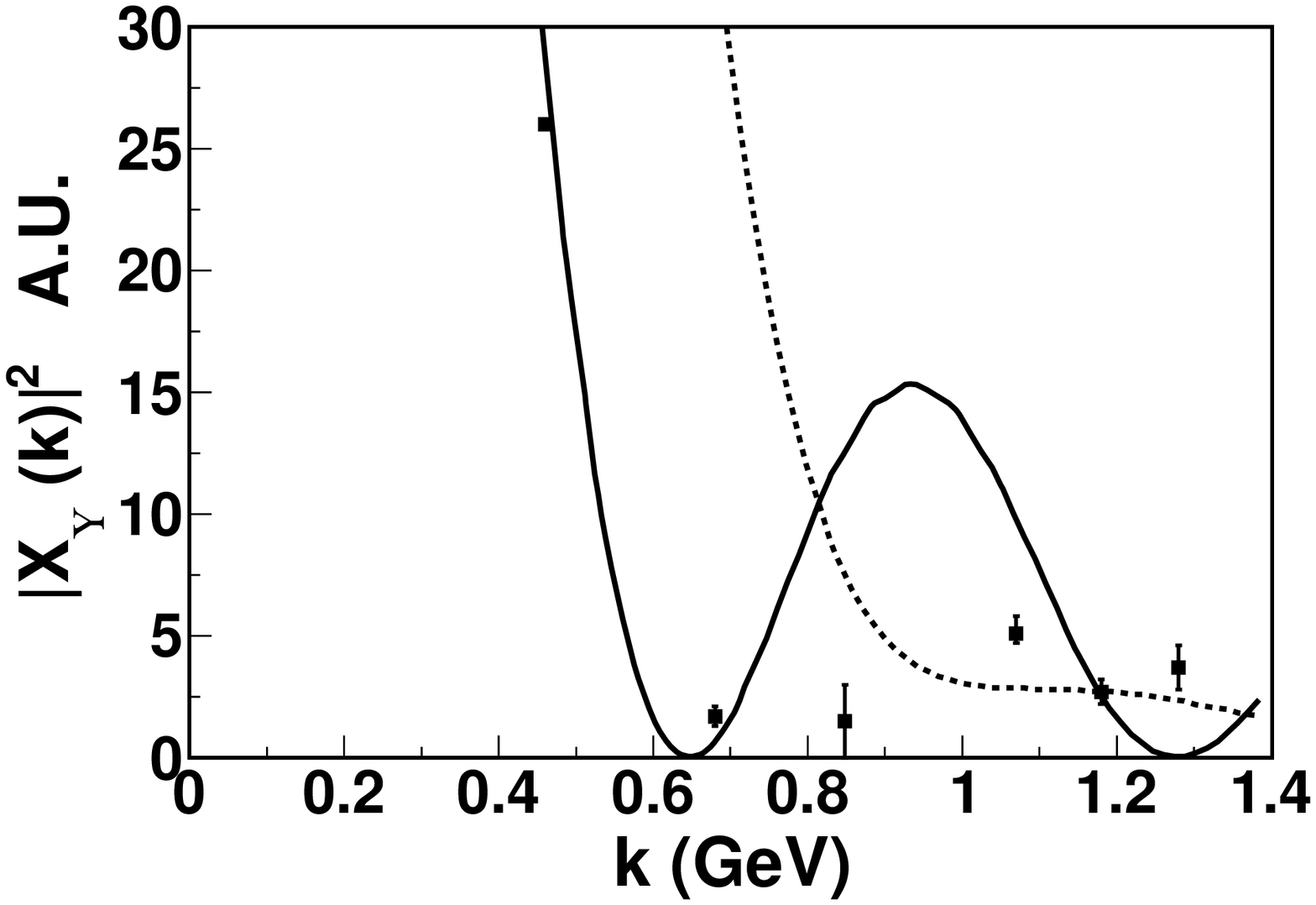} \hspace*{10mm}
\includegraphics[width=7.4cm]{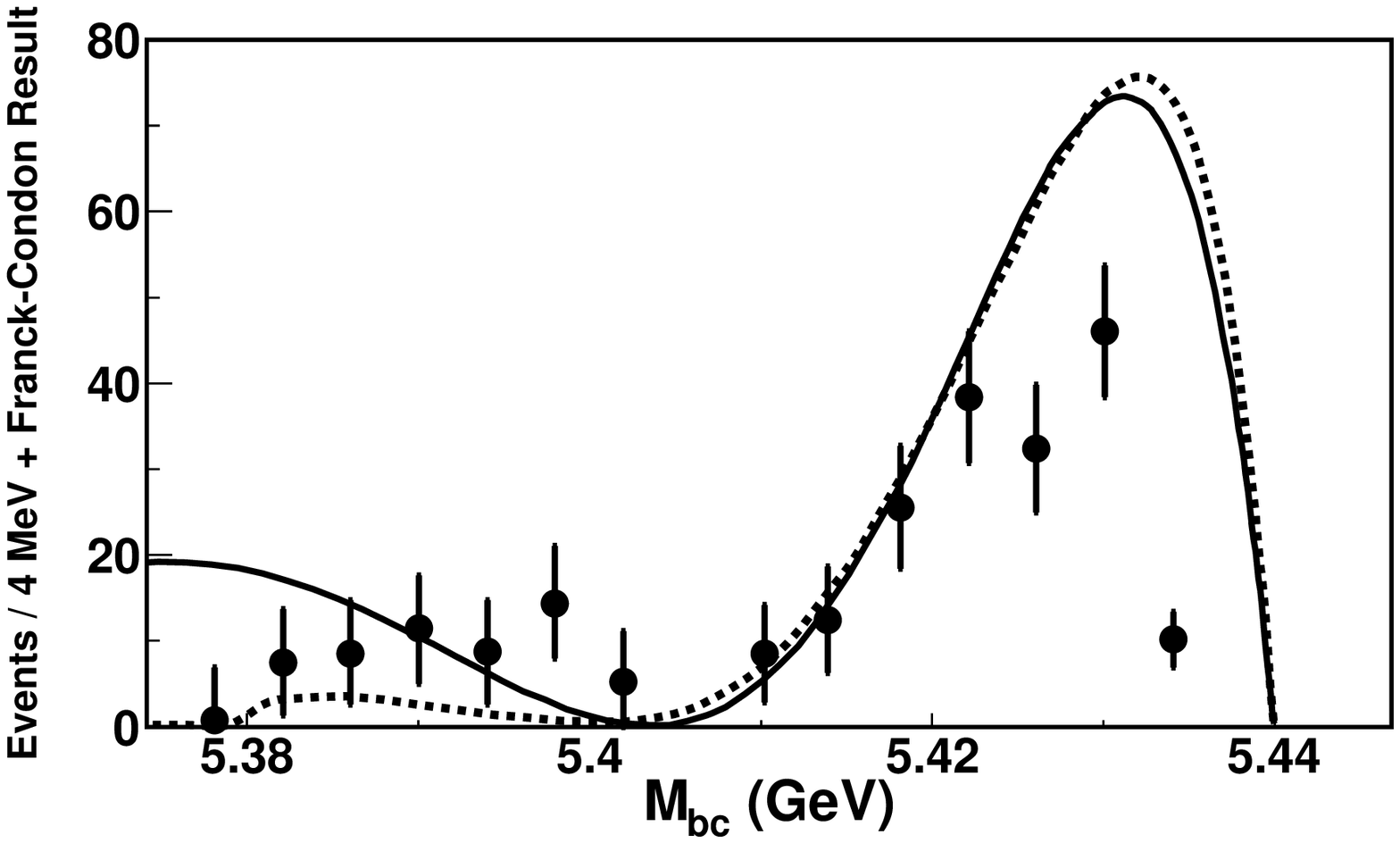}
\caption{Open-flavour decays of the $\Upsilon(5S)$ resonance as a probe of the $b$-quark
wave function squared (adapted from~\cite{TorresRincon:2010fu}).
Upper panel (two body decays): data points are, from right to left, for the channels
$B\bar{B}$, $B\bar{B}^*+B^*\bar{B}$, $B^*\bar{B}^*$, $B_s\bar{B}_s$, $B_s\bar{B}^*_s+B^*_s\bar{B}_s$,
$B^*_s\bar{B}^*_s$; the solid curve is for the Coulomb-gauge model wave function of
the $\Upsilon(5S)$. Lower panel (three body decays): data points are for $B^{(*)}\ov{B}^{(*)}\pi$
decays. Here $M_{bc} \equiv \sqrt{(M_{\Upsilon}/2)^2-k^2}$.
The experimental data, taken from \cite{Drutskoy:2009ci} and \cite{Drutskoy:2009ei},
are divided by the decay phase space factor, by the spin factor, and by an isospin
factor 2 for the non-strange decays. The remainder is proportional to the wave function
squared. The dashed curves include a $^3P_0$-vertex model computation.}\label{fig:dec:Upsilon}
\end{center}
\end{figure*}

Therefore, to extract information about the wave function squared
$|\psi_\Upsilon(k)|^2$ in bottomonium one needs to measure the $B$ meson momentum
distributions in various op\-en-flavour decays. In particular, in the two-body
$B_s^*\ov{B}_s^*$ channel, the $B_s^*$ meson possesses a relatively small momentum
$k$ and therefore this channel probes the central maximum of the wave function, as
depicted in Fig.~\ref{fig:dec:Upsilon} (left).

For the three-body channels (Fig.~\ref{fig:dec:Upsilon}, right), the zero
seen in the solid line is a Sturm-Liouville zero (a generic prediction of quantum
mechanics that requires orthogonality between all $nS$ bottomonia, and thus
requires the wave functions of excited states to vanish). Since this zero is
present in $\psi_\Upsilon(k)$, it should manifest itself in all open-flavour decays
and, in particular, one should expect a dip in the three-body $BB^*\pi$ channel
\cite{felipe}. With the high-statistics available at the future $B$-factories, it should be possible
to confirm or to refute the dip in the $B$ meson momentum plot.

\section{Non-{\boldmath $b\bar{b}$} vector states at {\boldmath $\Upsilon(5S)$} energy}\label{Hybridornot}

In this section we consider the possibility of existence for an alternative (non-conventional $b\bar{b}$) vector state
residing in the region of mass near 10860~MeV. We resort to the notation $Y_b$ for this vector state to distinguish it
from the $\Upsilon(5S)$ bottomonium discussed above. Notice that the vector $Y_b$ from this section should not be
confused with the hadronic molecules discussed in chapter~\ref{sec:xbyb}.

\subsection{{\boldmath $b\bar{b}g$} vector hybrid}

Since the very beginning of the QCD era, excitations of the glue attracted a lot of attention of both theorists and
experimentalists. Indeed, due to the non-Abelian nature of the interaction mediated by gluons, the latter can either
play a role of extra constituents of hadrons (hybrid mesons) or form a new type of compounds ---
glueballs made entirely of two, three, or more gluons. Although there is no general consensus on whether hybrids and
glueballs have indeed been observed experimentally, a number of promising candidates exist in various parts of the
spectrum
of hadrons. In particular, lattice simulations are known to predict the lowest bottomonium hybrid ($b\ov{b}g$) to lie
around 10900(100) MeV \cite{Morningstar}, that is exactly in the region accessible with $B$-factories. With the
high-statistics, high-resolution, and low-background data from future $B$-factories in perspective, especially in view
of the proximity of the expected mass of the vector bottomonium hybrid to the mass of the $\Upsilon(5S)$ state,
it is important to outline specific features inherent to bottomonium hybrids which would allow one to distinguish them
from conventional mesons.

\subsubsection{The radial wave function and the momentum distribution in open-flavour decays of the vector hybrid}

A number of theoretical approaches to hybrids has been discussed in the literature, such as
lattice simulations \cite{Morningstar,Michael,liu1,liu2,burch,dudek1,dudek2}, bag model
\cite{bag1,bag2}, flux tube model \cite{flux}, Coulomb-gauge QCD model
\cite{Cotanch:2001mc,General:2006ed,LlanesEstrada:2000hj}, potential quark model
\cite{bicudo},  con\-sti\-tu\-ent gluon model \cite{horn,orsay,orsay2}, QCD string model
\cite{hybrids1,hybrids21,hybrids22,yushybrdecay,hybrids3,Kalashnikova:2008qr,semay1,semay2}.

\begin{figure}[t]
\begin{center}
\begin{tabular}{c}
\hspace*{-22mm}\includegraphics[width=5cm]{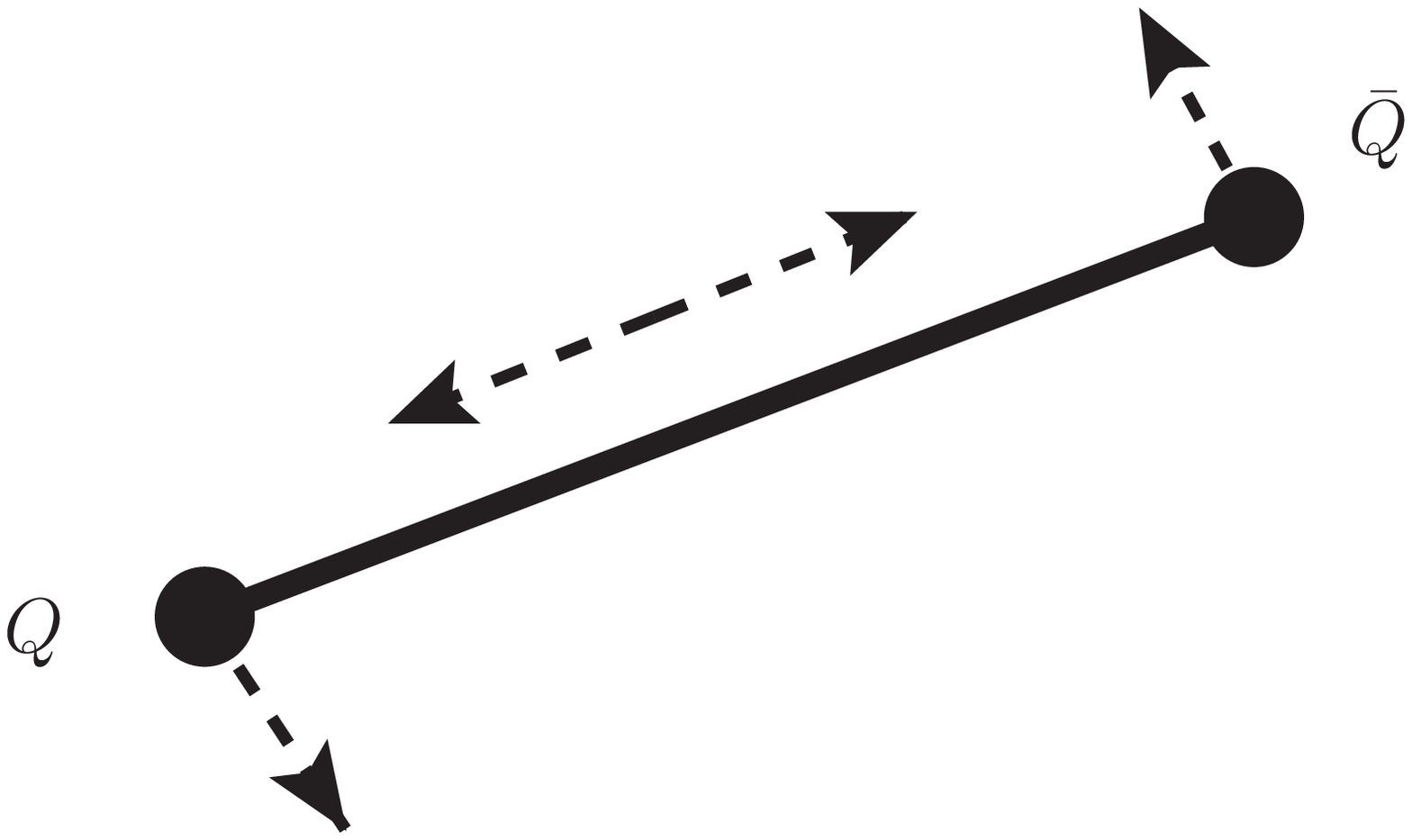} \\
\hspace*{30mm}\includegraphics[width=5cm]{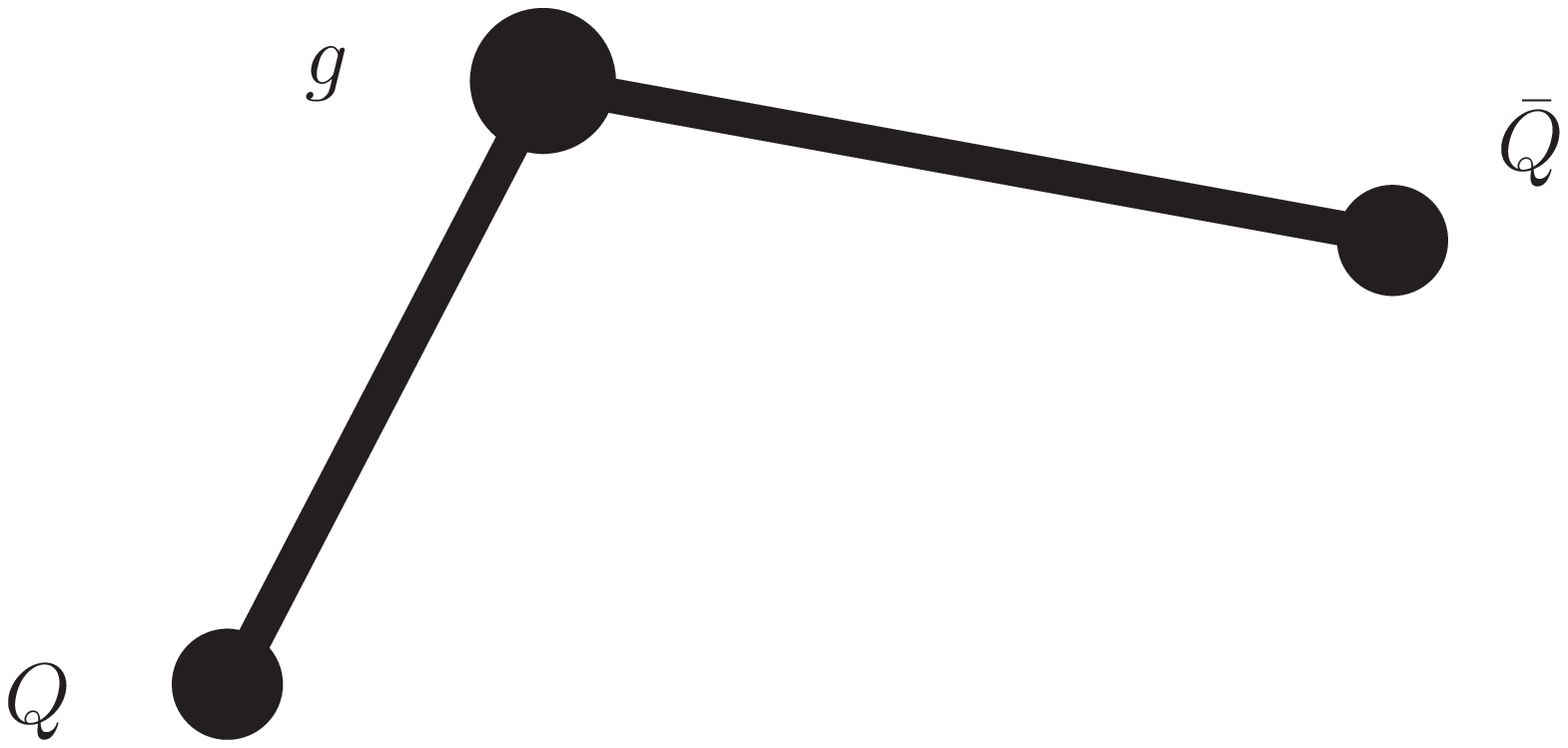}
\end{tabular}
\end{center}
\caption{Three possible types of excitations of the QCD string with quarks at the ends: rotation and radial motion
correspond to conventional mesons (first plot) while the string vibrations, described in terms of constituent gluons
attached to the string, correspond to hybrids (second plot).}\label{fig:hybrids}
\end{figure}

\begin{figure}[t]
\begin{center}
\includegraphics[width=7cm]{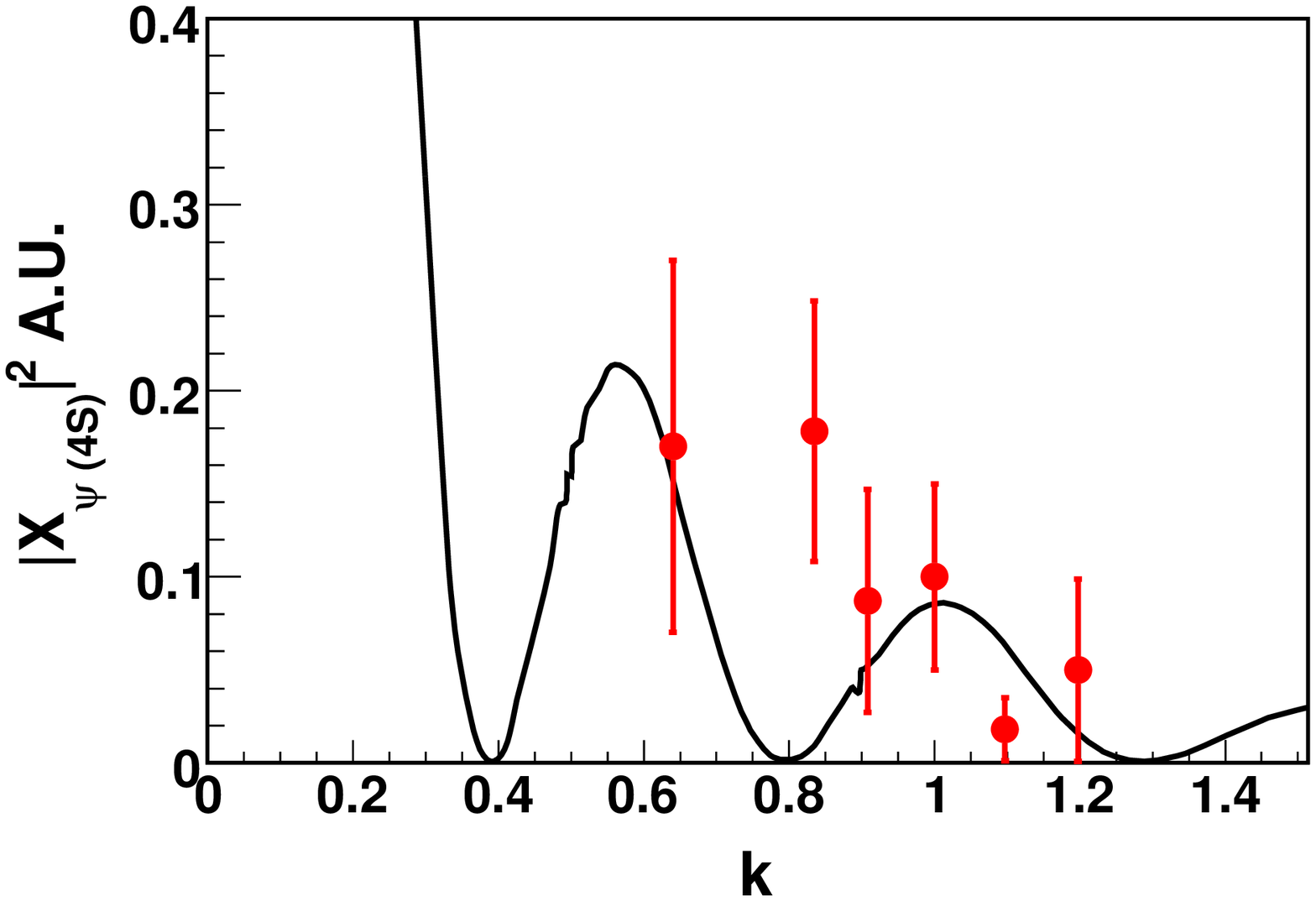}
\includegraphics[width=7cm]{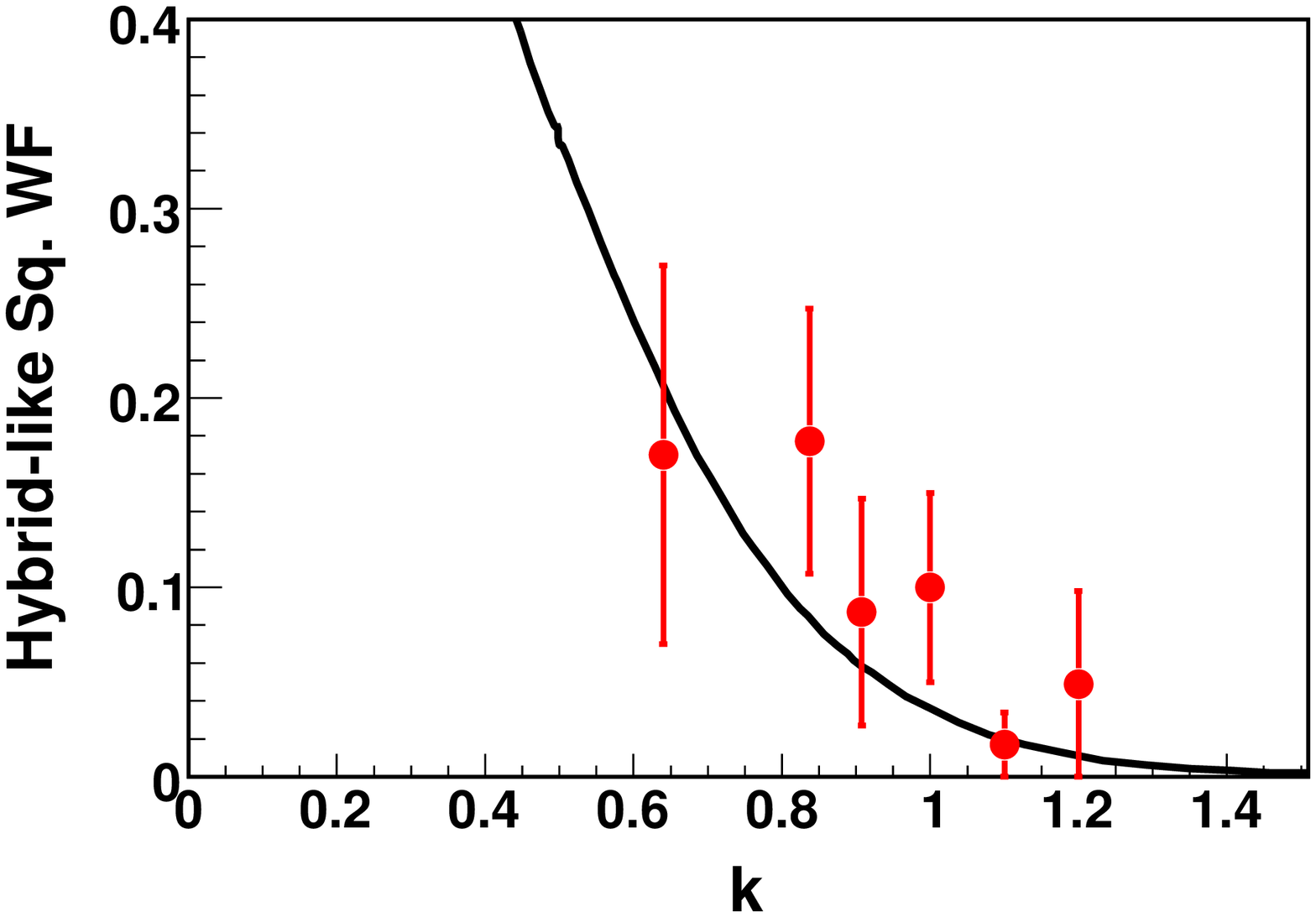}
\caption{Pure $c\ov{c}(4S)$ (first) versus charmonium hybrid (second) hypothesis (solid lines) for the $\psi(4400)$
decaying to
open-flavour. The solid lines are typical model-wave functions, and the coarse data are our adaption of various BaBar
and
Belle measurements of the 2-body branching fractions for $D\ov{D}\dots D_s^*\ov{D}_s^*$ dividing by known spin and
isospin factors.
\label{fig:ccbardecay}}
\end{center}
\end{figure}

For the sake of clarity, respective cartoons for the ordinary meson and for the
single-gluon hybrid in the framework of the QCD string model are shown in
Fig.~\ref{fig:hybrids}. Notice that excitation of the gluonic degree of freedom
brings a large contribution to the mass of the state, so that the lowest (not
excited radially) hybrid meson has a mass comparable with that of the highly
radially excited ordinary meson. Therefore, one can distinguish the radially
excited $\Upsilon(5S)$ quarkonium state and the hybrid $Y_b$ by the different
patterns of their radial quark-antiquark wave functions (4-node wave function
versus nodeless wave function). As argued in \cite{felipe}, the shape of the
$b\ov{b}$ wave function translates into the $B\ov{B}$ momentum distribution for the
three-body final $B\ov{B}\pi$ state, so that it is possible to make a conclusion
concerning the source of the $B\ov{B}$ pair: smooth distribution would identify the
source as the hybrid while the distribution with residual (after smearing due to
the quark recoil in $B$ mesons) structures would indicate a conventional quarkonium
as the corresponding source. In Fig.~\ref{fig:ccbardecay} this idea is exemplified
by the charmonium $\psi(4400)$ resonance, for which the existing Belle and BaBar
open-flavour $D_{(s)}^{(*)}\ov{D}_{(s)}^{(*)}$ decays data are used. A similar
analysis could be carried out for bottomonium resonances decaying to the
open-flavour $B_{(s)}^{(*)}\ov{B}_{(s)}^{(*)}$ channels.

\subsubsection{Sibling states}

One of the most prominent features of hybrids is the fact that exotic quantum numbers, such as $1^{-+}$ forbidden for a
conventional quarkonium, are accessible for hybrids. Indeed, once a hybrid possesses a gluon as an extra degree of
freedom as compared to the conventional quark--antiquark state, then this new degree of freedom contributes to the $C$-
and $P$-parities
of the system. In particular, for the constituent gluon or for the QCD string model, the quantum numbers of a one-gluon
hybrid are
\be
P=(-1)^{l_{q\ov{q}}+j},\quad C=(-1)^{l_{q\ov{q}}+s_{q\ov{q}}+1},
\label{JPCmag}
\ee
for the ``magnetic'' gluon ($l_g=j$), and
\be
P=(-1)^{l_{q\ov{q}}+j+1},\quad C=(-1)^{l_{q\ov{q}}+s_{q\ov{q}}+1},\\
\label{JPCel}
\ee
for the ``electric'' gluon ($l_g=j \pm 1$),
where $l_g$ is the relative angular momentum between the $q\ov{q}$ pair and
the gluon, $j$ is the total angular momentum of the gluon, $l_{q\ov{q}}$ is the
orbital momentum in the quark-antiquark subsystem, and $s_{q\ov{q}}$ is the spin of the quark-antiquark pair.

The vector quantum numbers of highest interest for $e^+e^-$ colliders can be achieved therefore both for the
electric gluon,
\be
s_{q\ov{q}}=1,\quad l_{q\ov{q}}=1,\quad l_g=0,2,\quad j=1,
\ee
as well as for the magnetic gluon,
\be
s_{q\ov{q}}=0,\quad l_{q\ov{q}}=0,\quad l_g=1,\quad j=1.
\ee

However, hybrids with an electric gluon couple too strongly to two $S$-wave
final-state mesons and, as estimated in \cite{orsay2}, do not exist as
resonances. On the contrary, for hybrids with a magnetic gluon, a selection rule is
established (see, for example, \cite{orsay,orsay2,yushybrdecay,kou,hybrdecayflux1,hybrdecayflux2})
which forbids its $B^{(*)}_{(s)}\ov{B}^{(*)}_{(s)}$ decay modes, so that the lowest
possible open-beauty modes are the ones with one $S$-wave and one $P$-wave $B$ meson (see
below).

The lowest states with magnetic gluons have $l_{q\ov{q}}=0$. Then the $1^{--}$ hybrid is a
spin-singlet state with respect to the quark spin,
\be
|1^{--}\rangle_m=\Phi(r,\rho)S_0(q\bar{q})\sum_{\nu_1\nu_2}C^{1m}_{1\nu_1
1\nu_2}
\rho Y_{1\nu_1}(\hat {\rho})S_{1 \nu_2}(g),
\label{vector}
\ee
and there are three other hybrid states, with $J^{-+}$, $J=0,1,2$, lying in the vicinity of the vector hybrid. These are
spin triplets,
\bea
|J^{-+}\rangle_m&=&\Phi(r,\rho)\sum_{\mu_1\mu_2}C^{Jm}_{1\mu_1 1\mu_2}S_{1\mu_1}(q\bar{q})\nonumber\\
&\times&\sum_{\nu_1 \nu_2}C^{1 \mu_2}_{1 \nu_1 1 \nu_2}\rho Y_{1 \nu_1}(\hat {\rho})S_{1 \nu_2}(g)\ , 
\label{siblings}
\eea
where $S_{1\nu}(g)$ is the spin wave function of the gluon, $S_0(q\bar{q})$ and
$S_{1\nu}(q\bar{q})$ are the singlet and triplet spin wave functions of the $q\bar{q}$ pair, $\Phi(r,\rho)$ is the
radial wave function
in momentum space. The four states (\ref{vector}) and (\ref{siblings}) are expected to be degenerate in the heavy-quark
limit, with the degeneracy removed by spin-dependent quark--gluon interactions.

The prediction for hybrids is therefore that, together with the vector hybrid, three more sibling hybrid states with
the quantum numbers $(0,1,2)^{-+}$ should exist in the same region of mass.

\subsubsection{Open-flavour decay pattern}

\begin{figure}
\begin{center}
\includegraphics[scale=0.4]{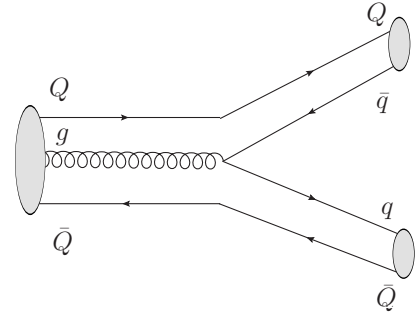}
\end{center}
\caption{Graphical representation for a $Q\bar{Q}g$ hybrid open-flavour decay.}\label{fig:decaydiag}
\end{figure}

Schematic representation of the single-gluon hybrid open-flavour decay is given in
Fig.~\ref{fig:decaydiag}. With the form of the wave functions (\ref{vector}) and
(\ref{siblings}) in hand, it is straightforward to calculate the corresponding
$Y_b\to \ov{B}(1S)B(1S)$ and $Y_b\to \ov{B}(1S)B(1P)$ recoupling coefficients
\cite{Kalashnikova:2008qr}\footnote{Notice that here and in what follows, where it
applies, charge conjugated components of the $Y_b$ wave function are omitted for
simplicity and an obvious shorthand notation is used, for example,
$\ov{B}(1S)B(1P)\equiv \frac{1}{\sqrt{2}}(\ov{B}(1S)B(1P)+B(1S)\ov{B}(1P))$. }. In
Table~\ref{t1} such coefficients are listed for the vector hybrid. Similar
coefficients for the siblings can be found in \cite{Kalashnikova:2008qr}. As was
mentioned above, all coefficients $Y_b\to B(1S)\ov{B}(1S)$ vanish.

\begin{table*}[t]
\begin{center}
\begin{tabular}{|c|ccccccccccc|}
\hline
Channel&$\ov{B}B$&$\ov{B}B^*$&$\ov{B}^*B^*$&$\ov{B}B_0$&$\ov{B}^*B_0$&$\ov{B}B_1({}^1P_1)$ $\vphantom{\ds\int_0^1}$
&$\ov{B}^*B_1({}^1P_1)$&$\ov{B}B_1({}^3P_1)$&$\ov{B}^*B_1({}^3P_1)$&$\ov{B}B_2$&$\ov{B}^*B_2$
$\vphantom{\ds\int_0^1}$\\
\hline
Coefficient&0&0&0&0&$\ds\frac{1}{\sqrt{6}}$&0&$\ds-\frac12$&$\ds\frac12$&$\ds\frac{1}{2\sqrt{2}}$&0&$\ds-\frac12\sqrt{
\frac56 }$ \\
\hline
\end{tabular}
\caption{Spin--recoupling coefficients for the vector hybrid.
Here $B^{(*)}$ is an $S$-wave $B^{(*)}$ meson and $B_J$ is a $P$-wave
$B$ meson with the total momentum $J$.}\label{t1}
\end{center}
\end{table*}

The vector hybrid decay modes into $\ov{B}B_J$ with $J=0,1,2$, followed by the decays $B_J\to B^{(*)}\pi$ (currently
known modes are $B_1\to B^{*+}\pi^-$, $B_2\to B^{*+}\pi^-$, and $B_2\to B^+\pi^-$), populate the final states
$\ov{B}^*B\pi$ and $\ov{B}B\pi$. Notice that decays into $\ov{B}^{(*)}B_J$ are not allowed because of the mass
threshold.

To summarise, the vector hybrid $Y_b$ is expected
\begin{itemize}
\item to contribute to the three-body $B\ov{B}^{^{(*)}}\pi$ final states;
\item to possess three hybrid siblings with the mass splittings about 20-40~MeV between each other and with the vector
$Y_b$ and with quantum numbers $J^{-+}$ $J=0,1,2$, including the exotic one $1^{-+}$;
\item to have a different $\ov{B}B$ momentum distributions in the final $\ov{B}B\pi$ state than conventional $\ov{b}b$
quarkonium.
\end{itemize}

\subsubsection{A scan between {\boldmath $\Upsilon(4S)$} and {\boldmath $\Upsilon(5S)$}}

The proliferation of vector charmonium or charmonium-like states
above threshold, that includes at least $\psi(3770)$, $\psi(4040)$, $\psi(4160)$,
$Y(4260)$, $\psi(4400)$, $Y(4360)$, $Y(4660)$, makes clear that
not all can be pure charmonium states, as the quark model cannot support such
a multiplicity of mesons.

The $b\bar{b}$ system, with better spaced narrow resonances, is a system where more
clarity can be achieved, especially concerning the existence or non-existence of
similar non-$b\bar{b}$ excitations. The BaBar Collaboration~\cite{Aubert:2008ab} has
conducted a scan between 10.5 and 11.2 GeV with 300 energy steps and about 25
inverse picobarn per step, with additional data around the $\Upsilon(6S)$ resonance.

Only the putative $\Upsilon(5S)$ and $\Upsilon(6S)$ states at 10860 MeV and 11020~MeV were revealed by
this scan. Thus, the room for hybrid mesons shrinks to two options. Either the
narrower vector hybrid is part of one of these two resonances, or if its mass is
elsewhere in the interval of interest (as predicted by theory), its $e^+e^-$
coupling is very small and higher beam intensity is needed to produce it. This is
not totally unexpected in view of the difficulty of exciting the flux tube by
an electromagnetic current. Thus, a rescan of this energy range with increased
luminosity would be welcome (the mass resolution reached by BaBar is sufficient for
this application and does not need to be increased).

Additionally, BaBar observed a cross-section dip due to interference at the
$B_s\bar{B}_s$ threshold, but no structure at around 11100-11200~MeV, which is the
energy where the pair of quarks $bs$ and antiquarks $\bar{b}\bar{s}$ can first be
produced in an S-wave. This could be rechecked, and it would be interesting to
examine whether any sign of the positive-parity $B_s$ mesons appears in the total
$b$ cross section.

Funally, it would be interesting to examine exclusive channels with closed
flavour, such as $\Upsilon^{(\prime)}\pi\pi$ and $\Upsilon^{(\prime)} KK$ in this
mass range, because the analogous $Y$ resonances in the charmonium spectrum
do not appear as bumps in the total charm cross section, but rather
exhibit themselves in the $\psi^{(\prime)}\pi\pi$ distributions. In
particular, some of the $Y$ states may be understood as
hadro-charmonia~\cite{Dubynskiy:2008mq}, such as the $Y(4660)$ as a $\psi'f_0(980)$
bound state~\cite{Guo:2008zg}. Such kind of states would be more visible in the
closed-flavour than the open-flavour final states. Analogously, there can be
hadro-bottomonia which can be searched for in the $\Upsilon^{(\prime)}\pi\pi$ and
$\Upsilon^{(\prime)} KK$ distributions using the initial-state radiation (ISR)
technique.

\subsection{Vector tetraquark}\label{tetrasec}

Tetraquark models \cite{tetraquark2,tetraquark1} result in a prediction of existence of two (light and heavy)
states, $Y[b,l/h]$,
which are linear superpositions of the $J^{PC}=1^{--}$ tetraquark states $Y[bu]=[bu][\ov{b}\ov{u}]$
and $Y[bd]=[bd][\ov{b}\ov{d}]$. If the mass difference between these two states, estimated as
$M(Y[b,h])-M(Y[b,l])=(5.6\pm 2.8)$ MeV \cite{tetraquark2,tetraquark1}, is neglected, the corresponding degenerate
tetraquark states can be considered as a candidate for the vector $Y_b$.

Presence of the light-quark pair in the wave function of the tetraquark $Y_b$ leads to a number of predictions:
\begin{itemize}
\item Enhancement of the dipion transitions $Y_b\to\Upsilon(nS)\pi\pi$ $(n<5)$ as compared to the transitions of the
conventional quarkonium state, $\Upsilon(5S)\to\Upsilon(nS)\pi\pi$. Predictions of the model are consistent (both for
the
distributions in the dipion invariant mass and for the helicity angle) \cite{tetraquark3} with the Belle data
\cite{Belle2pi}.
\item Model predictions exist for the $K^+K^-$ and $\eta\pi^0$ invariant mass distributions for the processes $e^+e^-\to
Y_b\to \Upsilon(1S)K^+K^-$, and $e^+e^-\to Y_b\to \Upsilon(1S)\eta\pi^0$, respectively. The model \cite{tetraquark2}
predicts the ratio
$$
\frac{\sigma(\Upsilon(1S)K^+K^-)}{\sigma(\Upsilon(1S)K^0\ov{K}^0)}=\frac14.
$$
\item The decay widths $Y_b\to h_b(1P)\eta$ and $Y_b\to h_b(2P)\eta$ are expected to be abnormally large, since the OZI
suppression does not apply.
\end{itemize}

It should be noted that the calculations \cite{tetraquark2} are based on an extra
assumption that the low-lying scalar $0^{++}$ states ($f_0(500)$, $f_0(980)$,
$a_0(980)$) are tetraquarks.

\section{Exploration of higher energies: towards the {\boldmath $\Upsilon(11020)$} state and above}

In this chapter we review some physics topics that can be addressed if a future $B$-factory can
access energies above the $\Upsilon(5S)$ resonance. It includes studies in the region of the
last currently known resonance of the bottomonium system, the $\Upsilon(11020)$, as
well as the production of triply charmed baryons (for which very clean theoretical
predictions are possible) and generally triple charm, and the $\Lambda_b^0 \bar{\Lambda}_b^0$ system
(that in addition to precision tests of pNRQCD will allow further studies of flavour
and $CP$-violation).

\subsection{Studies at the {\boldmath $\Upsilon(11020)$}}\label{11020}

It was argued in Sec.~\ref{Hybridornot} that, if hybrid bottomonium is discovered,
its exotic $J^{PC}$ quantum numbers can be used as a smoking gun, since both
flux-tube and quasi-particle models make very specific predictions for their
possible combinations.

However, if these states appear around the $\Upsilon(5S)$, their production with
the $1^{--}$ quantum numbers in the initial state requires running at a higher
energy, so they can appear as intermediate states in various decay chains. In
addition, the vector hybrid can be produced directly in ISR processes, that also
requires higher energies, around 11~GeV, to be accessible with a future $B$-factory, so it makes
sense to take data at 11020 MeV for spectroscopy studies. An
interesting question is whether the $B_s^{(*)}
\ov{B}_s^{(*)}$ branching fractions for this resonance can be a source of
bottom-strange mesons competitive with the $\Upsilon(5S)$ state.

In addition, one could think of reconstructing the $\omega$ meson in the reaction
$\Upsilon(11020)\to \omega\chi_b'$ \footnote{One should identify the $\omega$ in
the dilepton or triple pion channel, and then plot its momentum and energy spectra
in the initial $\Upsilon$ centre of mass frame. The recoiling positive parity meson
would appear as a resonance in this. The phase space being so tight, since
$11020-780=10240$~MeV, whether the positive parity state is detected will depend on
its precise mass, that might exceed the experimental window. At least a bound can
be set.}. This would open the second quadruplet of positive parity, excited
$\chi'_b$ mesons, that have been hinted~\cite{Artuso:2004fp} to be precisely at
about a mass of 10230 to 10250 MeV, at the end of phase space for the decay
involving an $\omega(780)$. Reconstruction of the $\omega$ would also allow to
search for the $\eta_b$ ground and first excited state $\eta_b'$ in a channel
different from radiative decays~\cite{galuska}.

As for testing the nature of the $\Upsilon(11020)$ state itself, it should be noted
that additional four-body open-flavour decay channels are open, and therefore
several thresholds allow for a detailed mapping of the wave function squared.

\begin{figure}
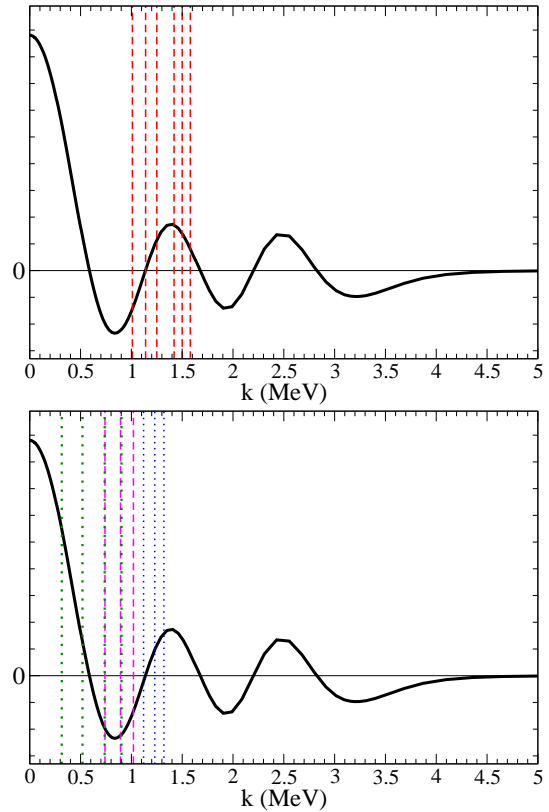

\begin{center}
\includegraphics*[width=7cm]{FIGS.DIR/WF6S2bodythresholds.eps}
\includegraphics*[width=7cm]{FIGS.DIR/WF6S3bodythresholds.eps}
\caption{Wave function of the $\Upsilon(6S)$ state (candidate assignment for the $\Upsilon(11020)$ bottomonium). The
upper plot shows the position of the six two-body thresholds as function of the final meson (initial quark) momentum.
The lower plot shows the three and four-body thresholds.
Interesting is, in spite of its small phase space, the rate of $B_s B_s \pi \pi$ decay, that resides near the central
wave function maximum.}\label{figwfs}
\end{center}
\end{figure}

A large number of three- and four-body decay channels would allow searches for Sturm-Liouville nodes in the final-state
$B$-meson momentum distributions (see Fig.~\ref{figwfs}). Particularly curious might prove the $B_s B_s \pi \pi$ that
has a very little phase space but resides near the central wave function maximum.

\begin{table*}[t]
\begin{center}
\begin{tabular}{|c|ccccccc|}
\hline
&$4S$&$3D$&$5S$&$4D$&$6S$&$5D$&$7S$\\
\hline
\cite{Badalian:2008ik}&10640&10700&10870&10920&11075&11115&-\\
\hline
\cite{Badalian:2009bu}&10645&10705&10880&10928&11084&11123&11262\\
\hline
\cite{Li:2009nr}&10611&10670&10831&10877&11023&11060&11193\\
\hline
\end{tabular}
\end{center}
\caption{Some recent theoretical predictions for the masses (in MeV) of radially excited $S$- and $D$-wave vector states
in the bottomonium spectrum.}\label{table:SD}
\end{table*}

Some of the three- and four-body final states will however be populated by
intermediate $B\ov{B}_J$ ($J=0,1,2$) resonances that need to be separated from the
pure multi-body decays. These quasi two-body decays make the $\Upsilon(11020)$ a
good starting point for studying excitations of the $B$ and $B_s$ mesons.

\subsection{Search for the Rydberg states of bottomonium}\label{Rydberg}

If energies in the range 11.2-11.4~GeV appear to be accessible at a future $B$-factory, further high lying states in the
$b\ov{b}$
spectrum can be searched for. In particular, quark model estimates show that
$\Upsilon(7S)$ (see Table~\ref{table:SD}) and $\Upsilon(8S)$ are expected to reside in this region. A relevant worry
is of course that these states are predicted broad, with the width of order 150-200~MeV, comparable with their energy
separation, so that even using high-resolution data and employing appropriate formulae for overlapping resonances
might not allow one to resolve them as separate states. In addition, the value of the wave function at the origin
decreases with the radial excitation number, production of higher $S$-wave states being suppressed accordingly.
Nevertheless, behaviour of the $R_b$ may still be sensitive to the existence of these highly excited bottomonia.

Another relevant issue is the problem of observation of vector $D$-wave states $\Upsilon(nD)$. Quark models predict a
few of them to reside in the region 10.7-11.2~GeV (see Table~\ref{table:SD}). Although pure $^3D_1$ quark-antiquark
states have tiny dileptonic widths, the $S$-$D$-wave mixing provides a mechanism which may increase substantially such a
width for the physical meson originated from the pure $^3D_1$ state. It has been argued in~\cite{Badalian:2009bu} that,
due to the $S$-$D$ mixing, the dielectron widths of the $\Upsilon(nD)$ bottomonia, with
$n=3,4,5$, may be as large as about 100~eV, that is compatible with similar widths of the $\Upsilon((n+1)S)$ states.
Together with a high-statistics expected to be available at future $B$-factories, this opens a possibility to
observe $\Upsilon(nD)$ bottomonia directly in $e^+e^-$ experiments.

\subsection{{\boldmath $B_J$} mesons}

\begin{figure*}
\begin{center}
\includegraphics*[width=7cm]{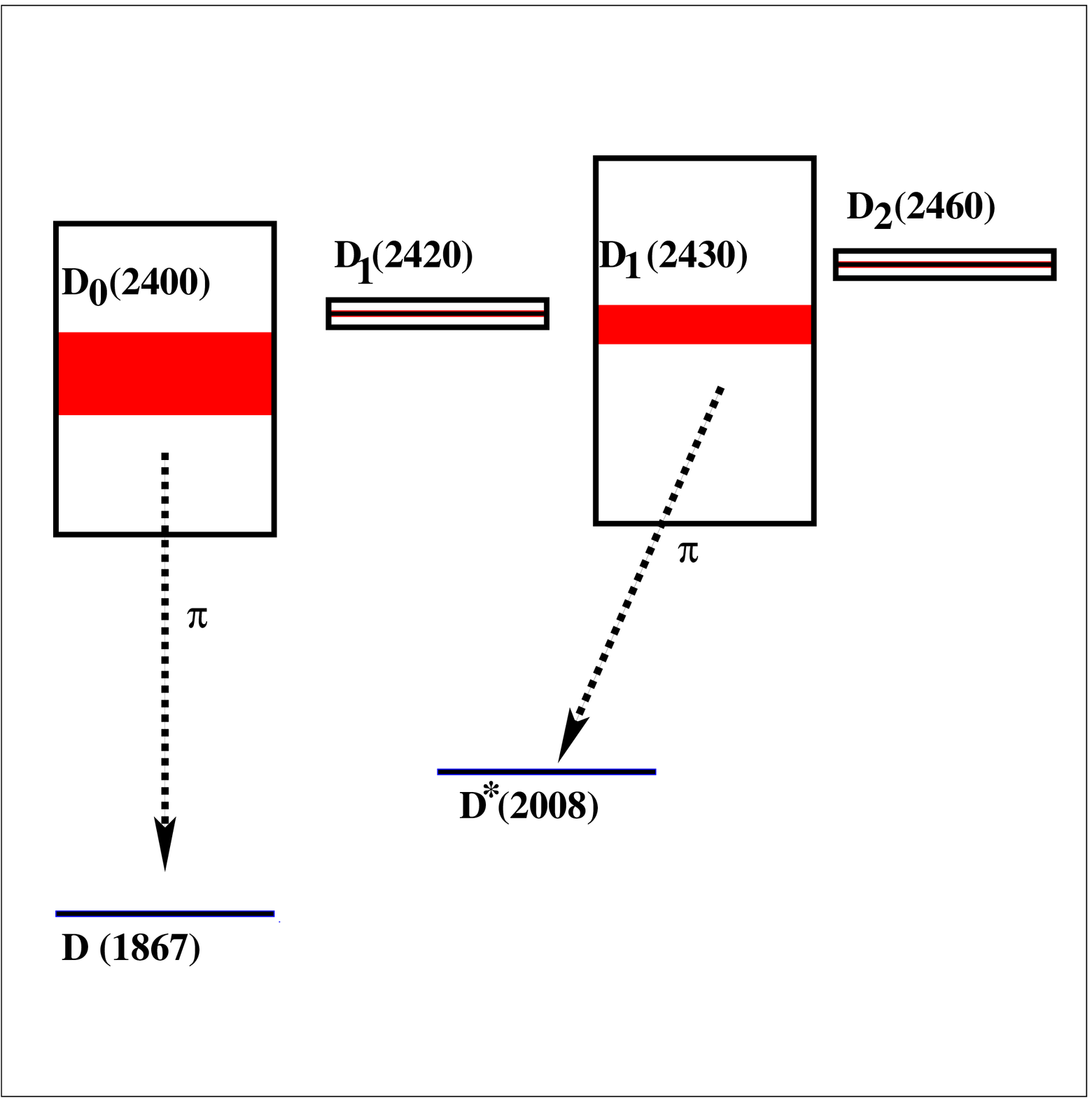}\hspace*{10mm}
\includegraphics*[width=7cm]{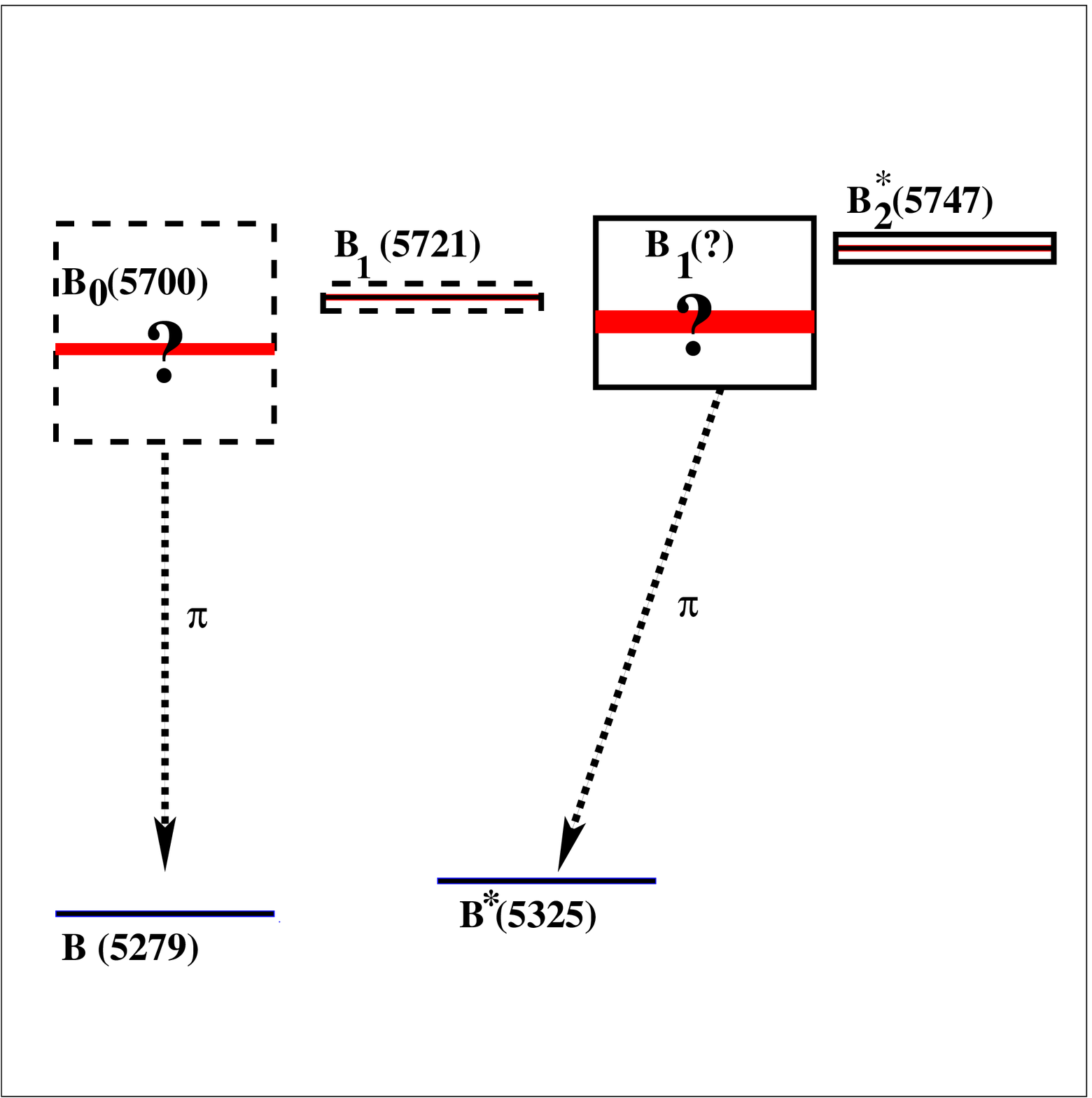}
\caption{Spectrum of positive-parity $D_J$ and $B_J$ mesons. Expected (but yet unknown) states in the $B$-meson spectrum
are tagged with question marks.} \label{fig:spectra}
\end{center}
\end{figure*}

\begin{table*}[t]
\begin{center}
\begin{tabular}{|c|cccccccc|}
\hline
&\cite{isgur2}&\cite{Kalashnikova:2001px}&\cite{Faustov}&\cite{Bardeen}&\cite{col}
&\cite{Falk}&\cite{lattice2}&\cite{PDG}\\
\hline
$B_0(P_{1/2})  $         & 5760 & 5722 & 5738 & 5627 & 5700 &  -   & 5754 &  - \\
$B_1(P_{1/2})  $& 5780 & 5741 & 5757 & 5674 & 5750 &  -   & 5730 &  - \\
$B_1(P_{3/2})  $& 5780 & 5716 & 5719 &  -   & 5774 & 5755 & 5684 & $5723.5\pm 2.0$ \\
$B_2^*(P_{3/2})$       & 5800 & 5724 & 5733 &  -   & 5790 & 5846 & 5770 & $5743\pm 5$\\
\hline
\end{tabular}
\end{center}
\caption{Theoretical predictions for the masses of the $P$-level $B_J$ mesons in
MeV (the first six columns). The last two columns contain predictions of lattice
QCD and the existing experimental data, respectively. The
subscripts $1/2$ and $3/2$ denote the light-quark total angular momentum,
that would be an experimental observable only in the limit of an infinite
heavy-quark mass.}\label{tab:BJmassesth}
\end{table*}

Another possible task for the $\Upsilon (11020)$ studies at the future $B$-factories is the
identification of bottomed positive-parity $B_J$ mesons;  the present experimental situation is somewhat obscure.

By analogy with other meson
families, such as the $D$ mesons, one expects a quadruplet of positive-parity
states $(0^+, 1^+, 1^+, 2^+)$, that in the quark model correspond to $P$-wave
quark-anti\-quark mesons. Heavy quark symmetry suggests to group them in two
doublets, $(0^+,1^+)$ and $(1^+,2^+)$, with an approximate mass degeneracy within
each doublet, as implemented in HQET. The corresponding states are known in the
$D$-meson spectrum (first panel in Fig.~\ref{fig:spectra}) and a similar pattern is
expected naturally for the spectrum of $B$ mesons (second panel in
Fig.~\ref{fig:spectra}). In Table~\ref{tab:BJmassesth} we overview theoretical
predictions for the masses of the positive-parity $B_J$ mesons (including
predictions of the lattice QCD). In particular, the mass of the $J=0$ state $B_0$
is estimated to be around 5.7-5.8~GeV or, from the effective chiral/HQET
Lagrangian approach (pending publication), to be smaller, about
5700~MeV, with some unitarised chiral Lagrangian studies putting it even lower at
5600~MeV~\cite{Abreu:2011ic,Kolomeitsev:2003ac1,Kolomeitsev:2003ac2,Guo:2006fu1,Guo:2006fu2} . Together with the mass of
5325~MeV for the vector $B^*$ meson, this sets
the threshold for the production of this family to be approximately 11~GeV.
Therefore, while open-bottom decays of $\Upsilon(5S)$ have all channels
$B^{(*)}\ov{B}_J$ kinematically closed, the $S$-wave decay $\Upsilon(11020)\to B^*
\ov{B}_0$ may appear to be allowed (it might be no coincidence that there is a resonance at this $S$-wave threshold).

In the meantime, the $B_0$ scalar state has not been observed, so we are forced to rely on theoretical predictions and
thus to place it at approximate\-ly 5700~MeV and to
expect it to be broad (see the second panel of Fig.~\ref{fig:spectra}).
The expected total width of $B_0$ is of order 100-200~MeV, that may make
this state difficult to identify experimentally.

The $2^+$ state,
$B_2^*(5747)$, has been observed experimentally, with a small width of $\Gamma = 23
(4)$~MeV and decaying to both $B^* \pi$ and $B \pi$ final states with similar
branching fractions~\cite{PDG}. Due to its width and the width of the
$\Upsilon(10120)$, which is $79\pm16$~MeV~\cite{PDG}, this state can in
principle also be produced at 11020~MeV energies (the nominal threshold being 11026
MeV).

If the pattern of $D_J$ mesons is taken as a true guide, one expects two
$B_1$ mesons, one narrow and one broad, producing two nearly degenerate
heavy-quark spin symmetry doublets with the $B_0$ and $B_2$ states. One of
these $J=1$ mesons is known as $B_1(5721)$~\cite{PDG}, however its width has not been
measured. It most probably is the narrow $P_{3/2}$ doublet partner of the $B_2^*$
meson. Then the other, not yet observed, $B_1$ state should be as light as its
$P_{1/2}$ doublet partner $B_0$, and it is expected to be broad. Therefore the
decays $\Upsilon(11020)\to B\ov{B}_1$ with both $B_1$'s in the final state
should be allowed. The above mentioned missing states are tagged in
Fig.~\ref{fig:spectra} with question marks.

The conclusion of this discussion is that the positive-parity $B_J$ states can be seen by a future $B$-factory operating
at and slightly above the $\Upsilon(11020)$ resonance, with the possibility of discovering two new states, and improving
the data quality for the other two.

Production of $B_J$'s is expected to be copious because of the $S$-wave open-bottom
threshold appearing near 11020 MeV and, given the fact that $\Upsilon(11020)$ resides exactly at the $B\ov{B}_J$
threshold, it may provide an interesting pattern of threshold phenomena.

The channels $B\bar{B}\pi$, $B\bar{B}\pi\pi$ are promising, and sufficient data may
be collected to perform an angular analysis of the $B^{(*)}\pi$
subsystem.

Finally, if data were taken at energies well above the $\Upsilon(11020)$
resonance, a second multiplet of positive parity $B$ mesons would open. The quark model
generically predicts a radial excitation of each member of the $B$ meson family at
a cost of about 600~MeV. The unitarised effective Lagrangian
method~\cite{Guo:2006fu1,Guo:2006fu2}, that incorporates $SU(3)$ symmetry among light quarks,
finds that there are two higher and narrow states in the $0^+$ and $1^+$ sectors.
Their masses are larger than the lowest positive-parity states by only 300 MeV. A
salient feature is that they strongly couple to channels with hidden strangeness,
that is $B_s^{(*)} K$ and $B^{(*)}\eta$. Hence, they are different from the
conventional quark model states, whose coupling to the $B^{(*)}\pi(\eta)$ and
$B_s^{(*)}K$ should possess an approximate $SU(3)$ flavour symmetry. Interestingly, it
is possible that these two higher states are slightly above the $B^{(*)}\eta$
thresholds, and in that case, due to the strong coupling, the decays into these
modes may even have a width comparable to the $B^{(*)}\pi$. A measurement of the
ratio of branching fractions into various channels related through the $SU(3)$ flavour
symmetry would be useful to identify the nature of these states.

\subsection{Triple charm and triply heavy baryons}

Double charm and charmonium production have been stu\-died for a long time. In 1987
the WA75 Collaboration start\-ed these studies at CERN. In the last decade, the Belle
and BaBar Collaborations produced double charm/ charmonium~\cite{Abe:2007jn}
reporting, in particular, a large production of the $J/\psi D^{(*)}\ov{D}^{(*)}$
final states. Given an expected high-luminosity, triple charm production may become
accessible for studies at the future $B$-factories. The minimum energy needed for triple $J/\psi$
production at an electron-po\-sit\-ron collider is 9300 MeV, well within the reach of
any $B$-factory. The cross section could potentially be calculated within NRQCD.

The next threshold with increasing energy would be the production of a
baryon-antibaryon pair. The relevant ground-state baryon is the $\Omega_{ccc}$, and
its spectrum of excitations is of great theoretical importance as this system is
expected to be well described in terms of the quark model, so that the given
baryon-antibaryon production would be useful to assess applicability of various
quark model features to QCD.

\begin{figure}
\begin{center}
\includegraphics*[width=8.4cm]{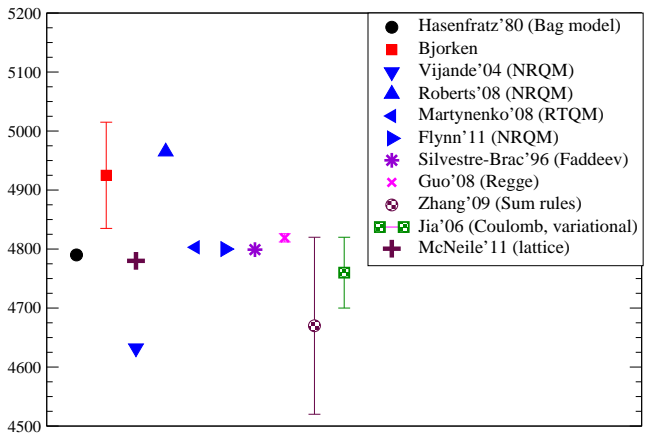}
\end{center}
\caption{Results of various computations of the ground-state triply charmed baryon.}\label{fig:3body}
\end{figure}

The ground-state $\Omega_{ccc}$ has not yet been discovered and its mass is
unknown. A spread of predictions obtained in various approaches can be seen in
Fig.~\ref{fig:3body}. Then, assuming the mass $M(\Omega_{ccc})\sim 4800-4900$ MeV,
the baryon-antibaryon threshold should lie at $9600-9800$ MeV. By the time of the
planned $B$-factories operation start, further results for the $\Omega_{ccc}$ mass are expected to appear to
better pin-down the mass range.

Another interesting threshold for $\Omega_{ccc}$ production is the open-flavour
threshold with three recoiling $\ov{D}$ mesons, via the reaction \be
\label{recoilOmega} e^+e^-\to \Omega_{ccc}\ov{N} \ov{D}\ov{D}\ov{D}
\ee
that opens up at about $11400\pm 200$ MeV with the current uncertainty in the
$\Omega_{ccc}$ mass. Such processes were studied in \cite{Baranov:2004er} with the
focus on LEP physics at the $Z$ pole and a prediction for the cross section of
about 0.4~fbarn was obtained. Translated to the properties of a future $B$-factory operating at
11.4 GeV, this estimate would imply cross sections of order of 3~fbarn, however
more theoretical work is clearly needed to make more accurate predictions.

Another very interesting channel is
\be\label{recoilOmega2}
e^+e^- \to \Omega_{ccc}^{++}\bar\Lambda_c^- \bar D^0 D^-\ .
\ee
All particles recoiling in the final state are weakly decaying particles so that
the reconstruction could be feasible. The threshold is around $10820-10920$~MeV,
the same energy range as the $\Upsilon(5S)$, assuming $M(\Omega_{ccc})\sim
4800-4900$~MeV.

Phenomenology of the $\Omega_{ccc}$ baryon is extensively discussed in \cite{Bjorken:1985ei}.
Recently the weak-decay chain
$$
\Omega_{ccc}\to \Xi_{ccs}\pi \to \Xi_{css}\pi\pi \to  \Omega_{sss} \pi \pi \pi
$$
was stressed in \cite{ChenWu} as the most promising avenue, but in view of low
statistics one should also consider additional channels with pions replaced by
kaons, with the corresponding baryon in the final state being then a nucleon,
$\Sigma$ or $\Xi$. Additionally one can employ the recoiling mass technique against
the $\ov{D}$ mesons to try to identify the thresholds for the
reactions~(\ref{recoilOmega},\ref{recoilOmega2}).

 Finally, an additional physics item can be highlighted, which is the QCD
three-body force.  QCD is a non-linear theory that features a three-body
gluon-gluon coupling not present in electrodynamics. This has been empirically
confirmed by the study of four-jet events at colliders. Three-heavy quark baryons
offer a window to this three-body force, since the triple gluon coupling also
forces a triple-quark potential $V(r_1,r_2,r_3)$ that cannot be decomposed in terms
of two-body potentials $V(r_1-r_2)$~\cite{Brambilla:2009cd}. This has been
demonstrated to be relatively small and of order 25-50 MeV for triply charmed
baryons, both in model terms~\cite{Flynn:2011gf} and in perturbation
theory~\cite{LlanesEstrada:2011kc}. Therefore, an electron-positron machine, in particular such
as SuperB, could deliver precise data needed to test this QCD feature. Further
tests should be developed by theorists by the time, when the planned $B$-factories will start operation.

\subsection{Pion transition form factor}

\begin{figure}[t]
\begin{tabular}{l}
\includegraphics[width=0.42\textwidth]{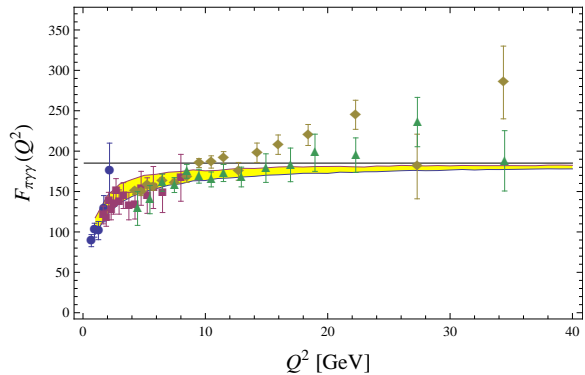} \\
\includegraphics[width=0.48\textwidth]{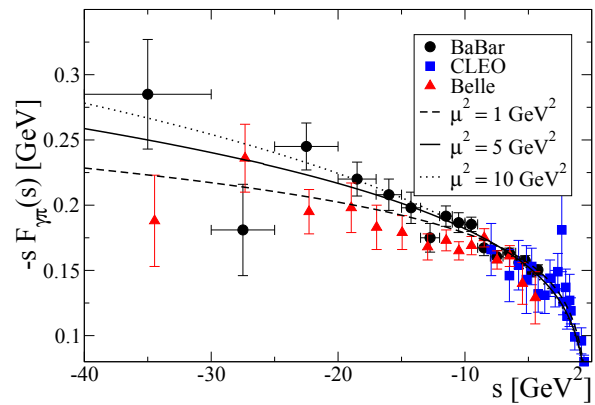}
\end{tabular}
\caption{The first plot (courtesy of E. Ruiz Arriola, P. Masjuan, and W. Broniowski) shows a canonical pQCD calculation
with two vector mesons to account for the form factor at low-momentum transfer. The second plot (courtesy of A.
Szczepaniak) is an additional theory calculation assuming that Regge behaviour
dominates the end point of the distribution amplitude. \label{fig:formfactor}}
\end{figure}

Perturbative QCD (pQCD) makes specific predictions for exclusive processes, such as
hadron form factors at large momentum transfer or exclusive cross sections at fixed
angle and asymptotically large momentum~\cite{BrodskyLepage}. These predictions
encode the Brodsky-Farrar counting rules~\cite{BrodskyFarrar} and entail that
relevant couplings in Yang-Mills theory are largely scale-invariant. They are
contingent on the hadron distribution amplitudes being finite at the end point when
one of the partons carries zero momentum.

The simplest example of the power-law behaviour of an exclusive amplitude is  the
pion form factor. Experimental data at squared momenta that should already be
considered high respect to the hadron 1 GeV$^2$ scale~\cite{ExpFormFactor1,ExpFormFactor2,ExpFormFactor3} is about
a factor of 3 larger than the pQCD prediction.

A much larger squared momentum has been achieved at BaBar and Belle for the
transition form factor that controls $\pi_0\to \gamma^*(Q^2)\gamma$. While the
BaBar data are in clear disagreement with the pQCD prediction, the Belle data are
less so. In particular, the quantity $Q^2 F(Q^2)$ should be a constant at
asymptotically large $Q^2$, as per the QCD counting rules, with the constant also
strictly determined by $F(Q^2) \to 2 f_\pi/Q^2$ to be $2f_\pi$ at leading order in
pQCD. Instead, the BaBar data are better compatible with a growing function rather
than with a constant one. This can be accommodated if the end point of the pion
distribution amplitude does not vanish because of low-$x$ Regge
behaviour~\cite{Gorchtein:2011xe}. However, the newer Belle data could be
compatible with both types of behaviour, although detailed
analysis~\cite{Noguera:2012aw} shows that the asymptotic regime has not been
reached even at 40 GeV$^2$ (quite a large scale for hadron structure).

One needs therefore a higher statistics study of the pion transition form factor in the reaction
\be
e^+e^-\to e^+e^-\gamma^*\gamma\to e^+e^- \pi^0,
\ee
for example, at the energy of the
$\Upsilon(4S)$ resonance to reduce the errors of the existing points. However,
since, at those high energies, the pion mass is negligible, the maximum $Q^2$
transferred through the virtual photon scales in proportion to the total
centre-of-mass energy squared $E_{cm}^2$ of the initial state.

Further data from the future high-luminosity $B$-facto\-ri\-es could clarify whether the pQCD counting rules and
specific predictions are valid in such
exclusive processes or one needs to resort to more sophisticated computations. In particular, increasing the energy from
10.579 GeV at the $\Upsilon(4S)$ to
the 11.020 GeV at the last $\Upsilon$ resonance entails a moderate 8.5\% increase in $Q^2$, so that an additional point
in Fig. \ref{fig:formfactor} at 50 GeV$^2$ is conceivable.

\subsubsection{Testing factorization of $\gamma\gamma^*\to \pi\pi$}

Similar tests can be conducted in the quite unexplored reaction
\be
e^- e^+ \to \gamma \gamma^*(Q^2) e^- e^+ \to \pi \pi e^- e^+ \ .
\ee
It is important to establish the high $Q^2$ virtuality of one of the photons by means of appropriate cuts on the final
state lepton momenta.

The currently untested pQCD factorization formula for the hadron tensor in this reaction
~\cite{Diehl:2000uv},
\be
T^{\mu\nu}_{\pi \pi} = -g_T^{\mu\nu}\sum_q \frac{e_q^2}{2}
\int_0^1 \frac{2z-1}{z(1-z)} \Phi_q^{\pi\pi}(z,\zeta,E^2_{\pi\pi})dz,
\ee
is analogous to the equivalent relation controlling the pion transition form factor in the same approximation
\be
T^{\mu\nu}_{\pi} = {\bf \epsilon}^{\mu\nu} \sum_q \frac{e_q^2}{2}
\int_0^1 \frac{dz}{z(1-z)} \phi_q^\pi(z)
\ee
but substituting the conventional pion distribution amplitude $\phi^\pi_q$ for quark $q$ inside the pion by a
Generalized Distribution Amplitude $\Phi_q^{\pi\pi}$ that characterizes the two-pion system (related by crossing to the
much discussed Generalized Parton Distributions).
Again, in the asymptotic regime where pQCD makes a statement, for $Q^2$ much larger than any other scale, the prediction
for the two--pion amplitude is scaling behaviour.
At fixed $\zeta$ and $W^2$ the amplitude is predicted to be $Q^2$ independent (in practice one has logarithmic
corrections from higher orders in perturbation theory, and subleading negative powers of $Q^2$ from higher-twist
corrections, but definitely no positive $Q^2$ power).

Should the test of scaling fail, and the amplitude behave as a positive power of $Q^2$, one should understand the
behaviour of $\Phi_q^{\pi\pi}$ near the end points $z=0,1$ to appreciate whether it was indeed finite as assumed, and
perhaps look for other parametrizations of data. Should scaling be satisfied, then there is a variety of studies that
can be performed to learn about hadron structure through these generalized amplitudes from the data~\cite{Diehl:2000uv}.

These investigations should help to settle the question of the applicability of pQCD reasoning in exclusive processes,
which is one of the outstanding theory predictions requiring testing.

\subsection{The {\boldmath $\Lambda_b\bar{\Lambda}_b$} threshold}

Should the beam energy allow to reach the 11.24 GeV region (in centre of mass),
one would encounter the $\Lambda_b\bar{\Lambda}_b$ threshold, where bottomed baryons are
first produced.
We can see two main reasons to produce a sample of these baryons.

The first is the pursuit of direct CP violation studies in decays. The Particle Data Group~\cite{PDG} collects two
measurements of such asymmetries,
\be
A_{CP} = \frac{B(\Lambda_b\to f)-B(\bar{\Lambda}_b\to \bar{f})}{B(\Lambda_b\to f)+B(\bar{\Lambda}_b\to \bar{f})}
\ee
$A_{CP}(\Lambda_b^0\to p\pi^-)=0.03\pm0.17\pm0.05$ (a very inconclusive measurement) and
$A_{CP}(\Lambda_b^0\to pK^-)=0.37\pm0.17\pm 0.03$ (that is already a very reasonable hint).
The theoretical interest in these and similar asymmetries is that they depend on different QCD matrix elements from
those computed for the $B$-meson decay programme, so they will continue overconstraining the CKM picture of quark
flavour with different systematics in theory computations.

Note also that the polarisation of the $\Lambda_b$ can be used to test time reversal symmetry $T$
violation~\cite{Ahmed:2011dd}. One way to proceed is to reconstruct the decay chain $\Lambda_b\to \Lambda_0 l^+ l^-$
where the two leptons have the invariant mass of a vector meson.  Identifying a non-vanishing transverse polarisation of
the resulting vector meson or hyperon measures a separation from time-reversal symmetry. Several tests of P and CP
violation can also be conducted.

The second major reason is that the dominant decay mode $b\to c$ will typically
leave as a remainder a singly-charmed baryon~\cite{Leibovich:1997az}, and since the
$\Lambda_b$ mass at 5619~MeV is so much higher than the $\Lambda_c$ mass at
2290~MeV and above, all charmed baryons are accessible and can be populated, which
can assist discovering some additional excited resonances in that spectrum, either
by direct reconstruction or by the missing mass technique. Note for example that,
while semileptonic decay $\Lambda_b\to \Lambda_c l \bar{\nu}_l$ has a branching
fraction of order 6-7\%, no single hadronic mode involving $\Lambda_c$ reaches 1\%.
Thus, many charmed baryons are being produced with a small branching fraction.

Historically, the detection of excited baryons has been pursued in order to assess
the ``missing resonance problem'' (quark models overpredict the number of excited
resonances in light baryons). Since baryons with one charm quark tend to be cleaner
probes of the excited degrees of freedom, there will be interest in the low-energy
hadron physics community to compare any possible new states with the Jefferson-Lab
program on light baryons. A theoretical prediction for the
$\Lambda_b\bar{\Lambda}_b$ production in $e^+e^-$ collisions can be found in
\cite{Simonov:2011jc}.

\subsection{The {\boldmath $B_c^+ B_c^-$} threshold}

So far CP violation studies have been carried out extensively with $B\ov{B}$ and $K\ov{K}$ meson pairs, as well as with
the $D\ov{D}$ and $B_s\ov{B}_s$ systems. Future studies can break ground in the $B_c\ov{B}_c$ system.
The currently measured mass of the $B_c$ meson puts the threshold for $B_c\ov{B}_c$ at 12550 MeV, a challenging energy
for a $B$-factory designed to operate at the energies up to around 11~GeV. Being realistic, this is perhaps a goal for a
machine upgrade after the first five-year running period, but it is worth exploring the implications early on.

Saliently, direct CP violation studies comparing the $B_c$ and $\ov{B}_c$ can help
further constrain the Standard Model CKM picture of CP
violation~\cite{Fleischer:2000pp} (since the $B_c$ mesons are charged, there is no
indirect CP violation in this system).

Furthermore, the $B_c$ system is theoretically cleaner than the $B$, $B_s$ systems.
Indeed, while one can exploit HQET in the latter two to parametrise theoretical
uncertainties into a few quantities, the $B_c$ meson contains two heavy quarks, so
that an additional effective theory, Non-Relativistic QCD, is of use, and its
potential version pNRQCD allows for perturbative (or semi-perturbative)
computations of several quantities of interest, reducing theory uncertainties.

For a taster, in a recent review of the flavour physics of
SuperB~\cite{Meadows:2011bk}, a stated goal is to measure $V_{bc}$ in the CKM
matrix  to a precision of 0.5-1\%. Such an extraction is theory dominated, and one
needs to trust theory to provide the necessary improvements to profit from the
future $B$-factories data. This makes the feasibility of the extraction more uncertain. An
alternative way to reach this precision, and better, would be to measure the
leptonic and semileptonic decays of the $B_c$ meson with reasonable accuracy.
Compare with the current situation, at the $2\%$ level, of the $V_{cs}$
measurement, limited by theory computations of the $D_s$ decay constant,
$f_{D_s}=249\pm 3$ MeV that enters the purely leptonic decays of this meson (and
analogously for semileptonic decays). Since the $B_c$ meson is much cleaner
theoretically, the requisite theory precisions will be off the shelf for lattice
collaborations addressing the issue of $V_{cb}$.

Since the $B_c^+ B_c^-$ system resides well above the open-bottom region, one does
not expect prominent resonances to be formed (and none has been found so far). To
increase the sample of $B_c$ pairs obtained in a small run, the near-threshold
region could be studied in first place, hoping to find there a cusp or a molecular
type resonance decaying into $B_c^+ B_c^-$. Observing this would on the other hand
be of tremendous interest for tetraquark studies (that tend to predict bound states
of two doubly heavy mesons \cite{Vijande:2003ki}).
Judging by the $B_s$ production off the $\Upsilon(5S)$ resonance, one should expect
about 0.01 units of R (the ratio to the $\mu\mu$ production cross section), or
about 5 pbarn, slightly above the $B_c\bar{B}_c$ threshold. With a modest 10
fb$^{-1}$ luminosity accumulated during a pilot run, a $B$-factory running at the energy close to the $B_c^+ B_c^-$ mass
threshold should produce 50 000
$B_c\bar{B}_c$ pairs. Taking into account that all existing experiments have
reconstructed only dozens of these mesons, it could break new ground enormously.

\begin{figure}
\centerline{\includegraphics[width=7cm]{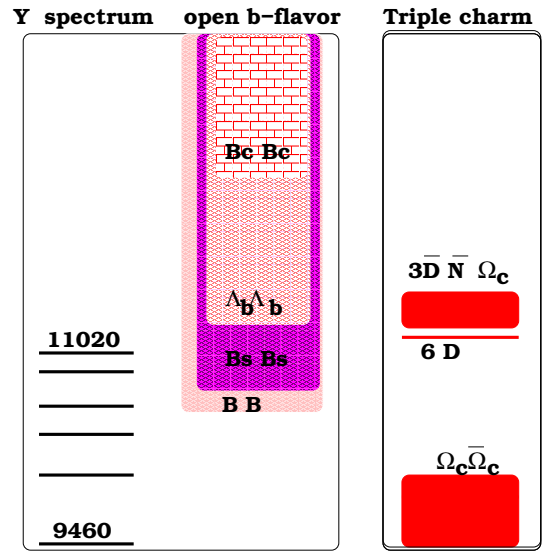}}
\caption{The $\Upsilon$ spectrum and open-$b$ flavour thresholds (left) as well as selected triple charm thresholds
(right; $\Omega_c$ here is the $\Omega_{ccc}$ triply charmed baryon).\label{fig:above11}}
\end{figure}

To summarise this section, the physics potential of extending the energy reach with
a future $B$-factory running above the $\Upsilon(5S)$ region is depicted in Fig.~\ref{fig:above11}.
Given the many interesting studies that can be performed, it is worth keeping the possibility open of running the future
$B$-factories at a higher energy, with a sensible rooftop at the $B_c^+ B_c^-$ threshold around 12550 MeV.

\section{Conclusion}

In this paper we outlined hadron physics potential of the planned electron-positron $B$-factories at the $\Upsilon(5S)$
resonance and above. In particular,
we stress that the opportunities of a $B$-factory operating at
slightly higher energies than the $\Upsilon(4S)$ for a moderate period of time are
manifold, involving both fundamental studies of the Standard Model and traditional
spectroscopy and hadron structure studies, much increasing the scientific impact of
the facility. The basic picture of particle physics between 10.5 and 12.5 GeV has
been glimpsed from past lepton accelerators such as LEP, and hadron colliders have
produced some of the new particles expected there, but not in sufficient amount nor
purity to carry out detailed studies. Some particles expected from theory to be
produced in this region remain undiscovered, such as a positive parity $B_1$ meson,
or the triply charmed $\Omega_{ccc}$. The search for QCD exotics will receive many
new constraints in this region of energy, where hybrid bottomonium resides,
according to theory. We look forward to the construction of these exciting machines,
and their fruitful operation at high energy.

\begin{acknowledgement}
The authors would like to thank Yu.S.~Kalashnikova, Yu.A.~Simonov, A.~Szczepaniak, M.B.~Voloshin for useful
discussions. This work is supported in part by the DFG and the NSFC
through funds provided to the Sino-German CRC 110 ``Symmetries and the Emergence of Structure in QCD'', the NSFC (Grant
No. 11165005), Spanish grants FPA2011-27853-01 and FIS2008-01323. 

\end{acknowledgement}


\begin{thebibliography}{99}
\bibitem{Aubert:2008ab}
  B.~Aubert {\it et al.}  [BABAR Collaboration],
  Phys.\ Rev.\ Lett.\  {\bf 102} (2009) 012001
  [arXiv:0809.4120 [hep-ex]].

\bibitem{Brambilla:2010cs}
N.~Brambilla, S.~Eidelman, B.~K.~Heltsley, R.~Vogt, G.~T.~Bodwin, E.~Eichten, A.~D.~Frawley and A.~B.~Meyer {\it et
al.},
  Eur.\ Phys.\ J.\ C {\bf 71} (2011) 1534
  [arXiv:1010.5827 [hep-ph]].

\bibitem{Chen:2008xia}
  K.~F.~Chen {\it et al.}  [Belle Collaboration],
  Phys.\ Rev.\  D {\bf 82} (2010) 091106
  [arXiv:0810.3829 [hep-ex]].

\bibitem{CLEO2007}
  D.~Cronin-Hennessy {\it et al.}  [CLEO Collaboration],
  Phys.\ Rev.\ D {\bf 76} (2007) 072001
  [arXiv:0706.2317 [hep-ex]].

\bibitem{Simonov:2008qy}
  Yu.~A.~Simonov and A.~I.~Veselov,
  Phys.\ Rev.\  D {\bf 79} (2009) 034024
  [arXiv:0804.4635 [hep-ph]].

\bibitem{Meng:2008dd}
  C.~Meng and K.-T.~Chao,
  Phys.\ Rev.\ D {\bf 78} (2008) 034022
  [arXiv:0805.0143 [hep-ph]].

\bibitem{Abe:2007tk}
  K.~F.~Chen {\it et al.}  [Belle Collaboration],
  Phys.\ Rev.\ Lett.\  {\bf 100} (2008) 112001
  [arXiv:0710.2577 [hep-ex]].

\bibitem{Kuang:1981se}
  Y.~P.~Kuang and T.~M.~Yan,
  Phys.\ Rev.\  D {\bf 24} (1981) 2874.

\bibitem{Aubert:2006bm}
  B.~Aubert {\it et al.}  [BABAR Collaboration],
  Phys.\ Rev.\ Lett.\  {\bf 96} (2006) 232001
  [arXiv:hep-ex/0604031].

\bibitem{Simonov:2008ci}
  Yu.~A.~Simonov and A.~I.~Veselov,
  Phys.\ Lett.\  B {\bf 671} (2009) 55
  [arXiv:0805.4499 [hep-ph]].

\bibitem{tetraquark2} A. Ali, C. Hambrock, and M. J. Aslam,
near the Upsilon(5S) resonance,''
  Phys.\ Rev.\ Lett.\  {\bf 104} (2010) 162001
   [Erratum-ibid.\  {\bf 107} (2011) 049903]
  [arXiv:0912.5016 [hep-ph]].

\bibitem{Ali:2010pq}
  A.~Ali, C.~Hambrock and S.~Mishima,
  Phys.\ Rev.\ Lett.\  {\bf 106}, 092002 (2011)
  [arXiv:1011.4856 [hep-ph]].

\bibitem{Chen:2011qx} D.-Y.~Chen, J.~He, X.-Q.~Li and X.~Liu,
production near the peak of $\Upsilon(5S)$,''
  Phys.\ Rev.\ D {\bf 84} (2011) 074006
  [arXiv:1105.1672 [hep-ph]].

\bibitem{Chen:2011zv} D.-Y.~Chen, X.~Liu and S.-L.~Zhu,
decay,''
  Phys.\ Rev.\ D {\bf 84} (2011) 074016
  [arXiv:1105.5193 [hep-ph]].

\bibitem{Adachi:2011gj}
  I. Adachi et al. [Belle Collaboration],
  arXiv:1105.4583 [hep-ex].

\bibitem{Belle:2011aa}
  A.~Bondar {\it et al.}  [Belle Collaboration],
  Phys.\ Rev.\ Lett.\  {\bf 108} (2012) 122001
  [arXiv:1110.2251 [hep-ex]].

\bibitem{CLEO1994}
  F.~Butler {\it et al.}  [CLEO Collaboration],
  Phys.\ Rev.\ D {\bf 49} (1994) 40.

\bibitem{Belle4S}
  K.~Abe {\it et al.}  [BELLE Collaboration],
  hep-ex/0512034;
  A.~Sokolov {\it et al.}  [Belle Collaboration],
  Phys.\ Rev.\ D {\bf 75} (2007) 071103
  [hep-ex/0611026].

\bibitem{VoloshinReview}
  M.~B.~Voloshin and Y.~.M.~Zaitsev,
  Sov.\ Phys.\ Usp.\  {\bf 30} (1987) 553
  [Usp.\ Fiz.\ Nauk {\bf 152} (1987) 361].

\bibitem{KuangReview}
  Y.-P.~Kuang,
  Front.\ Phys.\ China {\bf 1} (2006) 19
  [hep-ph /0601044].

\bibitem{Guo2004}
  F.-K.~Guo, P.-N.~Shen, H.-C.~Chiang and R.-G.~Ping,
  Nucl.\ Phys.\ A {\bf 761} (2005) 269
  [hep-ph/0410204].

\bibitem{VoloshinNew}
  M.~B.~Voloshin,
  Phys.\ Rev.\ D {\bf 74} (2006) 054022
  [hep-ph/0606258].

\bibitem{X1}
  M.~B.~Voloshin,
  JETP Lett.\  {\bf 37} (1983) 69
  [Pisma Zh.\ Eksp.\ Teor.\ Fiz.\  {\bf 37} (1983) 58].

\bibitem{X2} V.~V.~Anisovich, D.~V.~Bugg, A.~V.~Sarantsev and B.~S.~Zou,
Phys.\ Rev.\ D {\bf 51} (1995) 4619.

\bibitem{Guo2006} F.-K.~Guo, P.-N.~Shen, H.-C.~Chiang and R.-G.~Ping,
  Phys.\ Lett.\ B {\bf 658} (2007) 27
  [hep-ph/0601120].

\bibitem{Ablikim:2004qna}
  M.~Ablikim {\it et al.}  [BES Collaboration],
  Phys.\ Lett.\ B {\bf 598} (2004) 149
  [hep-ex/0406038].

\bibitem{Brown:1975dz}
  L.~S.~Brown and R.~N.~Cahn,
  Phys.\ Rev.\ Lett.\  {\bf 35} (1975) 1.

\bibitem{Mannel:1995jt}
  T.~Mannel and R.~Urech,
  Z.\ Phys.\ C {\bf 73} (1997) 541
  [hep-ph/9510406].

\bibitem{Lipkin:1988tg}
  H.~J.~Lipkin and S.~F.~Tuan,
  Phys.\ Lett.\ B {\bf 206} (1988) 349.

\bibitem{Moxhay:1988ri}
  P.~Moxhay,
  Phys.\ Rev.\ D {\bf 39} (1989) 3497.

\bibitem{Zhou:1990ik}
  H.-Y.~Zhou and Y.-P.~Kuang,
  Phys.\ Rev.\ D {\bf 44} (1991) 756.

\bibitem{FK:NREFT1} F.-K.~Guo, C.~Hanhart, U.-G.~Mei{\ss}ner,
  Phys.\ Rev.\ Lett.\  {\bf 103} (2009) 082003
  [Erratum-ibid.\  {\bf 104} (2010) 109901]
  [arXiv:0907.0521 [hep-ph]].

\bibitem{FK:NREFT2}
  F.-K.~Guo, C.~Hanhart, G.~Li, U.-G.~Mei{\ss}ner and Q.~Zhao,
  Phys.\ Rev.\ D {\bf 83} (2011) 034013
  [arXiv:1008.3632 [hep-ph]].

\bibitem{Adachi:2011ji} I.~Adachi {\it et al.}  [Belle Collaboration],
  Phys.\ Rev.\ Lett.\  {\bf 108} (2012) 032001
  [arXiv:1103.3419 [hep-ex]].

\bibitem{Bondar:2011ev}
  A.~E.~Bondar, A.~Garmash, A.~I.~Milstein, R.~Mizuk and M.~B.~Voloshin,
  Phys.\ Rev.\ D {\bf 84} (2011) 054010
  [arXiv:1105.4473 [hep-ph]].

\bibitem{Yang:2011rp}
  Y.~Yang, J.~Ping, C.~Deng and H.~-S.~Zong,
  J.\ Phys.\ G {\bf 39} (2012) 105001
  [arXiv:1105.5935 [hep-ph]].

\bibitem{Sun:2011uh}
  Z.~F.~Sun, J.~He, X.~Liu, Z.~G.~Luo and S.~L.~Zhu,
  Phys.\ Rev.\ D {\bf 84} (2011) 054002
  [arXiv:1106.2968 [hep-ph]].

\bibitem{Nieves:2011zz}
  J.~Nieves and M.~P.~Valderrama,
  Phys.\ Rev.\ D {\bf 84} (2011) 056015
  [arXiv:1106.0600 [hep-ph]].

\bibitem{Cleven:2011gp}
  M.~Cleven, F.-K.~Guo, C.~Hanhart and U.-G.~Mei{\ss}ner,
  Eur.\ Phys.\ J.\ A {\bf 47} (2011) 120
  [arXiv:1107.0254 [hep-ph]].

\bibitem{Bugg:2011jr}
  D.~V.~Bugg,
  Europhys.\ Lett.\  {\bf 96} (2011) 11002
  [arXiv: 1105.5492 [hep-ph]].

\bibitem{Ohkoda:2011vj}
  S.~Ohkoda, Y.~Yamaguchi, S.~Yasui, K.~Sudoh and A.~Hosaka,
  Phys.\ Rev.\ D {\bf 86} (2012) 014004
  [arXiv:1111.2921 [hep-ph]].

\bibitem{Danilkin:2011sh}
  I.~V.~Danilkin, V.~D.~Orlovsky and Y.~.A.~Simonov,
  Phys.\ Rev.\ D {\bf 85} (2012) 034012
  [arXiv:1106.1552 [hep-ph]].

\bibitem{Ali:2011ug}
  A.~Ali, C.~Hambrock and W.~Wang,
Implications,''
  Phys.\ Rev.\ D {\bf 85} (2012) 054011
  [arXiv:1110.1333 [hep-ph]].

\bibitem{Li:2012wf}
  M.~T.~Li, W.~L.~Wang, Y.~B.~Dong and Z.~Y.~Zhang,
  arXiv:1204.3959 [hep-ph].

\bibitem{simonovpions} Yu. A. Simonov, A. I. Veselov, Phys. Rev. D {\bf 79} (2009)
    034024

\bibitem{Adachi:2012cx}
  I.~Adachi {\it et al.}  [Belle Collaboration],
  arXiv:1209.6450 [hep-ex].


\bibitem{Voloshin:2011qa}
  M.~B.~Voloshin,
  Phys.\ Rev.\  D {\bf 84} (2011) 031502
  [arXiv: 1105.5829 [hep-ph]].

\bibitem{Mehen:2011yh}
  T.~Mehen and J.~W.~Powell,
  Phys.\ Rev.\  D {\bf 84} (2011) 114013
  [arXiv:1109.3479 [hep-ph]].
 
\bibitem{Voloshin:2012yq}
  M.~B.~Voloshin,
channel,''
  Phys.\ Rev.\ D {\bf 86} (2012) 034013
  [arXiv:1204.1945 [hep-ph]].

\bibitem{Lees:2011mx}
  J.~P.~Lees {\it et al.}  [BABAR Collaboration],
  Phys.\ Rev.\ D {\bf 84} (2011) 072002
  [arXiv:1104.5254 [hep-ex]].

\bibitem{Mizuk:2012pb}
  R.~Mizuk {\it et al.}  [Belle Collaboration],
  arXiv:1205.6351 [hep-ex].

\bibitem{FK} F.-K.~Guo, C.~Hanhart, U.-G.~Mei{\ss}ner,
  Phys.\ Rev.\ Lett.\  {\bf 105} (2010) 162001
  [arXiv:1007.4682 [hep-ph]].

\bibitem{Donoghue:1992ac}
  J.~F.~Donoghue and D.~Wyler,
  Phys.\ Rev.\  D {\bf 45}, 892 (1992).

\bibitem{Donoghue:1993ha}
  J.~F.~Donoghue, B.~R.~Holstein and D.~Wyler,
  Phys.\ Rev.\ Lett.\  {\bf 69}, 3444 (1992).

\bibitem{FK:NREFT3}
  F.-K.~Guo, C.~Hanhart, G.~Li, U.-G.~Mei{\ss}ner and Q.~Zhao,
  Phys.\ Rev.\ D {\bf 82} (2010) 034025
  [arXiv:1002.2712 [hep-ph]].

\bibitem{Friman:2002fs}
  B.~Friman, S.~H.~Lee and T.~Song,
  Phys.\ Lett.\ B {\bf 548} (2002) 153
  [nucl-th/0207006].

\bibitem{PDG}
  J. Beringer {\it et al.} [Particle Data Group], Phys. Rev. D {\bf 86} (2012)
  010001.

\bibitem{tetraquark3} A. Ali, C. Hambrock, I. Ahmed, and M. J. Aslam, Phys. Lett. B
    {\bf 684} (2010) 28.

\bibitem{Hwang:2008kb}
  D.~S.~Hwang and H.~Son,
  Eur.\ Phys.\ J.\ C {\bf 67} (2010) 111
  [arXiv:0812.4402 [hep-ph]].

\bibitem{felipe} I.~J.~General, S.~R.~Cotanch and F.~J.~Llanes-Estrada,
  Eur.\ Phys.\ J.\ C {\bf 51} (2007) 347
  [hep-ph/0609115].

\bibitem{TorresRincon:2010fu}
  J.~M.~Torres-Rincon and F.~J.~Llanes-Estrada,
Phys.\ Rev.\ Lett.\  {\bf 105} (2010) 022003 [arXiv:1003.5989 [hep-ph]].

\bibitem{Drutskoy:2009ci}
  A.~Drutskoy [Belle Collaboration],
PoS {\bf EPS-HEP2009} (2009) 056 [arXiv:0909.5223 [hep-ex]].

\bibitem{Drutskoy:2009ei}
  A.~Drutskoy,
arXiv:0905.2959 [hep-ex].

\bibitem{Morningstar} K.J. Juge, J. Kuti, and C.J. Morningstar, Phys. Rev. Lett. {\bf 82} (1999) 4400.

\bibitem{Michael} C. Michael, arXiv:hep-ph/0308293.

\bibitem{liu1} Y. Liu and X.-Q. Luo, Phys. Rev. D {\bf 73} (2006) 054510.

\bibitem{liu2} X.-Q.Luo and Y. Liu, Phys. Rev. D {\bf 74} (2006) 034502.

\bibitem{burch} T. Burch and C. Ehmann, Nucl. Phys. A {\bf 797} (2007) 33.

\bibitem{dudek1} J. J. Dudek, R. G. Edwards, N. Mathur, and D. G. RichardsPhys. Rev. D {\bf 77} (2008) 034501.

\bibitem{dudek2} J. J. Dudek, and E. Rrapaj, Phys. Rev. D {\bf 78} (2008) 094504.

\bibitem{bag1} M. Chanowitz and S. Sharpe, Nucl. Phys. B {\bf 222} (1983) 211, Erratum-ibid. B {\bf 228} (1983) 588.

\bibitem{bag2} T. Barnes, F. E. Close, and F. de Viron, Nucl. Phys. B {\bf 224} (1983) 241.

\bibitem{flux} T. Barnes, F. E. Close, and E. S. Swanson, Phys. Rev. D {\bf 52} (1995) 5242.
\bibitem{Cotanch:2001mc} S.~R.~Cotanch and F.~J.~Llanes-Estrada,
  Nucl.\ Phys.\ A {\bf 689} (2001) 481.

\bibitem{General:2006ed}
  I.~J.~General, S.~R.~Cotanch and F.~J.~Llanes-Estrada,
  Eur.\ Phys.\ J.\ C {\bf 51} (2007) 347
  [hep-ph/0609115].

\bibitem{LlanesEstrada:2000hj}
  F.~J.~Llanes-Estrada and S.~R.~Cotanch,
  Phys.\ Lett.\ B {\bf 504} (2001) 15
  [hep-ph/0008337].

\bibitem{bicudo} E. Abreu and P. Bicudo, J. Phys. G {\bf 34} (2007) 195207.

\bibitem{horn} D. Horn and J. Mandula, Phys. Rev. D {\bf 17} (1978) 898.

\bibitem{orsay} A. Le Yaouanc, L. Oliver, O. Pene, J.-C. Raynal, and S. Ono, Z. Phys. C {\bf 28} (1985) 309.

\bibitem{orsay2} F. Iddir, S. Safir, and O. Pene, Phys. Lett. B {\bf 433} (1998) 125.

\bibitem{hybrids1} Yu. A. Simonov, Nucl. Phys. B (Proc Suppl) {\bf 23} (1991) 283.

\bibitem{hybrids21} Yu. S. Kalashnikova and Yu. B. Yufryakov, Phys. Lett. B {\bf 359} (1995) 175.

\bibitem{hybrids22} Yu. A. Simonov, Phys. Atom. Nucl. {\bf 68} (2005) 1294.

\bibitem{yushybrdecay} Yu. S. Kalashnikova, Z. Phys. C {\bf 62} (1994) 323.

\bibitem{hybrids3} Yu. S. Kalashnikova and D. S. Kuzmenko, Phys. Atom. Nucl. {\bf 66} (2003) 955.

\bibitem{Kalashnikova:2008qr} Yu.~.S.~Kalashnikova and A.~V.~Nefediev,
Phys.\ Rev.\ D {\bf 77} (2008) 054025
[arXiv:0801.2036 [hep-ph]].

\bibitem{semay1} F. Buisseret, C. Semay, Phys.Rev. D {\bf 74} (2006) 114018.

\bibitem{semay2} F. Buisseret, V. Mathieu, C. Semay, B. Silvestre-Brac, Eur. Phys. J. A {\bf 32} (2007) 123.

\bibitem{kou} E. Kou and O. Pene, Phys. Lett. B {\bf 631} (2005) 164.

\bibitem{hybrdecayflux1} N. Isgur, R. Kokoski, and J. Paton, Phys. Rev. Lett. {\bf 54} (1985) 869.

\bibitem{hybrdecayflux2} F. E. Close and P. R. Page, Nucl. Phys. B {\bf 443} (1995) 233.

\bibitem{Dubynskiy:2008mq}
  S.~Dubynskiy and M.~B.~Voloshin,
  Phys.\ Lett.\ B {\bf 666} (2008) 344
  [arXiv:0803.2224 [hep-ph]].

\bibitem{Guo:2008zg}
  F.-K.~Guo, C.~Hanhart and U.-G.~Mei{\ss}ner,
  Phys.\ Lett.\ B {\bf 665} (2008) 26
  [arXiv:0803.1392 [hep-ph]].

\bibitem{tetraquark1} L. Maiani, F. Piccinini, A. D. Polosa, and V. Riquer, Phys. Rev. D {\bf 71} (2005) 014028.

\bibitem{Belle2pi} K. F. Chen et al. [Belle Collaboration], Phys. Rev. Lett. {\bf 100} (2008) 112001.

\bibitem{Artuso:2004fp}
  M.~Artuso {\it et al.}  [CLEO Collaboration],
  Phys.\ Rev.\ Lett.\  {\bf 94} (2005) 032001
  [hep-ex/0411068].

\bibitem{galuska} M. J. Galuska, undergraduate thesis submitted to the University of Giessen, 2008 (in German).

\bibitem{Badalian:2008ik}
  A.~M.~Badalian, B.~L.~G.~Bakker and I.~V.~Danilkin,
  Phys.\ Rev.\ D {\bf 79} (2009) 037505
  [arXiv:0812.2136 [hep-ph]].

\bibitem{Badalian:2009bu}
  A.~M.~Badalian, B.~L.~G.~Bakker and I.~V.~Danilkin,
  Phys.\ Atom.\ Nucl.\  {\bf 73} (2010) 138
  [arXiv:0903.3643 [hep-ph]].

\bibitem{Li:2009nr}
  B.~-QLi and K.~-T.~Chao,
  Commun.\ Theor.\ Phys.\  {\bf 52} (2009) 653
  [arXiv:0909.1369 [hep-ph]].

\bibitem{isgur2}S. Godfrey, N. Isgur, Phys. Rev. D {\bf 32} (1985) 189; S.
    Godfrey, R. Kokoski, {\it ibid.} D {\bf 43} (1991) 1679.

\bibitem{Kalashnikova:2001px}
Y.~.S.~Kalashnikova and A.~V.~Nefediev,
  Phys.\ Lett.\ B {\bf 530} (2002) 117
  [hep-ph/0112330].

\bibitem{Faustov}D. Ebert, V. O. Galkin, and R. N. Faustov, Phys. Rev. D {\bf 57}
(1998) 5663.

\bibitem{Bardeen} W.A. Bardeen, E.J. Eichten, and C.T. Hill, Phys. Rev. D {\bf 68}
(2003) 054024.

\bibitem{col} P. Colangelo, F. De Fazio, and R. Ferrandes, Nucl. Phys. (Proc.
Suppl.) {\bf 163} (2007) 177.

\bibitem{Falk} A.F. Falk and T. Mehen, Phys. Rev. D {\bf 53} (1996) 231.

\bibitem{lattice2} R. Lewis, R. M. Woloshyn, Phys. Rev. D {\bf 62} (2000) 114507.

\bibitem{Abreu:2011ic}
  L.~M.~Abreu, D.~Cabrera, F.~J.~Llanes-Estrada, J. M. Torres-Rincon,
  Annals Phys.\  {\bf 326} (2011) 2737
  [arXiv: 1104.3815 [hep-ph]].

\bibitem{Kolomeitsev:2003ac1}
  E.~E.~Kolomeitsev and M.~F.~M.~Lutz,
  Phys.\ Lett.\ B {\bf 582} (2004) 39
  [hep-ph/0307133].

\bibitem{Kolomeitsev:2003ac2}
  J.~Hofmann and M.~F.~M.~Lutz,
  Nucl.\ Phys.\ A {\bf 733} (2004) 142
  [hep-ph/0308263].

\bibitem{Guo:2006fu1}
  F.-K.~Guo, P.-N.~Shen, H.-C.~Chiang, R.-G.~Ping and B.-S.~Zou,
  Phys.\ Lett.\ B {\bf 641} (2006) 278
  [hep-ph/0603072].

\bibitem{Guo:2006fu2}
  F.-K.~Guo, P.-N.~Shen and H.-C.~Chiang,
  Phys.\ Lett.\ B {\bf 647} (2007) 133
  [hep-ph/0610008].

\bibitem{Abe:2007jn} K.~Abe {\it et al.}  [Belle Collaboration],
Phys.\ Rev.\ Lett.\  {\bf 98} (2007) 082001 [hep-ex/0507019].

\bibitem{Baranov:2004er}
  S.~P.~Baranov, V.~L.~Slad,
  Phys.\ Atom.\ Nucl.\  {\bf 67 } (2004)  808.
  [hep-ph/0603090].

\bibitem{Bjorken:1985ei}
  J.~D.~Bjorken, ``Is the CCC a New Deal for Baryon Spectroscopy?,''
FERMILAB-CONF-85-069, C85-04-20. Apr 1985.

\bibitem{ChenWu}
  Y.~Q.~Chen and S.~Z.~Wu,
  JHEP {\bf 1108} (2011) 144
  [Erratum-ibid.\  {\bf 1109} (2011) 089]
  [arXiv:1106.0193 [hep-ph]].

\bibitem{Brambilla:2009cd}
  N.~Brambilla, J.~Ghiglieri, A.~Vairo,
  Phys.\ Rev.\  {\bf D81 } (2010)  054031.
  [arXiv:0911.3541 [hep-ph]].

\bibitem{Flynn:2011gf}
  J.~M.~Flynn, E.~Hernandez and J.~Nieves,
  Phys.\ Rev.\ D {\bf 85} (2012) 014012
  [arXiv:1110.2962 [hep-ph]].

\bibitem{LlanesEstrada:2011kc}
  F.~J.~Llanes-Estrada, O.~I.~Pavlova and R.~Williams,
  Eur.\ Phys.\ J.\ C {\bf 72} (2012) 2019
  [arXiv:1111.7087 [hep-ph]].

\bibitem{BrodskyLepage}
  G. P. Lepage and S. J. Brodsky, Phys. Lett. B {\bf
  87} (1979) 359 ; {\it ibid.} Phys. Rev. D {\bf 22} (1980) 2157.

\bibitem{BrodskyFarrar} S. J. Brodsky and G. R. Farrar, Phys. Rev. Lett. {\bf 31} (1973)
1153.

\bibitem{ExpFormFactor1} J. Volmer {\it et al.}, Phys. Rev. Lett. {\bf 86} (2001) 1713.

\bibitem{ExpFormFactor2} T. Horn {\it et al.}, Phys. Rev. Lett. {\bf 97} (2006) 192001.

\bibitem{ExpFormFactor3} V. Tadevosyan {\it et al.}, Phys. Rev. C{\bf 75} (2007) 055205.

\bibitem{Gorchtein:2011xe}
  M.~Gorchtein, P.~Guo and A.~P.~Szczepaniak,
  arXiv: 1106.5252 [hep-ph]; arXiv:1102.5558 [nucl-th], Phys. Rev. C, in press.

\bibitem{Noguera:2012aw}
  S.~Noguera and V.~Vento,
  Eur.\ Phys.\ J.\ A {\bf 48} (2012) 143
  [arXiv:1205.4598 [hep-ph]].

\bibitem{Diehl:2000uv}
  M.~Diehl, T.~Gousset and B.~Pire,
  Phys.\ Rev.\ D {\bf 62} (2000) 073014
  [hep-ph/0003233];
 M.~Diehl {\it et al.}
 Phys.\ Rev.\ Lett.\  {\bf 81}, 1782 (1998)
 [hep-ph/9805380].

\bibitem{Ahmed:2011dd}
  A.~Ahmed,
  arXiv:1106.0740 [hep-ph].

\bibitem{Leibovich:1997az}
  A.~K.~Leibovich and I.~W.~Stewart,
  Phys.\ Rev.\ D {\bf 57}, 5620 (1998)
  [hep-ph/9711257].

\bibitem{Simonov:2011jc}
  Yu.~A.~Simonov,
  Phys.\ Rev.\ D {\bf 85} (2012) 105025
  [arXiv: 1109.5545 [hep-ph]].

\bibitem{Fleischer:2000pp}
  R.~Fleischer and D.~Wyler,
Phys.\ Rev.\ D {\bf 62} (2000) 057503 [hep-ph/0004010].

\bibitem{Meadows:2011bk}
B.~Meadows, M.~Blanke, A.~Stocchi, A.~Drutskoy, A. Cer\-velli, M.~Giorgi, A.~Lusiani and A.~Perez {\it et al.},
  arXiv:1109.5028 [hep-ex].

\bibitem{Vijande:2003ki}
  J.~Vijande, F.~Fernandez, A.~Valcarce and B.~Silvestre-Brac,
  Eur.\ Phys.\ J.\ A {\bf 19}, 383 (2004)
  [hep-ph/0310007].

\end{thebibliography}
\end{document}